\pdfoutput=1
\documentclass[reprint,amsmath,amssymb,prd,nofootinbib]{revtex4-1}

\usepackage{aas_macros}
\usepackage{graphicx}
\usepackage{dcolumn}
\usepackage{bm}
\usepackage{fontawesome}
\usepackage[normalem]{ulem}

\usepackage{xcolor}

\definecolor{colorLink}{rgb}{0.9,0,0} 
\definecolor{colorCite}{rgb}{0,0.7,0} 
\definecolor{colorURL} {rgb}{0,0,0.8} 
\usepackage[colorlinks=true,linktocpage=true,linkcolor=colorLink,citecolor=colorCite,urlcolor=colorURL]{hyperref}
\newcommand{\HI}{H\textsc{i}\,\,} 
\begin{document}

\title{Gas-rich dwarf galaxies as a new probe of dark matter interactions\\ with ordinary matter}

\author{Digvijay Wadekar}
 \email{jay.wadekar@nyu.edu}
\author{Glennys R. Farrar}
\email{gf25@nyu.edu}
\affiliation{
Center for Cosmology and Particle Physics, Department of Physics, New York University, New York, NY 10003, USA
}
\date{June 28, 2020}

\begin{abstract}
We use observations of gas-rich dwarf galaxies to derive constraints on dark matter scattering with ordinary matter. 
We require that heating/cooling due to DM interacting with gas in the Leo~T dwarf galaxy not exceed the ultra-low radiative cooling rate of the gas. This enables us to set $(i)$~stronger bounds than all the previous literature on ultra-light hidden photon DM for nearly all of the mass range $10^{-23}\lesssim m_\textup{DM} \lesssim 10^{-10}$ eV, $(ii)$~limits on sub-GeV millicharged DM which add to the constraints on the recent EDGES 21\,cm absorption anomaly, and $(iii)$~constraints on DM-baryon interactions directly at low relative velocities $v_\textup{rel}\sim 17$ km/s. Our study opens a new direction at using observations of gas-rich dwarf galaxies from previous, current and upcoming optical and 21cm surveys to probe physics beyond the standard model. \href{https://github.com/JayWadekar/Gas_rich_dwarfs}{\faGithub}



\end{abstract}
\maketitle

\section{Introduction}
The particle nature of Dark Matter (DM) and its origin is still a mystery.
The EDGES collaboration reported observations suggesting the temperature of gas during the cosmic dawn of the Universe was roughly half the expected value \cite{bowman18}. To explain the anomalous observation, Ref. \cite{barkana18} proposed that DM exchanges heat with the neutral hydrogen (H\textsc{i}) gas by non-standard Coulomb-like interactions of the form $\sigma \propto v_\textup{rel}^{-4}$ between DM and baryons. A physically motivated model that can lead to Coulomb-like DM gas interactions is the millicharged model \cite{milicharge85}. Because the relative velocity between DM and baryons at the cosmic dawn, $v_\textup{rel}$, is much lower ($\lesssim0.3$ km/s) than in standard astrophysical systems ($\gtrsim 200$ km/s in the Milky Way), the explanation by \cite{barkana18} is effective at cooling the gas during the cosmic dawn and simultaneously evading the traditional astrophysical bounds on DM.

Dwarf galaxies have low velocity dispersion and can therefore be used to constrain DM-baryon interactions directly at low $v_\textup{rel}$, $\sim \mathcal{O}(10 $ km/s). 
Nearby dwarf galaxies (e.g. Draco, Fornax) have been used as powerful probes of self interactions or decay of DM particles \cite{Col07,Bal07,Atw09,McC12,Zav13,Drl15_DES2,Reg15,Reg17,Cap18,Reg20}. However, they cannot be directly used to probe interactions of DM with ordinary matter due to the absence of gas inside them. 
This is due to gas being stripped in dwarfs located within the virial radius of large halos like the Milky Way, and is caused by pressure of the surrounding ionized medium (i.e. ram pressure stripping \cite{GunGot72,Grc09}). Recent high-resolution optical and 21cm surveys have been able to probe dwarfs beyond the virial radius of the Milky Way, and have discovered and characterized a large number of dwarfs which are gas-rich \cite{Wal08_Things,BegChe08_Figgs,Can11_Sheild,Ott12_Angst,Hun12_LittleThings,Ada13,Sau12,Sim19,Put21}.
 In this paper, we use a particular well-studied gas-rich dwarf galaxy called Leo~T;
we require that the DM heating/cooling rate not exceed the radiative cooling rate of the \HI gas in Leo~T and obtain, for the first time, constraints from dwarf galaxies on DM scattering with ordinary matter.  Our constraints on millicharged DM are complementary to the early universe ($z\gtrsim 1000$) constraints  \cite{DubGorRub04,McDermott11_millicharge,dvorkin14,boddy18,Kov18_MClimits,Xu18_CMB,Slatyer18_CMB,DavHanRaf00_BBN,KamKohTak17} (which are also at a low value of $v_{\rm rel}, \sim 30$ km/s \cite{TseHir10}), because the Leo~T constraints are not affected by the various assumptions about cosmology. 
Moreover, the systematics involved in the analysis of Leo~T are entirely different from the systematics in the early universe CMB/BBN analysis. 

In addition to using Leo~T to add to the constraints on DM interactions in the cosmic dawn and on millicharged DM more generally, we report constraints on ultra-light hidden photon DM (HPDM). In this scenario, the dark sector is comprised of vector bosons (no corresponding particle) in the ultra-light regime: $m_\chi \lesssim 10^{-11}$ eV. These bosons (called hidden photons) kinematically mix with the standard model photon. Such dark photons arise naturally in many theoretical setups and could account for the entire DM abundance \cite{AgaKitRee18,DroHarNar19,BasSanUba19, CoPieZha19, DubHer15,DarkPhotonCMB_Ari12,AnPosPra13_HPDM,DarkPhotonCMB_Jae10,GraMad16,AnPosPra15_HPDM,NelSch11,ChaGraIrw15,PigCleGui15_HPDM,Kov18_DPDM,ArvDub11, BhoBramante19, LonWan19,McDWit20,CapLiuMis20,CapLiuMis20b,GraKapMar18,WitRos20}. There are also planned experiments to test this mass regime such as DM Radio \cite{ChaGraIrw15,Silva-Feaver:2016qhh,HarnikSRF,GrassellinoSRF,PieRilZha18,BarHuaLas18,GraKapMar18,osti_1409838}.
 Ref.~\cite{DubHer15} showed that the hidden photons can induce an electric current in an astrophysical plasma. Dissipation of this current would cause heating of the gas in Leo~T, which enables us to place stronger constraints on HPDM than all the previous literature. 
 
Gas-rich dwarf galaxies like Leo~T have been a subject of large recent interest in 21cm and optical surveys, in order to populate the low-mass end of the baryonic Tully-Fisher relation and the galaxy luminosity function (to address the ``missing satellite problem" \cite{BucPet18}). Numerous past surveys have specifically targeted such galaxies:
 THINGS \cite{Wal08_Things}, FIGGS \cite{BegChe08_Figgs}, SHIELD \cite{Can11_Sheild}, VLA-ANGST \cite{Ott12_Angst} and LITTLE THINGS \cite{Hun12_LittleThings}, among many others. Ongoing and future 21cm surveys (like WALLABY \cite{Kor20_Wallaby} and SKA \cite{SKA}), and also optical surveys (like DES \cite{Kop15_DES,Bec15_DES,Drl15_DES,Drl20}, SAGA survey \cite{Geh17,Mao21}, DESI \cite{DESI}, HSC \cite{Aih18}, MSE \cite{McC16}, Rubin observatory \cite{LSST,Drl19_LSST} and the Roman telescope \cite{WFIRST}) will find and characterize an even larger number of these galaxies. This motivates finding new ways to use gas-rich dwarfs to probe physics beyond the standard model.

Apart from using dwarf galaxies to constrain the heat exchange due to DM-gas interactions, we also use diffuse neutral clouds in the Milky Way (MW).
The reason we select Leo~T and certain MW gas clouds for this study is because they have very low astrophysical radiative cooling rates; this makes them very sensitive to energy transfer by a non-standard source.
The radiative cooling rate of \HI gas decreases as the temperature, density, or metallicity of the gas decreases;  Leo~T has low metallicity, and the MW gas clouds that we analyze are cold ($T<500$ K).

A variety of MW gas clouds have already been used for constraining DM-gas interactions \cite{Chi90,DubHer15, BhoBramante18,BhoBramante19,BhoBra20}. However,
\citet{BhoBramante18} (B18 hereafter) chose clouds discovered by \cite{McClureLock13} which are entrained in the hot, high-velocity nuclear outflow (HVNO) --- a stream of gas at T $\sim 10^{6-7}$ K moving at $\sim 330$ km/s outward from the Galactic Center \cite{McClureLock13,DiTeodoroLock18}. Clouds in such an extreme environment cannot be assumed to be stable over the long timescales associated with their radiative cooling rates. The clouds are subject to a number of destructive effects due to their environment \cite{cooper08_WindSim, ScaBru15_WindSim, SchRob17_WindSim, ArmFra17_WindSim,MelGou13_WindSim,McCourt18_WindSim}, some of which occur at comparatively much shorter timescales (see Appendix~\ref{apx:OtherClouds} for a further discussion).
Therefore, we eschew use of the HVNO clouds and instead use the robustly observed gas clouds of \cite{PidLock15} which are in tranquil environments  co-rotating with the galactic disk and not close to the Galactic center.


\section{Bounds from DM heat exchange}

In an astrophysical system, let $\dot{Q}=\frac{dE}{dt dV}$ be the volumetric rate of energy transfer to the gas due to collisions of DM with different species (electrons, ions or atomic nuclei) present in the gas:
\begin{equation}
\dot{Q} = \sum_i \int d^3 v_i f(v_i) \int d^3 v_\chi f(v_\chi) n_i n_\chi v_{\textup{rel}}\, E_{\textup{T}} (v_\textup{rel})\, \sigma^T_{i\chi}\, ,
\end{equation}
where $n_i$ is the number density, $f(v_i)$ is the velocity distribution, $E_\textup{T}(v_\textup{rel})$ is the energy transferred in the DM scattering and $\sigma^T_{i\chi}$ is the cross section of the scattering species $i$.
Let $\dot{C}$ and $\dot{H}$  be the volumetric radiative cooling rate and the astrophysical heating rate, respectively. For a system to be in a steady state, we need $|\dot{Q}| = |\dot{C}-\dot{H}|$. In this paper, we set conservative bounds on the DM interaction cross section by requiring $|\dot{Q}|\leq \dot{C}$. Note that more stringent bounds could in principle be placed by including the astrophysical heating rate $|\dot{Q}|\leq |\dot{C}-\dot{H}|$. \smallskip

\emph{\underline{Radiative cooling}---}
To compute $\dot{C}$ for \HI gas in Leo~T and the MW clouds, we use
\begin{equation}
\dot{C}=n^2_\textup{H}\, \Lambda (T)\, ,
\label{eq:grackle}
\end{equation}
where $n_\textup{H}$ is the hydrogen number density and $\Lambda(T)$ is the cooling function which depends on the temperature and metallicity of the gas. We obtain $\Lambda(T)$ from the astrophysical radiative cooling library Grackle \cite{grackle}.
The cooling function scales with metallicity of \HI gas as $\Lambda(T) \propto 10^{[\textup{Fe}/\textup{H}]}$ where the metallicity is defined as: $[\textup{Fe}/\textup{H}] \equiv \log_{10} (n_\textup{Fe}/n_\textup{H})_\textup{gas} -\log_{10} (n_\textup{Fe}/n_\textup{H})_{\textup{Sun}}$. See Appendix~\ref{apx:AstroHeat} for further details on the astrophysical cooling and heating rates.

\section{Properties of astrophysical systems used}
\subsection{Leo~T galaxy}
Leo~T is an ultra-faint dwarf irregular galaxy located about 420 kpc from the center of the Milky Way. Leo~T is both dark-matter dominated and gas-rich, which makes it an ideal astrophysical system to study DM scattering with ordinary matter.
Moreover Leo~T is well-studied observationally and has garnered modelling attention.  Ref. \cite{leoObsOld08} analyzed high-resolution Giant Meterwave Radio Telescope (GMRT) and Westerbork Synthesis Radio Telescope (WSRT) observations to determine the \HI gas distribution.  Ref. \cite{Kir13} found the mean spectroscopic metallicity to be [Fe/H] $\sim -1.74 \pm 0.04$ \footnote{\cite{weisz12} found the isochronal metallicity [M/H] of Leo~T using photometric observations to be $\sim -1.8$ to $-1.6$. However, we use the more robust spectroscopic metallicity: [Fe/H]=$-1.74$ in our calculations.}; Leo~T is therefore metal poor (like the majority of ultra-faint dwarfs), and we need only consider H and He with the cosmic number density fraction $n_\textup{He}/n_\textup{H}=0.08$ for calculating DM-gas interactions.
\begin{figure}
\centering
\includegraphics[scale=0.6,keepaspectratio=true]{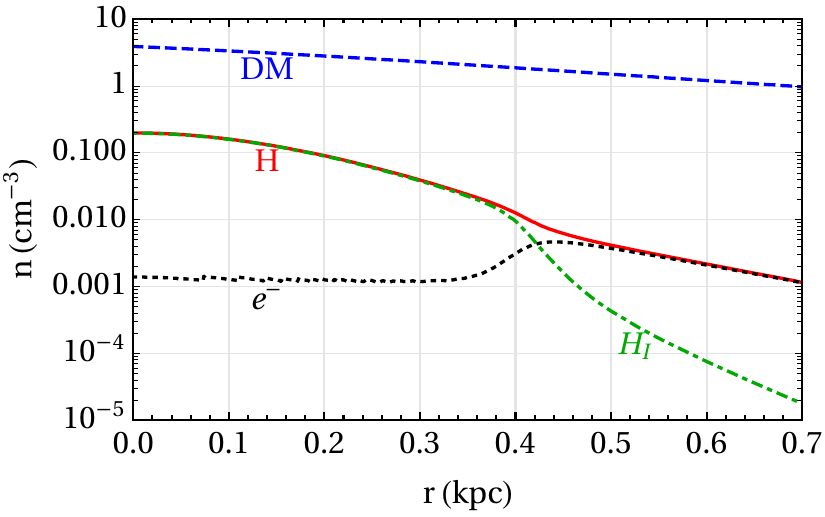}
\caption{Number density of DM (for $m_\chi=1$ GeV), atomic hydrogen (H\textsc{i}), electrons (e$^-$) and total hydrogen (H) components of the gas-rich dwarf galaxy Leo~T given by the model of \cite{faerman13}.} 
\label{fig:LeoTobs}
\vspace{-0.1in}
\end{figure}

Ref. \cite{faerman13} modelled the DM halo of Leo~T by fitting a well-motivated flat-core (Burkert) DM profile to the \HI distribution of \cite{leoObsOld08}.
We adopt  the model of \cite{faerman13} for our analysis and show the best-fit individual components in this model in Fig.~\ref{fig:LeoTobs}
(for reference, the halo mass is $M(r<0.3\, \textup{kpc}) \sim 8 \times 10^6\, M_\odot$). See Appendix~\ref{apx:LeoT} for a further discussion on the DM model.

Leo~T is one of the few dwarf galaxies for which the kinematics are well measured using both stars and gas. The rms line-of-sight velocity dispersions of both the stars (which are collisionless and trace the underlying gravitational potential) and the \HI gas are similar, $\sim$ 7 km/s (T $\simeq$ 6000 K) \cite{leoObsOld08,SimonGeha07,adams17}. Consequently the velocity dispersion of the gas and DM are equal ($v_\chi \simeq v_{\textup{H}_\textup{I}}$).  As a result, DM heats or cools the gas depending on the DM-proton mass difference because $T_\chi-T_{\textup{H}_\textup{I}}\simeq (m_\chi-m_p) v^2_{\textup{H}_\textup{I}}$; $T_{\textup{He}} = T_{\textup{H}}$ due to the large atomic cross-section. There is also no evidence of a substantial coherent magnetic field or recent supernova activity in Leo~T which could change the trajectories of DM particles if DM is charged. Note that this is a major advantage of using dwarf galaxies like Leo~T over large galaxies like the Milky Way or galaxy clusters, which have stronger magnetic fields potentially leading to a non-trivial distribution of charged DM \cite{chuzhoy09,McDermott11_millicharge,MunLoeb18,KadSek16_MilliConstraint,DunHal18_MilliConstraint}.

The outer part of Leo~T ($r>$ $0.35$ kpc) is ionized. It is difficult to find a robust measure of the rate at which the outer ionized region cools. Therefore, we restrict our study to the inner region $0<r<0.35$ kpc (designated \emph{Region-1} below), which is largely neutral and shielded from the metagalactic UV background. 
For calculating the bound on DM heat exchange in Leo~T,  we integrate the volumetric radiative cooling rate and the DM energy transfer rate over Region-1 and then require  $|\int_{\textup{Region-1}} \dot{Q} dV|\leq |\int_{\textup{Region-1}} \dot{C} dV|$. The gas in Region-1 has a very low $\dot{C}$ with the average value being $\sim 3.8 \times 10^{-30}$ erg cm$^{-3}$ s$^{-1}$.\\

\subsection{Milky Way gas clouds}
We use cores of diffuse neutral clouds which were observed at high Galactic latitudes with both the Very Large Array (VLA) and the Green Banks Telescope (GBT) \cite{PidLock15}.
These clouds are co-rotating with the MW disk and their mean velocity relative to DM is $\sim$220 km/s \cite{Lock02,Ford08,Ford10}.  They are at a distance $0.4-1$ kpc from the disk and are considered to
be representative of typical clouds and to have near-solar or lower metallicities \cite{Leh04,CloudMetal_Wak01,CloudMetal_Rich01,CloudMetal_Rich01b,CloudMetal_Semb04,CloudMetal_HerLock13}. 
Among the co-rotating clouds in \cite{PidLock15}, we find that the core of the G33.4$-$8.0 cloud gives the strongest constraint on DM interactions. G33.4$-$8.0 has $n_{\textup{H}_\textup{I}}$ = $0.4\pm0.1$ cm$^{-3}$, $T$ = $400\pm90$ K, and using $[\textup{Fe}/\textup{H}]\sim 0$ gives the average value of $\dot{C}$ to be $\sim 2.1 \times 10^{-27}$ erg cm$^{-3}$ s$^{-1}$. Further details on G33.4$-$8.0 and on other clouds which give similar bounds are in Appendix~\ref{apx:OtherClouds}.

\section{DM heat exchange rate}

In Leo~T, the heat exchange due to collisional DM-gas interactions is similar to a system of two fluids with different temperatures in thermal contact without any relative bulk velocity between them.  As noted earlier, for DM mass above (below) the proton mass, the DM heats (cools) the \HI gas.
In the case of MW clouds, in addition to the temperature difference, there is a high relative bulk velocity ($220$ km/s) between the gas and DM which leads to frictional heating of the two fluids. Let us now briefly discuss the heating mechanisms for specific DM models.

\subsection{Millicharge DM}
In the millicharged DM model, the DM particle effectively carries a small electric charge $Q=\epsilon e$, where $e$ is the charge on an electron \cite{milicharge85}. Such a model naturally leads to 
Coulomb-like DM-gas interactions $\sigma \propto v_\textup{rel}^{-4}$. There are two possibilities for the origin of such DM-gas interactions: a light $U(1)$ gauge boson (hidden photon) kinematically mixed with the Standard Model photon, or DM particle carrying a tiny charge. Our constraints are applicable for both scenarios.

There are two ways in which charged DM particles can interact with the Leo~T galaxy and the MW clouds. The first is that the DM particles interact with free electrons and ions. As seen in Fig.~\ref{fig:LeoTobs} for Leo~T, a fraction of Hydrogen is ionized even in Region-1 because of penetrating metagalactic background.
The MW cloud cores we use are at comparatively lower temperatures, so we conservatively consider that only carbon, silicon and iron in the clouds are ionized by the metagalactic background \cite{Maio07,BhooCool13}.
The second scenario arises when the de Broglie wavelength of DM becomes smaller than the screening length of the atom. Charged DM can then interact with the nucleus $A$ of neutral atoms in the astrophysical systems for nuclear recoil energy $E_\textup{nr} \gtrsim 1/(2m_A a^2_0)$ because the nucleus is screened over a distance $\sim a_0$ (Bohr radius). For DM in Leo~T interacting with the Hydrogen nucleus, this happens for $m_\chi \gtrsim 0.07$ GeV. The number density of neutral atoms in Leo~T is enough greater than that of electrons or ions, that the dominant contribution to heat exchange by DM for $m_\chi \gtrsim 0.07$ GeV comes from neutral H and He atoms. For details on the formalism of the heat exchange rate, see Appendices~\ref{apx:cross-section} and~\ref{apx:DMheat}.

\subsection{Ultra-light hidden-photon DM (HPDM)}
HPDM transfers heat to the gas in Leo~T in a different manner as compared to particle DM with collisional interactions. The hidden photons induce an oscillating electric field which accelerates the residual free electrons in the plasma of Leo~T. These electrons collide with ions and thus dissipate heat in the gas. We adopt the formalism of Ref.~\cite{DubHer15} for calculating the heating rate of the gas and refer the reader to Appendix~\ref{apx:HPDM} for further details.


\section{Results}

\begin{figure}
\centering
\includegraphics[scale=0.47,keepaspectratio=true]{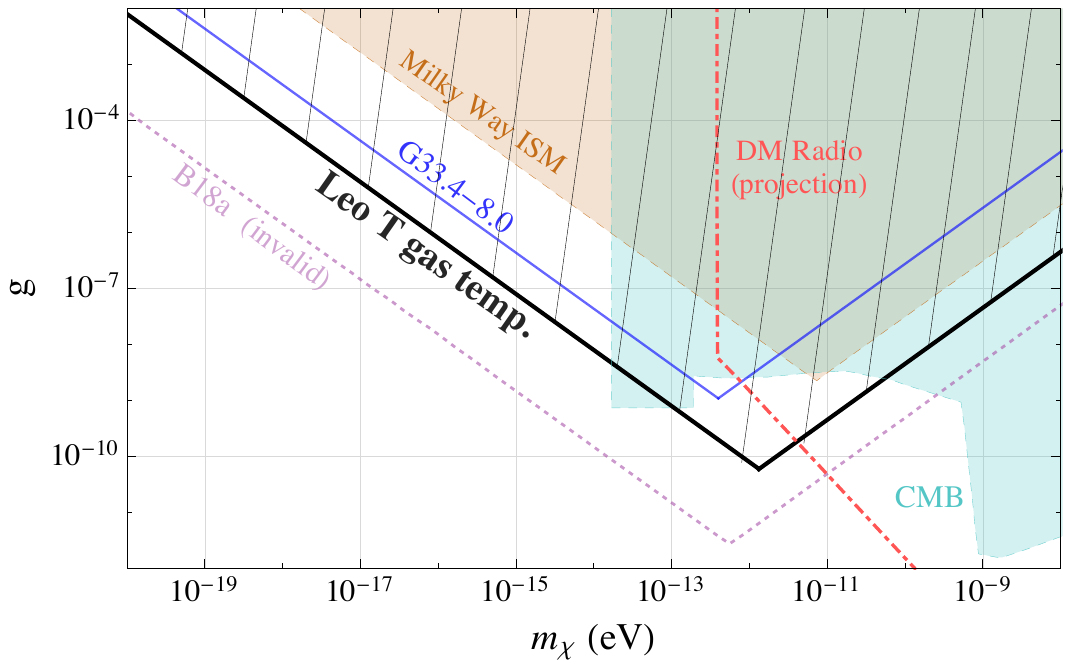}
\caption{Bounds on ultra-light hidden-photon DM with mass $m_\chi$ and kinetic mixing parameter $g$ from the Leo~T dwarf galaxy (black), the best MW co-rotating gas cloud, G33.4$-$8.0 (blue), heating of MW's interstellar medium (ISM) \cite{DubHer15}, CMB (cyan) \cite{DarkPhotonCMB_Ari12,DarkPhotonCMB_Jae10}. We also show the projected upper bound from stage 3 of the DM Radio experiment (red) \cite{ChaGraIrw15,Silva-Feaver:2016qhh,osti_1409838}.
The (invalid) constraint reported by \cite{BhoBramante19} based on HVNO clouds is shown by the purple dotted line (see Appendix~\ref{apx:OtherClouds} for details). \emph{Note added.}---Subsequent to our work, Refs. \cite{McDWit20,WitRos20,CapLiuMis20,CapLiuMis20b} appeared with strong limits on HPDM 
from anomalous heating of the intergalactic medium.}
\label{fig:ULDP}
\end{figure}

In both the aforementioned models, our analysis of Leo~T and robust MW gas clouds produces stringent new restrictions on the interaction strength or mixing parameter.  The limits for HPDM from Leo~T heating are shown in Fig.~\ref{fig:ULDP}.  As can be seen, these Leo~T limits are significantly more restrictive than any other constraint, for all DM masses below $10^{-10}$ eV, while for higher mass HPDM, CMB limits are the most constraining\footnote{The CMB limits are due to resonant conversion of dark photons into ordinary photons, and the limits disappear when the HPDM mass becomes smaller than the plasma mass of photons in the early universe.}. We reiterate that---for the reasons discussed earlier---HVNO clouds cannot be used to place limits, invalidating the reported limit of \cite{BhoBramante19}.  We have not shown the forecasts from future 21cm surveys of the cosmic dawn \cite{Kov18_DPDM}.

\begin{figure}
\centering
\includegraphics[scale=0.42,keepaspectratio=true]{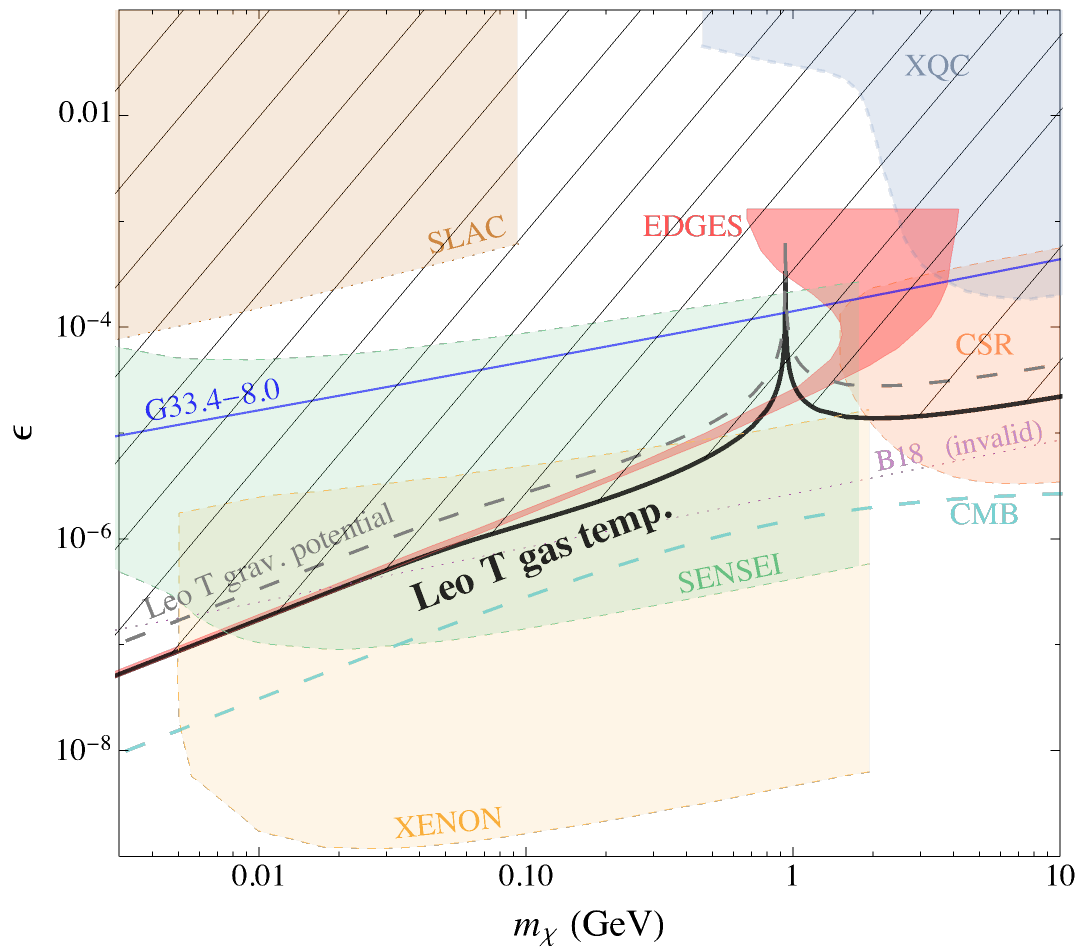}
\caption{Upper bound on the charge of DM from the Leo~T galaxy (black);  these constraints can be extrapolated to the left and right using $\epsilon_\textup{max}\sim10^{-4.7} (m/$GeV) and $\epsilon_\textup{max}\sim 10^{-5.2}\sqrt{(m/\textup{GeV})}$. The dashed grey line approximately shows the fundamental constraint of the very existence of the Leo~T DM halo. The blue line shows the upper bound from the best MW co-rotating gas cloud, G33.4$-$8.0, while the (invalid) constraint reported by B18 is shown by the purple dotted line.
We show in red the parameter range which explains the EDGES 21cm anomaly taken from \cite{Berlin18_MClimit} (see also \cite{barkana+18,MunLoeb18,Kov18_MClimits}). Note that the Leo~T bounds cannot be trivially extrapolated for the case when only a fraction of DM ($f_{\textup{mDM}}$) is charged; see the text for details and for the other constraints shown.}
\label{fig:epsilon}
\vspace{-0.1in}
\end{figure}

We also derive two new constraints on millicharged DM using Leo~T. 
The first qualitative constraint comes from the very existence of Leo~T's DM halo.  If $m_\chi<m_p$ and the DM-gas interactions are frequent enough, DM would evaporate rather than remaining gravitationally bound.  On the other hand, if $m_\chi>m_p$ and the interactions are strong enough, the DM profile would be visibly more concentrated than observed due to being cooled by its interactions with gas.  Proper evaluation of these constraints requires simultaneously modeling the DM and gas distributions which we leave for the future.  In lieu of that, we indicate with the dashed gray line of Fig.~\ref{fig:epsilon}, the value of $\epsilon$ such that the characteristic DM energy loss/gain time is comparable to the Leo~T lifetime, $(d {\rm ln} E/dt) ^{-1} \approx 10$ Gyr;  the millicharge coupling must be smaller than this.

The second constraint, which is stronger than the one from the existence of the DM halo, comes from the observed temperature of the \HI gas in Leo~T and is shown by the black solid line in Fig.~\ref{fig:epsilon}. This limit lies below the grey line, and gas cooling/heating rather than DM evaporation/collapse is the dominant effect.  The Leo~T temperature constraints are weakest when $m_\chi$ is close to $m_p$, where DM and gas are at similar temperatures and there is no heat exchange.   This is precisely the DM mass regime for which Earth can have a significant DM atmosphere (translating the constraints of \cite{nfm18} into the $\epsilon - m_\chi$ parameter space is underway \cite{XFinprep20}).  Fig.~\ref{fig:epsilon} also shows constraints taken from \cite{mahdawi18,SENSEI18} provided by the SLAC millicharge experiment \cite{SLAC98}, SENSEI \cite{SENSEI18}, XENON10 \cite{Xenon12,Xenon17}, XQC Rocket \cite{XQC02} and CRESST Surface Run (CSR) \cite{CRESST17}. The dashed cyan line labeled as CMB is our translation of the DM-proton interaction cross section limits of \cite{boddy18}. The XQC and CSR exclusion regions \cite{mahdawi18} assume a nuclear recoil thermalization efficiency of  2\% and are merely suggestive of the bounds that may be possible, because the efficiency still needs to be calibrated experimentally \cite{mahdawi18}.  

One source of uncertainty in our Leo~T bounds comes from the uncertainties in modeling its DM halo. Ref. \cite{faerman13} reports a range of fitted halo parameters with $3 \sigma$ errors; the parameter set within this range which gives the worst possible weakening of our bounds, changes our bounds on the mixing parameter $g$ or the DM charge $\epsilon$ only by $\lesssim 5$\%. The scenario where a fraction of dark matter ($f_{\textup{mDM}}$) is millicharged has recently gained popularity \cite{barkana+18,MunLoeb18,Kov18_MClimits,Berlin18_MClimit,boddy18,LiuOutRedVol19}. For $m_\chi > m_p$, the constraints from Leo~T in Fig.~\ref{fig:epsilon} can be scaled as $\sqrt{f_{\textup{mDM}}}$. In the case $m_\chi < m_p$ and $f_{\textup{mDM}}\ll1$, the millicharged subcomponent can interact with gas and evaporate from the halo, thereby evading our constraints.

 Stronger limits on DM could be set by modeling the observed \HI luminosity and temperature profiles self-consistently with the DM distribution, including the effects of astrophysical heating and cooling along with possible heat exchange due to DM-gas interactions.
Such self-consistent modeling would map out more precisely the constraints imposed by the existence of the observed gravitational potential, 
possibly leading to constraints on a subcomponent of light DM which evades current constraints (the case of $f_{\textup{mDM}}<1$ for example). 

\begin{figure}
\centering
\includegraphics[scale=0.42,keepaspectratio=true]{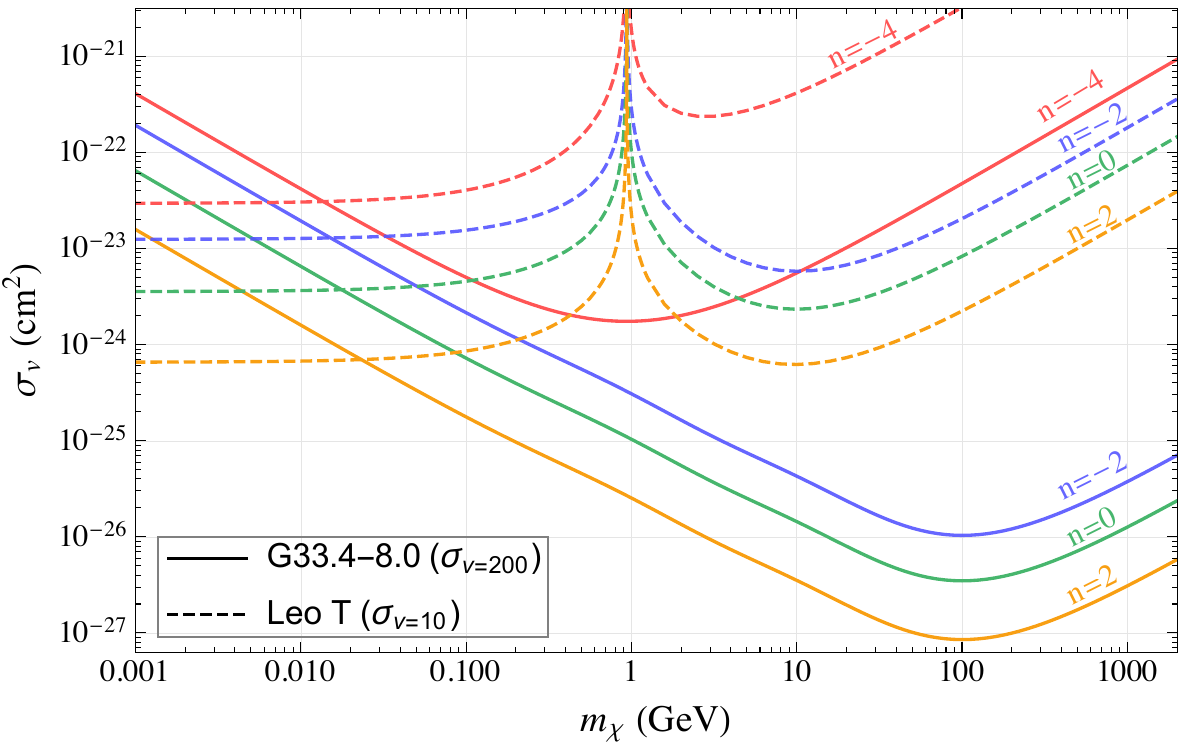}
\caption{Upper bounds on the DM-nucleon scattering cross-section from the gas temperature of the cloud G33.4$-$8.0 (solid) and Leo~T (dashed), for velocity dependence $\sigma_{n\chi} (v) = \sigma_{v_0} (v/v_0 )^n$ for $n= 0, \pm 2, -4$. Note that the cross section limits are shown at different velocities, namely the typical values constrained by the systems ($v_0 = 10$ km/s for Leo~T and $v_0 = 200$ km/s for the MW cloud). For $n=-4$, the bound from Leo~T for sub-GeV DM is comparable to the strongest bound in the current literature, which is from the CMB analysis by \cite{boddy18} (see Fig.~\ref{fig:sigma-4}).} 
\label{fig:sigma}
\vspace{-0.1in}
\end{figure}

It could be that DM interacts with ordinary matter but not through a photon. Therefore in Fig.~\ref{fig:sigma}, we show upper limits on the DM-nucleon momentum-transfer cross section for a variety of cross section dependences on relative velocity $\sigma_{\textup{N}\chi} (v_\textup{rel}) = \sigma_{v_0} (v_\textup{rel}/v_0)^n$.  The limits from Leo~T are shown at $v_0=$ 10 km/s, and those from G33.4$-$8.0 are shown at $v_0=$ 200 km/s.
We show the details of the calculations in Appendices~\ref{apx:cross-section} and~\ref{apx:DMheat}. We consider DM scattering with H and He for the case of Leo T. For the MW clouds, however, we also consider scattering with the elements \{Fe, O, Ne, Si\}. We use the Born approximation relation in Eq.~(\ref{eq:BornApprox}) for calculating the DM-nucleus cross section. Note that, for a Yukawa interaction, the cross section limits in some cases could fall in the non-perturbative regime, weakening the resulting limits by up to an order of magnitude \cite{XFinprep20,XFinprep20b}.
We have also probed the case of velocity independent and elastic DM-electron interactions, and we show the corresponding constraints in Fig.~\ref{fig:sigma_e}.

\section{Discussion and Conclusions}

Observational searches for dwarf galaxies in the local group and beyond have recently gained a huge momentum as a result of numerous high-precision surveys. 
Traditionally, observations of dwarf galaxies been used to probe DM self-annihilation or DM decay. We have shown here, for the first time, that dwarf galaxies can also be used to probe DM interactions with ordinary matter.
We used observations of a particular gas-rich dwarf galaxy called Leo~T and required that DM-gas interactions transfer energy at a lower rate than the gas radiative cooling rate. Leo~T sets stronger bounds than all the previous literature on ultra-light hidden photon DM for nearly all of the mass range $10^{-23}\lesssim m_\textup{DM} \lesssim 10^{-10}$ eV.  Our constraints on other models, e.g., millicharged DM, complement those of the CMB as the systematics and assumptions in our analysis are entirely different from those in the CMB analysis (e.g., the assumption of a quasi-linear prescription for the relative bulk velocity in the nonlinear Boltzmann calculation \cite{dvorkin14,boddy18,Kov18_MClimits,Xu18_CMB,Slatyer18_CMB}). 
The bounds on millicharged DM from Leo~T are also a valuable complement to direct detection experiments since they {\it i)} are independent of possible uncertainty in charged DM distribution due to magnetic fields from various sources in the Milky Way \cite{chuzhoy09,McDermott11_millicharge,MunLoeb18,DunHal18_MilliConstraint}, {\it ii)} directly probe much lower $v_\textup{rel}$ than previously tested, {\it iii)} avoid the problem for CRESST Surface Run and XQC that the efficiency of thermalization in the detectors has not yet been measured and must be calibrated experimentally before their results can be translated into DM limits \cite{mahdawi18}, and {\it iv)} are independent of the uncertainties in the velocity distribution and number density of DM at Earth \cite{necib+18,AntHel18,ManMajRen19}.
We also present first limits from robust Milky Way gas clouds, complementing Leo~T with greater sensitivity to higher DM-baryon relative velocities.

It is worth noting that, after our study, Leo T has been used to place stringent limits on gas heating due to primordial black holes (PBHs) \cite{LuTak21,Kim20,LahLu20,TakLu21}.  We also plan to report limits from Leo T on gas heating due to other DM candidates like Axion like particles (ALPs), primordial magnetic black holes and s-wave annihilation of DM in an upcoming work \cite{WadFarinprep20}. There have been recent observations of gas-rich dwarfs similar to Leo~T (e.g. Leo~P, Pheonix), and limits from them will also be reported. Furthermore, upcoming optical and 21cm surveys will find and characterize a much larger number of gas-rich dwarfs than present. Our analysis opens a new way of using their data to probe interactions of DM with ordinary matter.

The mathematica code associated with this paper and the data files for the plots are publicly available online \href{https://github.com/JayWadekar/Gas_rich_dwarfs}{\faGithub}.\footnote{\href{https://github.com/JayWadekar/Gas_rich_dwarfs}{\textcolor{blue}{https://github.com/JayWadekar/Gas\_rich\_dwarfs}}.}
\acknowledgments

We thank A. Sternberg, F. J. Lockman, C. McKee, M. S. Mahdawi, M. Baryakhtar, N. Outmezguine, A. Kravtsov, K. Van Tilburg, Y. Ali-Haimoud, E. Kovetz, K. Boddy, C. Mondino, M. Muzio, K. Schutz, A. Maccio, J. Ruderman, M. Kulkarni and S. Dubovsky for useful discussions, and Y. Faerman, X. Xu and  M. S. Mahdawi for contributions to the plots.
The research of GRF has been supported by NSF-PHY-1212538 and NSF-AST-1517319.

\appendix

\section{Properties of astrophysical systems used}
\label{apx:properties}
In the main text we reported the constraints on DM using Leo~T and MW gas clouds. In this section, we discuss the properties of these two systems in further detail and some of the caveats involved in the analysis.
\subsection{Leo~T galaxy}
\label{apx:LeoT}
We have adopted the model of the Leo~T DM halo by Ref.~\cite{faerman13} (FSM13 hereafter) for calculating the bounds on DM in this paper.
FSM13 assumed the \HI in Leo~T to be an isothermal gas sphere (T$\sim$6000 K) and adopted the flat-core (Burkert) profile for modeling the DM halo. The choice of the Burkert profile is motivated by observations of constant density cores in most of the low-mass dwarf galaxies.
FSM13 also assumed that the gas in Leo~T is hydrostatically supported as there is no evidence of coherent rotation in Leo~T. Low-mass dwarf galaxies like Leo~T as a class do not exhibit rotation \cite{BegChe06}, so we can discount the possibility of coherent rotation which eludes detection because $\hat{J}$ is along our line-of-sight. It is worth noting that, apart from the warm neutral medium, there is also a small amount of cold neutral medium in Leo~T at a much lower temperature ($\sim$500K) \cite{leoObsOld08};  including it in our analysis might make our bounds stronger but the cold medium has not been observed with appropriate resolution to be able to model it, so we ignore it in this study.

The more recent paper \cite{Patra18} (P18 hereafter) also models the DM halo of Leo~T but does not assume an isothermal model for the gas. One of the major differences between FSM13 and P18 is that P18 does not include effects of the metagalactic background \cite{hm12} in their analysis.
The metagalactic background determines the transition from neutral gas in the central region to ionized gas in the outer region of Leo~T and is therefore necessary for properly interpreting the \HI column density profile and also for determining the distribution of free electrons. We therefore adopt the FSM13 model in our analysis. Furthermore, the mass of the Leo~T DM halo in P18 is inconsistent with the halo mass from other Leo~T DM halo models in the literature \cite{StrBulKap08,faerman13,leoObsOld08,SimonGeha07,adams17}. 
A comprehensive modeling effort including the latest observations of \cite{adams17} and utilizing the iterative-unfolding technique of \cite{Patra18}, while also modeling the metagalactic ionization profile and known astrophysical heating and cooling, will result in a yet better description of Leo~T and enable more accurate determination of the limits on DM interactions using this system.
\subsection{Milky Way gas clouds}
\label{apx:OtherClouds}

\begin{figure*}
\centering
\includegraphics[scale=0.45,keepaspectratio=true]{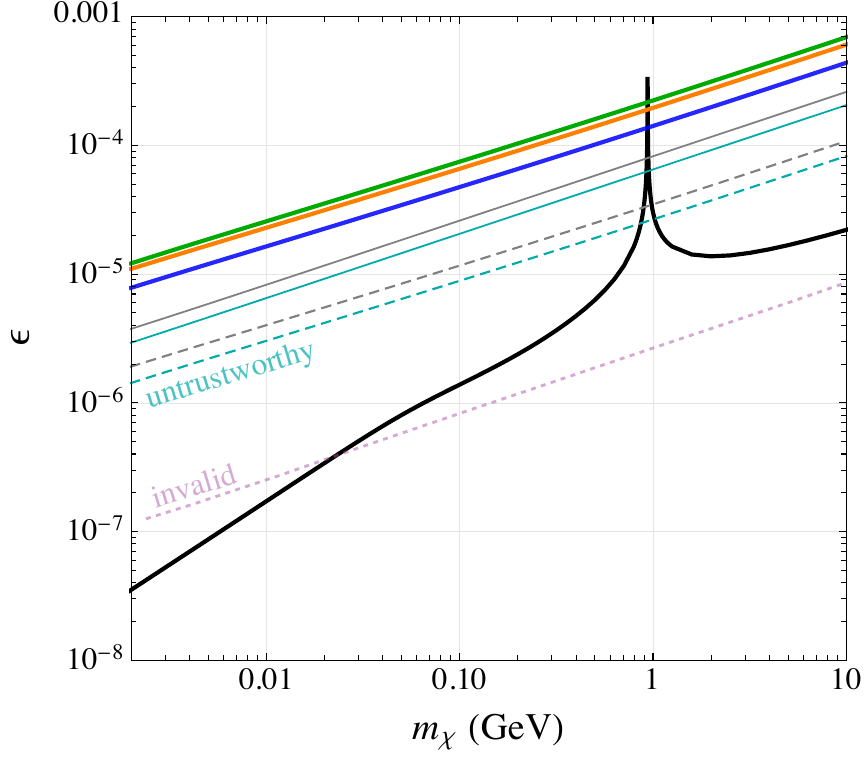}
\includegraphics[scale=0.45,keepaspectratio=true]{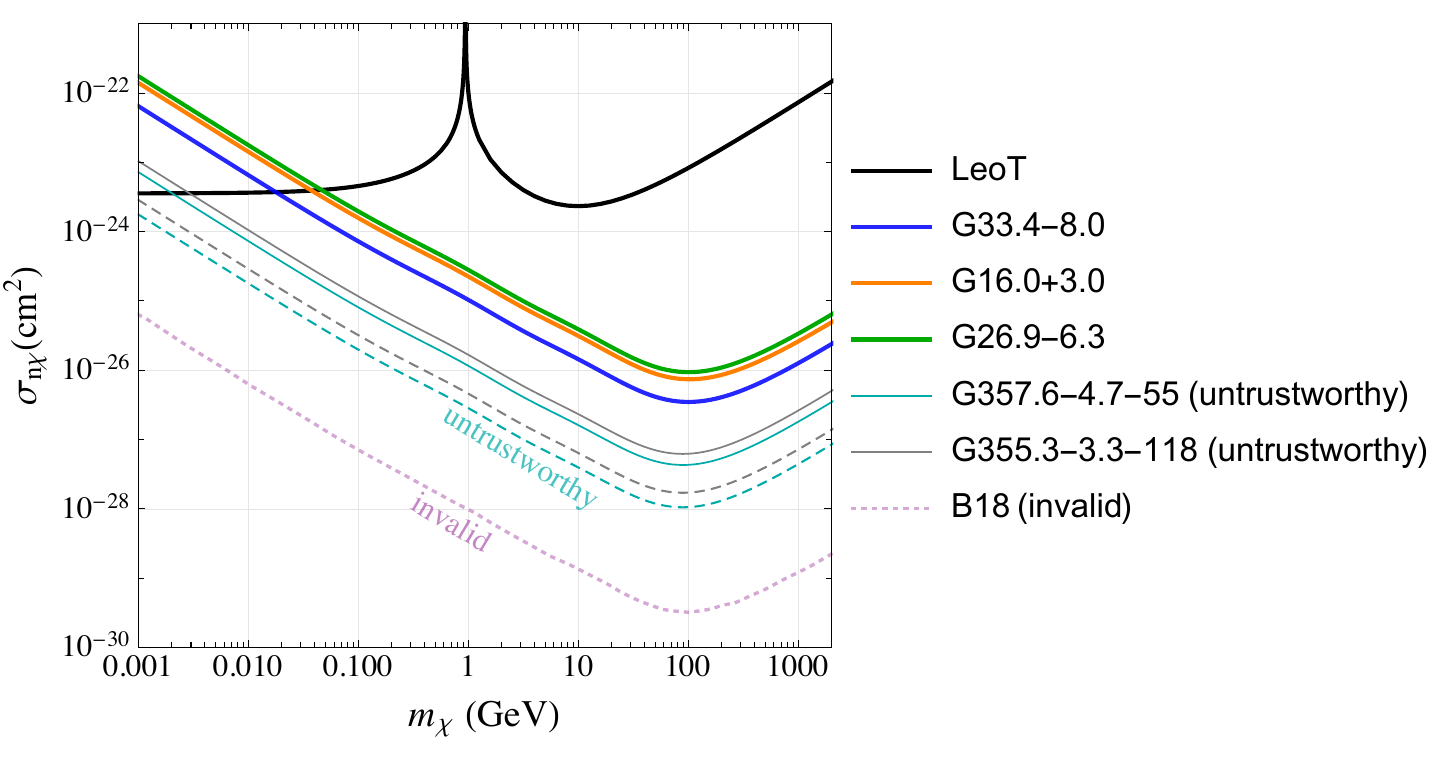}
\caption{Limits on $\epsilon$ (left panel) and the DM-nucleon velocity independent cross section $\sigma_{n\chi}$ (right panel).
The bound from Leo~T is shown in black and the bounds from the three best co-rotating MW gas clouds are shown as solid blue, orange, and green lines.  The bounds from the two best Galactic high velocity nuclear outflow clouds are shown as thin cyan and gray lines; the solid lines are the bounds when the best fit Burkert profile from \cite{Nes13_BurkertFit} is used for the DM halo and the dashed lines when the NFW profile from \cite{Galpy} is used. For the co-rotating clouds, the change in bounds when the Burkert profile is used instead of NFW is insignificant. The dotted line is the bound claimed by B18. Solar metallicity is assumed for all the clouds.}
\label{fig:AllClouds}
\end{figure*}

In the main text we reported the best robust constraints on DM-gas interactions using cooling rates of diffuse neutral clouds in the Milky Way (MW), based on G33.4$-$8.0.  In this section we discuss other clouds in addition to G33.4$-$8.0 which we have considered. The parameters for the most relevant clouds are reported in Table \ref{table:clouds}. For the parameters reported with error bars, we use the central values for calculating our results.

Among the clouds in \cite{PidLock15} which co-rotate with the MW disk and are therefore potentially long-lived, the cores of G26.9$-$6.3 and G16.0$+$3.0 give the next best constraints after G33.4$-$8.0. For all three cloud cores, the radiative cooling time ($\tau_\textup{cool} \lesssim 0.7$ Myr) is much smaller than the free-fall or dynamical time-scale ($\tau_{\textup{dynamic}} \sim 20-50$ Myr) \cite{PidLock15}. Therefore the cores can be taken to be in thermal equilibrium. The constraints on DM obtained from these clouds are shown in Fig.~\ref{fig:AllClouds}.

To find the local DM density at the position of these clouds, we adopt the NFW profile $\rho_\chi = \rho_0/[(r/r_0)(1+r/r_0)^2]$ with $\rho_0=0.32$ GeV/cm$^3$, scale radius $r_0=16$ kpc and virial radius $180$ kpc for the Milky Way DM halo from \cite{Galpy}. Using the best-fit Burkert profile from \cite{Nes13_BurkertFit} gives a similar but slightly higher value, so the NFW choice is conservative. We adopt an isotropic DM velocity dispersion profile from \cite{Hoeft04_DMdispersion}. The DM density and velocity dispersion at the location of G33.4$-$8.0 thus derived are 0.64 GeV/cm$^3$ and 124.4 km/s respectively. In the case when DM is charged, the magnetic fields in the Milky Way could cause charged DM to have a non-trivial distribution \cite{chuzhoy09,McDermott11_millicharge,MunLoeb18,DunHal18_MilliConstraint}. The bounds on charged DM derived using the Milky Way DM halo therefore suffer from some uncertainty. 
In the case of DM which interacts very strongly with ordinary matter, there is a possibility that DM would be unable to heat the inner regions of the MW clouds because the DM particles lose most of their energy on collisions with the outer layers \cite{BhoBramante19}.
We have checked that the DM bounds in the range shown in the plots in this paper are not affected by such shielding of the inner regions.
\begin{table}
\caption{\label{table:clouds}Parameters for all the analyzed clouds}
\begin{ruledtabular}
\begin{tabular}{llllll}
Cloud Name          & n (cm$^{-3}$)   & T (K)   & $R$ (kpc)   & $|z|$ (kpc) \\ \hline
G33.4$-$8.0       & 0.4$\pm$0.1 & 400$\pm$90 & 4.68$\pm$0.41 & 1$\pm$0.28    \\
G16.0$+$3.0       & 1.7$\pm$0.2 & 480$\pm$20 & 2.34$\pm$0.2 & 0.43$\pm$0.05 \\
G26.9$-$6.3       & 2.5$\pm$0.5 & 200$\pm$13 & 3.85$\pm$0.36 & 0.84$\pm$0.19 \\
G357.6-4.7-55\footnote{\label{note1}Clouds in the Galactic high velocity nuclear outflow.}   &0.43&136.4& 0.36  & 0.70  \\
G355.3-3.3-118\textsuperscript{\ref{note1}}  &0.24 &366.8& 0.72    & 0.48  \\
G1.4-1.8-87\textsuperscript{\ref{note1}}  &0.3 &15441\footnote{B18 uses the incorrect value 22 K for the temperature of G1.4$-$1.8+87, while the corrected value is 15441 K.}& 0.24    & 0.27  \\
\end{tabular}
\end{ruledtabular}
\end{table}\\

\begin{figure*}
\centering
\includegraphics[trim = 0.4in 0.1in 0.3in 0.2in,width=0.38\textwidth, keepaspectratio=true]{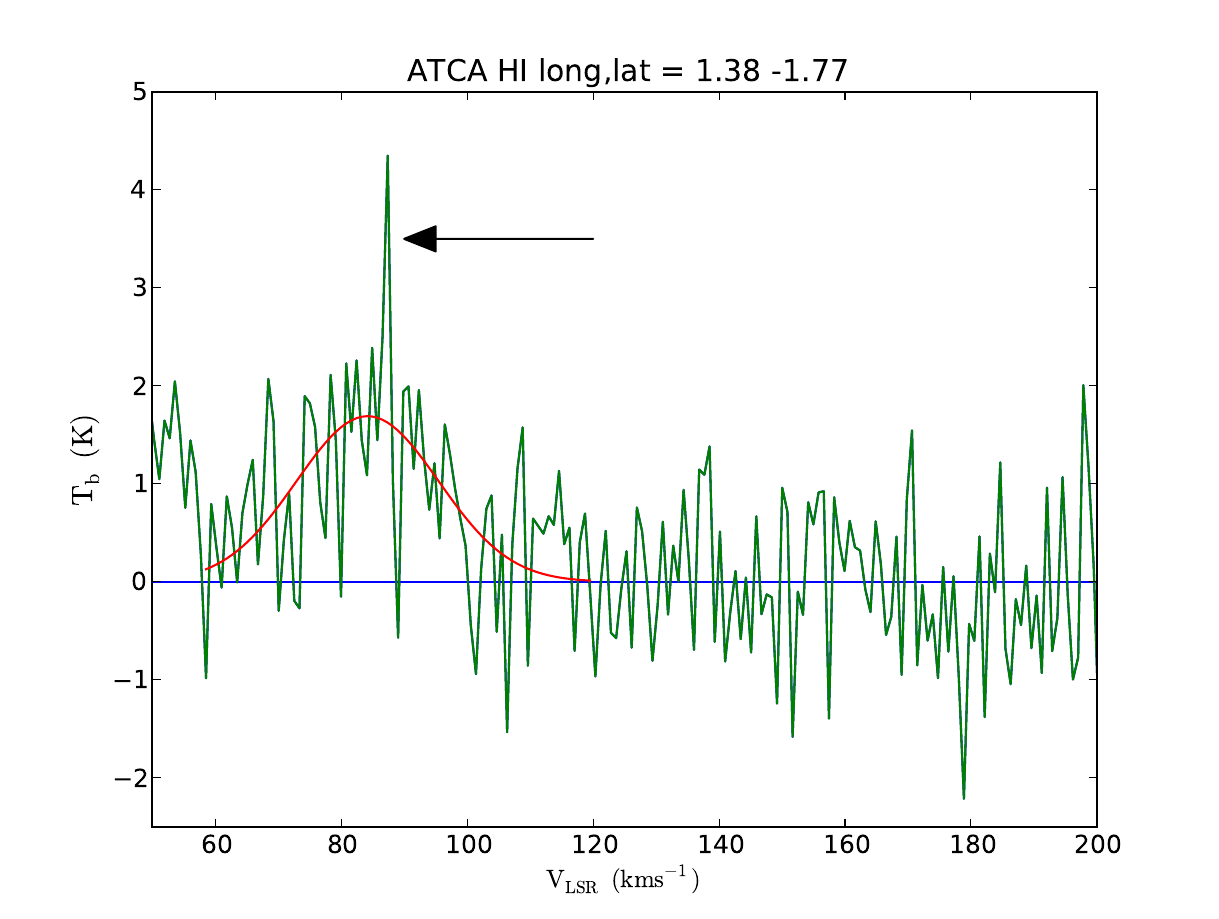}
\includegraphics[trim = 0.4in 0.1in 0.3in 0.1in,width=0.38\textwidth,keepaspectratio=true]{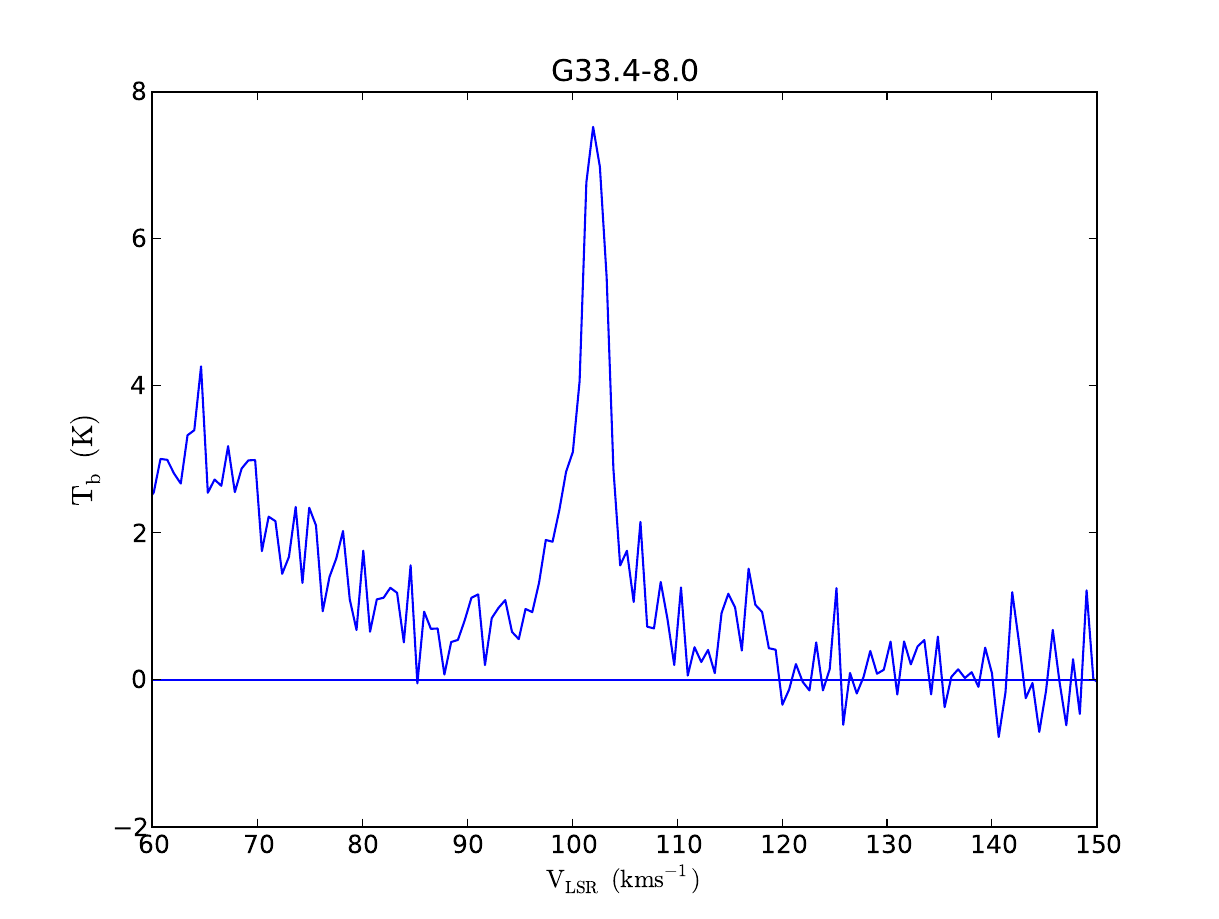}

\caption{{\it Top:} The \HI brightness temperature spectrum in the direction of G1.4-1.8+87. 
An arrow marks the extremely narrow line quoted in the table in \cite{McClureLock13}, while the
smooth curve shows a Gaussian fit to the emission
feature. {\it Bottom:} The corresponding spectrum for G33.4-8.0 \cite{PidLock15}. Figures and fit courtesy F. J. Lockman.}
\label{fig:CloudSpectra} 
\end{figure*}

\emph{\underline{HVNO clouds}---}
Compared to the co-rotating clouds, one could potentially obtain stronger constraints on DM using clouds which are near the Galactic center where the DM density is higher. A number of clouds entrained in the high velocity nuclear outflow (HVNO) originating in the Galactic center were discovered in \HI data from the Australia Telescope Compact Array (ATCA) and the Green Bank Telescope (GBT) \cite{McClureLock13, DiTeodoroLock18}.  Ref.~\cite{BhoBramante18} (B18) used the cloud G1.4$-$1.8+87 reported in \cite{McClureLock13} 
to constrain millicharge DM, but with incorrect parameters for the cloud. Correct parameters are listed in Table \ref{table:clouds}.   The temperature of G1.4$-$1.8+87 quoted in Table \ref{table:clouds} is determined by fitting the public online \HI brightness temperature spectrum data from \cite{MCGLock12}, shown in the top panel of Fig.~\ref{fig:CloudSpectra}.   For comparison, the spectrum of the co-rotating cloud G33.4$-$8.0 used in our analysis, is shown in the bottom panel of Fig.~\ref{fig:CloudSpectra}.


Correcting the parameters of G1.4$-$1.8+87, one finds that among the HVNO clouds, G357.6$-$4.7$-$55 and G355.3$-$3.3$-$118 would yield the strongest constraints. If the use of HVNO clouds for constraining DM were valid, the corresponding limits would be those shown by the thin solid cyan and grey lines in Fig.~\ref{fig:AllClouds}.  We also show for comparison, the limits reported by B18 based on incorrect parameters for G1.4$-$1.8+87. The correct limits on the DM cross section derived from that cloud are a factor $\sim 10^6$ weaker than reported by B18; for further details see \cite{comment}. 
The limits shown in Fig.~\ref{fig:AllClouds} are obtained assuming a uniform density profile for the clouds of \cite{McClureLock13,DiTeodoroLock18} and taking the clouds to be moving radially outward from the Galactic center with a speed of $\sim 330$ km/s, as inferred from simulations of the ensemble of clouds performed by \cite{DiTeodoroLock18}. Following  \cite{DiTeodoroLock18}, we deduce the 3D position of the cloud using the cloud's latitude and longitude, its radial velocity following the HVNO flow \cite{DiTeodoroLock18}, and the individual cloud's line-of-sight velocity toward the local standard of rest, $v_{\rm LSR}$. We give the cylindrical radial distance $R$ from the Galatic center and height $z$ from the disk in Table \ref{table:clouds}.

So close to the Galactic center, the DM density is quite uncertain.  An isothermal core  is more plausible than the NFW cusp. We adopt a recent parameterization of the Burkert profile $\rho_\chi=\rho_B/[(1+r/r_B)(1+r^2/r^2_B)]$ by \cite{Nes13_BurkertFit} with the core radius $r_B\sim 9.26$ kpc and the central density $\rho_B \sim 1.57$ GeV/cm$^3$.
Assuming an NFW profile would give stronger constraints shown by the dashed grey and cyan lines in Fig.~\ref{fig:AllClouds}, but would be neither well-motivated nor conservative.

However, there is strong reason to be skeptical about the general strategy of deriving bounds on DM from the HVNO clouds because such analysis requires the clouds to be stable at their current temperature over the long timescales associated with their radiative cooling rate. Being entrained in a high velocity, presumably turbulent wind, such clouds are subject to shocks, hydrodynamic instabilities (in particular, the Kelvin-Helmholtz and Rayleigh-Taylor instabilities), surface ablation and evaporation, hydrodynamic drag force and ram pressure due to the hot wind \cite{cooper08_WindSim, ScaBru15_WindSim, SchRob17_WindSim, ArmFra17_WindSim,MelGou13_WindSim}.

The radiative cooling timescales of G357.6$-$4.7$-$55 and G355.3$-$3.3$-$118 are $\sim 0.6-1.2$ Myr.   For a cloud of density $n_c$ and radius $r_c$ entrained in a wind with density $\rho_w$, hot phase velocity $V_{hw}$ and temperature $T_w$, the cloud crushing time \cite{cooper08_WindSim} is $t_\textup{cr} = \rho_c r_c/\rho_w V_{hw} \sim 0.2-0.4$ Myr for the two best HVNO clouds we consider, based on the hydrodynamical models of the wind in \cite{McClureLock13} ($n_w \sim 9 \times 10^{-3}$ cm$^{-3}$, $V_{hw} \sim 1600$ km/s and $T_w \sim 10^{6-7}$K).  Furthermore, for clouds entrained in a high-velocity hot wind, shocks occur on their surface at the temperature $T_\textup{cl,sh}=3/16\, T_w\, (n_w/n_c)$ \cite{cooper08_WindSim}. The cooling time of such shocked regions on the cloud surface is on the order of a few hundred years \cite{cooper08_WindSim,McClureLock13}. The shock cooling time is short compared to the cloud-shock crossing time and could cause the radiative shock driven into the cloud to form a high-density shell which might prevent further disruption \cite{cooper08_WindSim}. However, more recent studies suggest that radiative cooling causes pressure-confined gas clouds to fragment into cloudlets with sizes $0.1-1$ pc \cite{McCourt18_WindSim,Gronke18_WindSim,Sparre18_WindSim}. Thus the long-term stability of the HVNO clouds is subject to great uncertainty.

Note that the simple Cloudy \cite{cloudy} simulation of the HVNO clouds employed in \cite{BhoBramante19} ignores the effect on the clouds of the hot, high-velocity wind in their surroundings and assumes a nearly-uniform density profile for the clouds. Therefore it cannot be expected to accurately model the cloud properties.






\section{Astrophysical heating and cooling processes}
\label{apx:AstroHeat}

\begin{figure}
\centering
\includegraphics[scale=0.55,keepaspectratio=true]{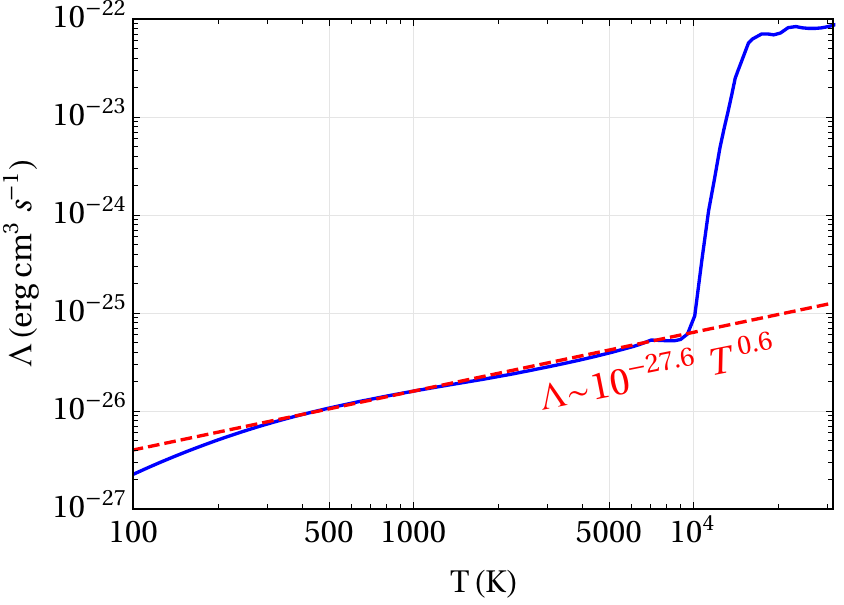}
\caption{The radiative cooling function from \cite{grackle} for \HI gas with solar metallicity (solid blue line). A convenient approximation for the cooling function $\Lambda(T)=10^{-27.6}\, T^{0.6}$ for $T\sim(300-8000)$ K is shown as the dashed line.}
\label{fig:grackle_rates}
\end{figure}

In this section we present additional details on the astrophysical radiative cooling and gas heating rates for the interested reader.\bigskip

\emph{\underline{Radiative cooling}---}
For calculating the radiative cooling rate of \HI gas, we use the chemistry and radiative cooling library for astrophysical simulations called Grackle \cite{grackle}. It allows us to include metal cooling and other collisional cooling effects and to take into account the ionization due to the metagalactic background \citep{hm12} and the self-shielding from the UV background \cite{Rahmati13}.
The cooling function $\Lambda(T)$ for a gas with solar metallicity is shown in Fig.~\ref{fig:grackle_rates}. For solar metallicity gas within the parameter domain $T\sim (300-8000)$ K and $n_\textup{H}\sim(0.1-10^3)$ cm$^{-3}$, the dominant contribution to radiative cooling comes from the fine structure transitions of oxygen, carbon, silicon and iron \cite{Maio07}. An approximation which we found applicable over most of the range of the cooling curve is also shown; it is  convenient for quickly assessing a cloud's likely utility for constraining DM interactions (roughly speaking the product $n_{\textup{H}_\textup{I}} \times 10^{[\textup{Fe}/\textup{H}]}\times T^{0.6}$ should be minimized for best constraints).  Note that we use the exact, solid curve from \cite{grackle} in our analysis for obtaining limits on DM interactions.\smallskip

\emph{\underline{Astrophysical heating}---}
The uniform background radiation comprised of the integrated emission from all the active nuclei and star-forming galaxies throughout the history of the universe is called the metagalatic background \cite{hm12}. The UV and X-ray component of the metagalactic background and the radiation from local stars causes photo-ionization heating in Leo~T and MW gas clouds. A substantial contribution to the heating of the MW clouds comes from photo-ejection of electrons due to UV radiation impinging on dust \cite{WolMck95,WolMckee03}. The dust content of the MW clouds is quite uncertain but can easily span the range required to balance the calculated cooling rate. Moreover, the high density of cosmic rays near the MW disk can also contribute to the heating of MW gas clouds owing to CR-gas collisions.

In Leo~T, the situation is different: Leo~T has low metallicity so its dust content is low and Leo~T is located far from the MW galactic center so the local cosmic ray density is also very low. The inner region of Leo~T has a very low free electron density so the X-ray Compton heating is sub-dominant  \cite{MadEfs99}. Furthermore, there is no reported AGN activity and very little on-going star formation which can contribute to any substantial ionizing radiation (SFR $< 10^{-5}$ M$_\odot$/yr for Leo~T \cite{weisz12}).

The primary source of astrophysical heating in Leo~T is due to photo-ionization heating of \HI gas due to radiation from local stars and the metagalactic background.
Let us first separately consider the heating due to the UV and X-ray components of the metagalactic background.
UV photons have high interaction cross section with \HI and because the \HI density in the \emph{Region-1} of Leo~T is high, the gas becomes optically thick to the UV radiation and is therefore largely self-shielded (the fitting functions of \cite{Rahmati13} give that only $\sim$ 0.2 \% of the UV photons penetrate \emph{Region-1}). The case is different for the X-ray photons which have relatively lower interaction cross section with \HI and therefore some of the X-ray photons penetrate the inner region of Leo~T and cause photoheating which further leads to some residual ionization ($\sim$ 2 \% of the gas as seen in the Fig.~1 of the main text). 
 
Another source of heating is the photo-ionizing radiation from old stars inside Leo~T \cite{Can10}; it can be calculated using the spectral energy distribution of a standard stellar profile from \cite{BruChar03}. However, the effect of local stars can dominate over the metagalactic background only for massive ellliptical galaxies and should be sub-dominant for Leo~T \cite{KanMac14}.  We have till now discussed all the ingredients needed to calculate the astrophysical heating rate for Leo~T, but an exact calculation requires using radiative transfer codes and is beyond the scope of this paper. We have checked that none of the effects mentioned in this section leads to a heating rate which is larger than the magnitude of the cooling rate ($|\dot{H}| > |\dot{C}|$). As the calculation of $|\dot{C}|$ is relatively much easier and more robust, we used $|\dot{C}|$ to put constraints on the DM heat exchange rate in this paper. If we include astrophysical heating in our analysis, we could get stronger bounds on DM interactions which is left to a future study.
\section{Momentum-transfer cross sections employed}
\label{apx:cross-section}

In this section, we provide a provide a formula for the transfer cross section for various scenarios in which a DM particle interacts with gas in the astrophysical systems.\bigskip

\emph{\underline{Charged DM interacting with plasma}---}
For a DM particle interacting with a charged particle $i$ present in the plasma, the differential Rutherford scattering cross section is
\begin{equation}
\frac{d\sigma_{i\chi}}{d\Omega_{\textup{CM}}} = \frac{Z_i^2\alpha^2_{\textup{em}}\epsilon^2}{4 \mu_i^2 v_{\textup{rel}}^4 \sin^4(\theta_{\textup{CM}}/2)}\, ,
\label{eq:rutherford}
\end{equation}
where $\mu_i$ is the DM-particle reduced mass, $Z_i$ is the particle charge, $n_i$ is the number density and $\theta_\textup{CM}$ is the scattering angle in the center of mass frame.
The ions in a thermal plasma are screened at distances greater than the Debye length
\begin{equation}
\lambda_D = \sqrt{\frac{T}{4\alpha_\textup{em}\pi (\sum_i Z^2_i n_i)}}\, .
\label{eq:Debye}
\end{equation}
The momentum-transfer cross section (defined in the first equality) is
\begin{equation}
\begin{split}
\sigma_{i\chi}^T(v)&\equiv \int_{\Omega^\textup{min}_\textup{CM}} d\Omega_\textup{CM} \frac{d\sigma_{i \chi}}{d\Omega_\textup{CM}} (1-\cos\theta_\textup{CM})\\
&=\frac{4 \pi Z^2_i \alpha^2_{\textup{em}}\epsilon^2 }{\mu_i^2 v_{\textup{rel}}^4} \ln \left(\frac{2 \mu_i v_\textup{rel}}{1/\lambda_D}\right)\, ,
\end{split}\label{eq:DMplasma_sigT}
\end{equation}
where the angular integral is cut-off when the scattering angle becomes the Debye angle $\theta^{\textup{min}}_\textup{CM} = 1/(\lambda_D\, \mu_i v_\textup{rel})$ for a non-relativistic plasma \cite{DubGorRub04,DvoLinSch19, Raffelt86,DavHanRaf00_BBN}.\bigskip

\begin{figure}
\centering
\includegraphics[scale=0.5,keepaspectratio=true]{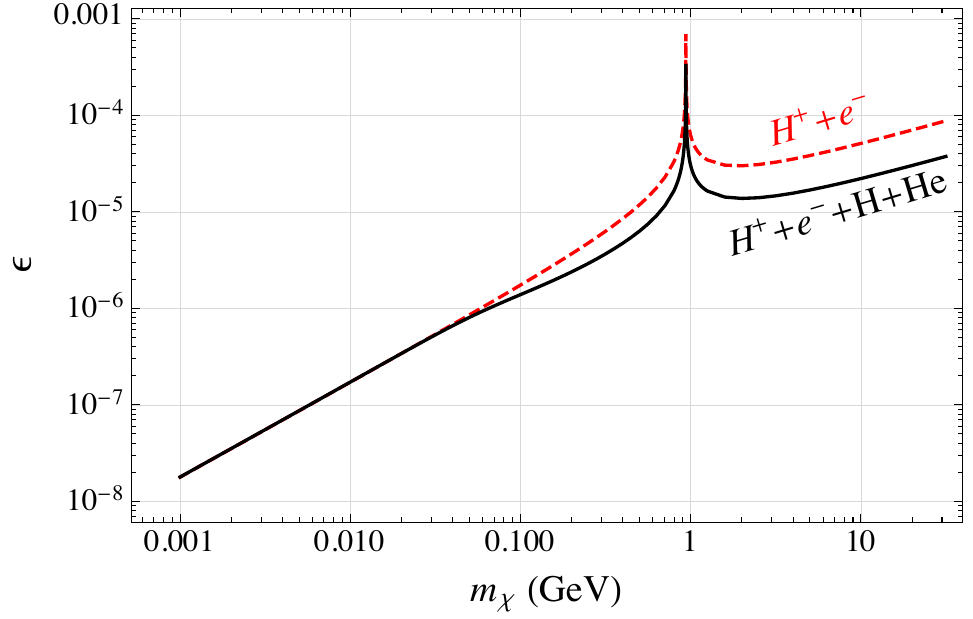}
\caption{The dashed red curve is the upper bound on millicharged DM obtained by considering only the DM interactions with the H$^+$ ions and free electrons in Leo~T. Upon including the interaction of DM with neutral H \& He atoms in Leo~T, we get a stronger upper bound as the solid black line.}
\label{fig:epsilon_NeutralAtom}
\end{figure}

\emph{\underline{Charged DM interacting with neutral atoms}---}
Following \cite{KouShoe14}, we use the screened Coulomb potential ($V = q_1q_2 e^{-r/a}/r$, where $q_{1,2}$ are the charges on DM and the nucleus) for incorporating the effect of screening. The nucleus $A$ is screened at distances greater than the Thomas-Fermi radius $a=0.8853$  $a_0/Z_A^{1/3}$, where $a_0$ is the Bohr radius and
the differential DM-nucleus cross section is
\begin{equation}
\frac{d\sigma_{A\chi}}{d E_\textup{nr}} = \frac{8\pi \alpha^2_{\textup{em}}\epsilon^2Z_A^2m_A a^4}{v^2(2 a^2 m_A E_\textup{nr}+1)^2}\, ,
\end{equation}
where $E_\textup{nr}$ is the nuclear recoil energy. Therefore,
\begin{equation}
\begin{split}
&\sigma_{A\chi}^T(v)\equiv \int_0^{E^{\textup{max}}_{\textup{nr}}} dE_\textup{nr} \frac{d\sigma_{A}}{dE_\textup{nr}} (1-\cos\theta_\textup{CM})\\
=&\frac{2\pi \alpha^2_{\textup{em}}\epsilon^2Z_A^2}{\mu_A^2 v_\textup{rel}^4}\left[\ln\left(1+4\mu_A^2 v_\textup{rel}^2a^2\right)-\frac{1}{1+(4\mu_A^2v_\textup{rel}^2a^2)^{-1}}\right]\, ,
\end{split}
\label{eq:DMneutral_sigT}
\end{equation}
where $E^{\textup{max}}_{\textup{nr}} = (2\mu_A v)^2 /2m_A$ is the maximum possible recoil energy.
A more refined way of treating the electron screening would be to use the form factor of the electron cloud surrounding the nucleus and not use the Born approximation (see \cite{LiuOutRedVol19} for a discussion). This should make our bound stronger but we use the conservative expression in Eq. (\ref{eq:DMneutral_sigT}) in this paper.
Although the cross section in Eq. (\ref{eq:DMplasma_sigT}) for charged DM interacting with ions/electrons is higher than that of neutral atoms, the number density of neutral atoms in Leo~T is much greater than that of electrons or ions in Region-I. Therefore including the neutral atoms as well as ions/electrons in Leo~T gives us a stronger bound as seen in Fig.~\ref{fig:epsilon_NeutralAtom}. \bigskip

\emph{\underline{DM-baryon interactions}---}
For a more general interaction of DM with nucleus $A$, we consider the momentum-transfer cross section to have a power law dependence of the form $\sigma^T_{i\chi}(v_\textup{rel})=\sigma_0 v_\textup{rel}^n$ where $v_\textup{rel}$ is the velocity of DM relative to baryons in units of $c$.
$n= -4$ arises for a massless mediator as in Rutherford scattering, $n=-2$ for DM having an electric dipole moment \cite{sigurdson_DMdipole}, and $n=0$ applies to much of the parameter space for Yukawa interactions (although there could be cases when Yukawa interaction produces more complex $v$-dependence not generally described by a simple power law \cite{TulYuZur13,Khr03,FarWanXu20,XFinprep20}).
 For $n=-4$, we assume the interaction is Coulomb-like so the DM-nucleon cross section scales similar to Eq. (\ref{eq:rutherford}), so 
\begin{equation}
\sigma_{A \chi} \simeq \left( \frac{Z_A}{Z_\textup{H}} \frac{\mu_\textup{H}}{\mu_A} \right)^2  \sigma_{\chi\textup{H}}\, ,
\end{equation}
where $Z$ is the atomic number.
For $n \in \{-2, 0, 2\}$, we assume a heavy mediator for the DM-nucleon interaction such that the cross section in Born approximation is given by 
\begin{equation}
\sigma_{A\chi}= \sigma_{\textup{N}\chi} \left( \frac{\mu_A}{\mu_p}\right)^2 A^2 F^2_A (E_\textup{nr})\, .
\label{eq:BornApprox}
\end{equation}
We adopt the nuclear form factor proposed by \cite{Helm56}
\begin{equation}
F_A (E_\textup{nr})=3\left(\frac{\sin(q r_A)-q r_A\cos(q r_A)}{(q r_A)^3}\right) e^{-s^2 q^2/2}\, ,
\end{equation}
where $q=\sqrt{2m_AE_\textup{nr}}$ is the momentum transfer and $r_A$ is the effective nuclear radius given by $r_A^2 = c^2 +\frac{7}{3}\pi^2 a^2 -5s^2$ with parameters $c\simeq (1.23 A^{1/3}-0.6)$ fm, $a\simeq 0.52$ fm and $s=0.9$ fm. We have neglected the nuclear form factor in Coulomb-like scattering due to smallness of the momentum transfer. 

\begin{figure}
\centering
\includegraphics[scale=0.47,keepaspectratio=true]{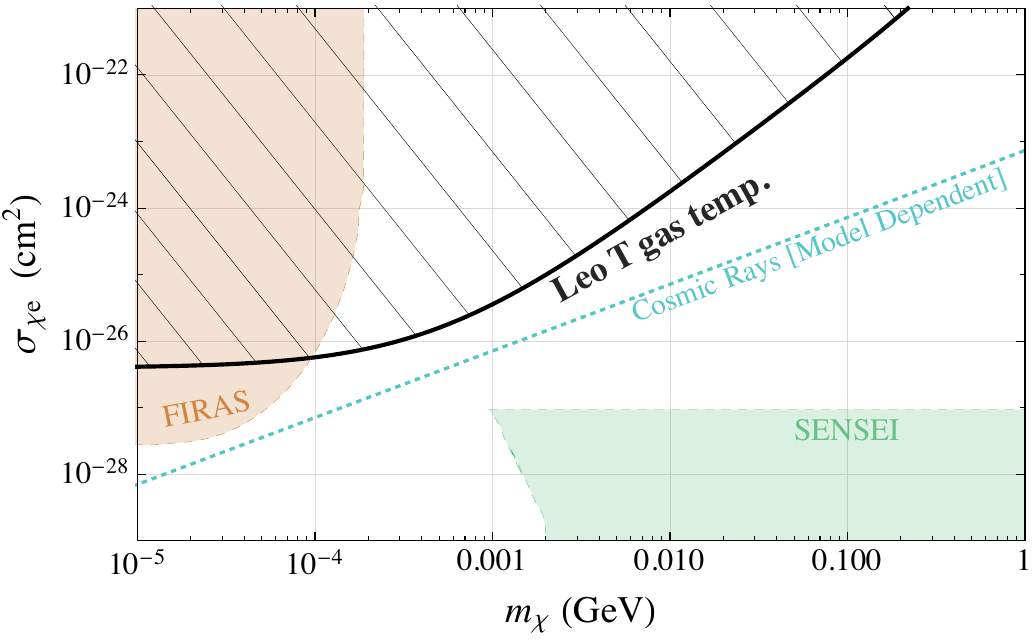}
\caption{Upper bounds on velocity independent DM-free electron scattering from Leo~T (black), FIRAS (brown) \cite{HaiChlKam15} and indirect, model-dependent bounds from Cosmic Rays \cite{CapNgBea18} (cyan). The exclusion region from SENSEI \cite{SENSEI18} is also shown.}
\label{fig:sigma_e}
\vspace{-0.1in}
\end{figure}

In a similar way as for DM-baryon scattering, we also compute limits on velocity independent and elastic DM-electron interactions and show the results in Fig.~\ref{fig:sigma_e}.

\section{DM heat exchange rate}
\label{apx:DMheat}
In the main text, we require that the DM-gas heat exchange rate is less than the gas radiative cooling rate and thus obtain stringent constraints on the DM interaction strength. In this section, we compute the heat exchange rate when particle DM interacts with gas.
The mechanism for heating of plasma by ultra-light hidden photon DM is different and will be considered in Appendix~\ref{apx:HPDM} separately. 
We consider that the momentum-transfer cross section for particle DM interactions
has the form $\sigma^T_{i\chi}(v_\textup{rel})=\sigma_0 v_\textup{rel}^n$, where $v_\textup{rel}$ is in units of c.
This is consistent with the charged DM scenario (Eqs. (\ref{eq:DMplasma_sigT}) and (\ref{eq:DMneutral_sigT})) because we can ignore the $v_\textup{rel}$ dependence in the argument of the logarithm to a good approximation. We therefore replace $v_\textup{rel}$ in the argument of logarithms in Eqs. (\ref{eq:DMplasma_sigT}) and (\ref{eq:DMneutral_sigT}) by an average value when computing the heat exchange rate. \bigskip

\begin{figure}
\centering
\includegraphics[scale=0.5,keepaspectratio=true]{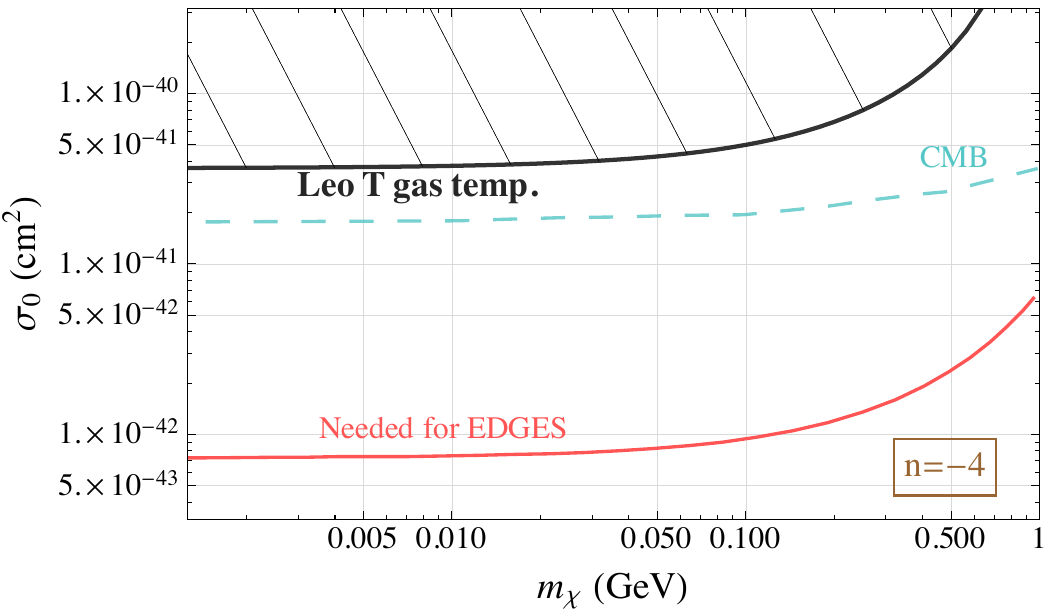}
\caption{The upper bound from Leo~T on DM-baryon interaction cross section corresponding to $n=-4$ is shown in black. The strongest current constraint is from CMB analysis by \cite{boddy18}, which is shown in cyan. The minimum cross section required to explain the EDGES 21cm anomaly (red) is taken from \cite{Kov18_MClimits}; it currently remains unexcluded.}
\label{fig:sigma-4}
\end{figure}

\emph{\underline{Leo~T}---}
In Leo~T, both DM and a component $i$ of the gas have the same Maxwell-Boltzmann velocity distributions due to being in equilibrium in the same gravitational potential, and correspondingly have temperatures $T_\chi$ and $T_i$ respectively. Energy is exchanged between them if $T_i \neq T_\chi$. For the case when there is no relative bulk velocity between gas and DM (as is the case for Leo~T), the rate of energy transfer to a component $i$ of the gas per unit time per unit volume is \cite{dvorkin14}
\begin{equation}
\begin{split}
\dot{Q}_i= \frac{2^{\frac{n+5}{2}}\Gamma(3+\frac{n}{2})}{\sqrt{\pi}(m_i+m_\chi)^2} \rho_i \rho_\chi \sigma_0\, (T_{\chi} - T_{i})\, u^{n+1}_\textup{th}\, ,
\end{split}
\label{eq:v^n_constraint}
\end{equation}
where $u^2_{\textup{th}} = T_\chi/m_\chi + T_i/m_i$ is the thermal sound speed of the DM-target fluid. We require that $|\dot{Q}_i|$ is less than the astrophysical heating or cooling rate. The bound from Leo~T on the cross section for $n=-4$ case is comparable to the strongest bound in the current literature, which is from the CMB analysis by \cite{boddy18}. We show the comparison of these bounds in Fig.~\ref{fig:sigma-4} and also show the parameter space needed to explain the EDGES anomaly. \bigskip

\emph{\underline{MW gas clouds}---}
Treatment of heat exchange in gas clouds is different than for Leo~T because, in addition to gas and DM being at different temperatures, they also have a non-zero relative bulk velocity between them. 
This bulk motion causes heating of gas and DM even if $T_g= T_\chi$.
The energy transferred to a particle $i$ in the cloud by a DM collision is $E_\textup{T}=\mu^2_{i\chi}v^2_{\textup{rel}} (1-\theta_\textup{CM})/m_i$, where $\theta_\textup{CM}$ is again the scattering angle in the center of mass frame.

For the millicharged case, using the cross sections from Eqs. (\ref{eq:DMplasma_sigT}) \& (\ref{eq:DMneutral_sigT}), the rate of energy transfer to an OM particle of type $i$ as a result of a DM interactions becomes $\dot{Q}_i \propto Z^2_i/m_i$. Therefore free electrons make the dominant contribution towards the energy exchange with millicharged DM in the MW gas clouds. The gas heating scales linearly with the free electron density and thus linearly with the cloud metal fraction. Because the radiative cooling rate also scales linearly with the metal fraction, the bounds on millicharged DM are not affected by the cloud metallicity. Additional ionization in the gas clouds due to cosmic ray scattering, dust and UV radiation from stars should also increase the heating and cooling rate together and therefore their inclusion should not have a significant effect on the millicharge DM bounds from clouds.

The gas clouds that we have considered have low temperatures $T_g\lesssim500$ K, which means that the velocity dispersion of atoms or ions in the clouds is $v \lesssim 2$ km/s which is much less than the relative bulk velocity $V_{\chi i}$ between DM and the cloud ($V_{\chi i}$ is 220 and 330 km/s for the co-rotating and the HVNO clouds respectively).
Thus we can neglect the velocity dispersion to make the following simplification when computing the heat exchange when DM scatters with nuclei or ions in the clouds:
\begin{equation}
\begin{split}
\dot{Q}_A &= \int d^3 v_A f(v_A) \int d^3 v_\chi f(v_\chi) n_A n_\chi  \frac{\mu^2_A}{m_A} \sigma_{0} v_\textup{rel}^{n+3}\\
&\simeq \int d^3 v_\chi f(v_\chi) n_A n_\chi  \frac{\mu^2_A}{m_A} \sigma_{0} v_\textup{rel}^{n+3}\, .
\end{split}
\end{equation}

For DM nucleon scattering in the case $n=-4$, metals in MW clouds have negligible contribution to DM heating whereas for $n \in \{-2, 0, 2\}$, metals in the MW clouds contribute dominantly to the DM heating for $m_\chi \gtrsim 10$ GeV. The bounds on DM nucleon cross section for $n \in \{-2, 0, 2\}$ \& $m_\chi \gtrsim 10$ GeV are independent of the cloud metallicity because both the radiative cooling rate and DM heating scale linearly with the metal fraction. While for the remaining cases ($n=-4$ or $m_\chi \lesssim 10$ GeV) the bounds would change by a factor $10^{[\textup{Fe}/\textup{H}]}$ for metallicity different from solar (not expected). 
The solar mass fractions for various elements are taken from \cite{solarAbun_Lod10} 
\begin{align}
r_A &= [r_H, r_{He},r_{O},r_{C},r_{Ne},r_{Fe},r_{Si}]\\ \nonumber
&=[0.739,0.2469,0.0063,0.0022,0.0017,0.0012,0.0007].\nonumber
\end{align}

For the case when DM scatters with electrons via Coulomb scattering ($n=-4$), the velocity dispersion of free electrons $v \lesssim 90$ km/s cannot be neglected as compared to the relative bulk velocity.
In such a case we need to generalize Eq. (\ref{eq:v^n_constraint}) to include the drag force between gas and DM alongside their thermal velocity distributions.
Such a generalized form of Eq. (\ref{eq:v^n_constraint}) for the $n=-4$ case is \cite{MunYacine15}:

\begin{equation}
\begin{split}
\dot{Q}_i = &\frac{\rho_i \rho_\chi \sigma_0}{(m_\chi+m_i)^2  u_{\textup{th}}}\\
&\times \left[\sqrt{\frac{2}{\pi}} \left(\frac{T_\chi-T_i}{u^2_{\textup{th}}}-m_\chi \right) \exp\left(-\frac{1}{2}\frac{V^2_{\chi i}}{u^2_{\textup{th}}}\right)\right.\\
&\ \ \ \ \ \ \left.+m_\chi\frac{ u_{\textup{th}}}{V_{\chi i}}\textup{Erf}\left(\frac{V_{\chi i}}{\sqrt{2}u_{\textup{th}}}\right)\right]\, ,
\end{split} \label{eq:DragEnergy}
\end{equation}
where $u^2_{\textup{th}} = T_\chi/m_\chi + T_i/m_i$ is the same thermal sound speed of the DM-target fluid as defined earlier.

\section{Heating due to ultra-light hidden photon DM}
\label{apx:HPDM}
We had earlier showed in Fig.~\ref{fig:ULDP} the constraints from Leo T for the case of HPDM.
In this section we present the formalism which was adopted from Ref.~\cite{DubHer15} and used in our calculations. We only show
 the relevant steps of the calculation here and we encourage the reader to refer to \cite{DubHer15} for further details. We consider that the hidden photon is kinematically mixed with the standard model photon and the coupling strength is determined by a dimensionless parameter $g$.
The resulting Lagrangian is 
\begin{equation}
\begin{split}
\mathcal{L} =& -\frac{1}{4}F_{\mu\nu}F^{\mu\nu} -\frac{1}{4}\tilde{F}_{\mu\nu}\tilde{F}^{\mu\nu} +\frac{m_\chi^2}{2}\tilde{A}_{\mu}\tilde{A}^{\mu}\\
&- \frac{e}{(1+g^2)^{1/2}}J^\mu\left(A_\mu + g\tilde{A}_\mu\right),
\end{split}
\label{Lagr}
\end{equation} 	
where $A_\mu$ ($\tilde{A}_\mu$) and  ${F}_{\mu\nu}$ ($\tilde{F}_{\mu\nu}$) stand for the visible (dark) photon gauge fields and field strengths respectively. 
As discussed in the main text, some part of Leo~T is ionized due to the penetrating metagalactic X-ray background \cite{hm12} and acts as a non-relativistic plasma, with the plasma frequency being
\begin{equation}
w_p = \sqrt{\frac{4\pi n_e \alpha}{m_e}}\, ,
\label{eq:PlasmaFreq}
\end{equation}
where $n_e$ is the number density of electrons. 
Very light hidden photons produce an oscillating electric field which induces an electric current in the ionized plasma in Leo~T. 
The induced current is dissipated because the free electrons which are accelerated by the oscillating electric field collide with ions. The frequency $\nu$ of electron-ion collisions is given by
\begin{equation}
\nu = \frac{4\sqrt{2\pi}\alpha^2n_e}{3m_e^{1/2}T_e^{3/2}} \ln\Lambda_C\, ,
\end{equation}
where $T_e$ is the electron temperature and the Coulomb logarithm is $\ln \Lambda_C=0.5 \log [4\pi T_e^3/(\alpha^3n_e)]$.
Due to the electron-ion collisions in the plasma, the hidden photon potential energy is transformed into the kinetic energy of charged particles in Leo~T
and the resulting heating rate of the gas in Leo~T per unit volume is given by
\begin{equation}
\dot{Q} = 2 |\gamma_h| \rho_\chi\, ,
\end{equation}
where $\rho_\chi$ is the hidden photon gravitational energy density in GeV/cm$^{-3}$ and 
$\gamma_h$ is the imaginary part of the frequency of the hidden photon modes ($w = w_h + i\gamma_h$) and is given by
\begin{equation}
\gamma_h = \left\lbrace\begin{array}{ll}
-\nu\frac{m_\chi^2}{2\omega_p^2}\frac{g^2}{1+g^2}  \text{   for $m_\chi \ll \omega_p$}\\
-\nu\frac{\omega_p^2}{2m_\chi^2}\frac{g^2}{1+g^2} \text{  for $m_\chi \gg \omega_p$}\;.
\end{array}\right. 
\end{equation}

\bibliographystyle{apsrev4-1}
\bibliography{leoT}

\begin{thebibliography}{163}%
\makeatletter
\providecommand \@ifxundefined [1]{%
 \@ifx{#1\undefined}
}%
\providecommand \@ifnum [1]{%
 \ifnum #1\expandafter \@firstoftwo
 \else \expandafter \@secondoftwo
 \fi
}%
\providecommand \@ifx [1]{%
 \ifx #1\expandafter \@firstoftwo
 \else \expandafter \@secondoftwo
 \fi
}%
\providecommand \natexlab [1]{#1}%
\providecommand \enquote  [1]{``#1''}%
\providecommand \bibnamefont  [1]{#1}%
\providecommand \bibfnamefont [1]{#1}%
\providecommand \citenamefont [1]{#1}%
\providecommand \href@noop [0]{\@secondoftwo}%
\providecommand \href [0]{\begingroup \@sanitize@url \@href}%
\providecommand \@href[1]{\@@startlink{#1}\@@href}%
\providecommand \@@href[1]{\endgroup#1\@@endlink}%
\providecommand \@sanitize@url [0]{\catcode `\\12\catcode `\$12\catcode
  `\&12\catcode `\#12\catcode `\^12\catcode `\_12\catcode `\%12\relax}%
\providecommand \@@startlink[1]{}%
\providecommand \@@endlink[0]{}%
\providecommand \url  [0]{\begingroup\@sanitize@url \@url }%
\providecommand \@url [1]{\endgroup\@href {#1}{\urlprefix }}%
\providecommand \urlprefix  [0]{URL }%
\providecommand \Eprint [0]{\href }%
\providecommand \doibase [0]{http://dx.doi.org/}%
\providecommand \selectlanguage [0]{\@gobble}%
\providecommand \bibinfo  [0]{\@secondoftwo}%
\providecommand \bibfield  [0]{\@secondoftwo}%
\providecommand \translation [1]{[#1]}%
\providecommand \BibitemOpen [0]{}%
\providecommand \bibitemStop [0]{}%
\providecommand \bibitemNoStop [0]{.\EOS\space}%
\providecommand \EOS [0]{\spacefactor3000\relax}%
\providecommand \BibitemShut  [1]{\csname bibitem#1\endcsname}%
\let\auto@bib@innerbib\@empty
\bibitem [{\citenamefont {Bowman}\ \emph {et~al.}(2018)\citenamefont {Bowman},
  \citenamefont {Rogers}, \citenamefont {Monsalve}, \citenamefont {Mozdzen},\
  and\ \citenamefont {Mahesh}}]{bowman18}%
  \BibitemOpen
  \bibfield  {author} {\bibinfo {author} {\bibfnamefont {J.~D.}\ \bibnamefont
  {Bowman}}, \bibinfo {author} {\bibfnamefont {A.~E.~E.}\ \bibnamefont
  {Rogers}}, \bibinfo {author} {\bibfnamefont {R.~A.}\ \bibnamefont
  {Monsalve}}, \bibinfo {author} {\bibfnamefont {T.~J.}\ \bibnamefont
  {Mozdzen}}, \ and\ \bibinfo {author} {\bibfnamefont {N.}~\bibnamefont
  {Mahesh}},\ }\href {\doibase 10.1038/nature25792} {\bibfield  {journal}
  {\bibinfo  {journal} {Nature}\ }\textbf {\bibinfo {volume} {555}},\ \bibinfo
  {pages} {67} (\bibinfo {year} {2018})}\BibitemShut {NoStop}%
\bibitem [{\citenamefont {Barkana}(2018)}]{barkana18}%
  \BibitemOpen
  \bibfield  {author} {\bibinfo {author} {\bibfnamefont {R.}~\bibnamefont
  {Barkana}},\ }\href {\doibase 10.1038/nature25791} {\bibfield  {journal}
  {\bibinfo  {journal} {Nature}\ }\textbf {\bibinfo {volume} {555}},\ \bibinfo
  {pages} {71} (\bibinfo {year} {2018})},\ \Eprint
  {http://arxiv.org/abs/1803.06698} {arXiv:1803.06698 [astro-ph.CO]}
  \BibitemShut {NoStop}%
\bibitem [{\citenamefont {Holdom}(1986)}]{milicharge85}%
  \BibitemOpen
  \bibfield  {author} {\bibinfo {author} {\bibfnamefont {B.}~\bibnamefont
  {Holdom}},\ }\href {\doibase 10.1016/0370-2693(86)91377-8} {\bibfield
  {journal} {\bibinfo  {journal} {Phys. Lett.}\ }\textbf {\bibinfo {volume}
  {166B}},\ \bibinfo {pages} {196} (\bibinfo {year} {1986})}\BibitemShut
  {NoStop}%
\bibitem [{\citenamefont {Colafrancesco}\ \emph {et~al.}(2007)\citenamefont
  {Colafrancesco}, \citenamefont {Profumo},\ and\ \citenamefont
  {Ullio}}]{Col07}%
  \BibitemOpen
  \bibfield  {author} {\bibinfo {author} {\bibfnamefont {S.}~\bibnamefont
  {Colafrancesco}}, \bibinfo {author} {\bibfnamefont {S.}~\bibnamefont
  {Profumo}}, \ and\ \bibinfo {author} {\bibfnamefont {P.}~\bibnamefont
  {Ullio}},\ }\href {\doibase 10.1103/PhysRevD.75.023513} {\bibfield  {journal}
  {\bibinfo  {journal} {Phys. Rev. D}\ }\textbf {\bibinfo {volume} {75}},\
  \bibinfo {pages} {023513} (\bibinfo {year} {2007})},\ \Eprint
  {http://arxiv.org/abs/astro-ph/0607073} {arXiv:astro-ph/0607073} \BibitemShut
  {NoStop}%
\bibitem [{\citenamefont {{Baltz}}\ \emph {et~al.}(2008)\citenamefont {{Baltz}}
  \emph {et~al.}}]{Bal07}%
  \BibitemOpen
  \bibfield  {author} {\bibinfo {author} {\bibfnamefont {E.~A.}\ \bibnamefont
  {{Baltz}}} \emph {et~al.},\ }\href {\doibase 10.1088/1475-7516/2008/07/013}
  {\bibfield  {journal} {\bibinfo  {journal} {\jcap}\ }\textbf {\bibinfo
  {volume} {2008}},\ \bibinfo {eid} {013} (\bibinfo {year} {2008})},\ \Eprint
  {http://arxiv.org/abs/0806.2911} {arXiv:0806.2911 [astro-ph]} \BibitemShut
  {NoStop}%
\bibitem [{\citenamefont {{Atwood}}\ \emph {et~al.}(2009)\citenamefont
  {{Atwood}} \emph {et~al.}}]{Atw09}%
  \BibitemOpen
  \bibfield  {author} {\bibinfo {author} {\bibfnamefont {W.~B.}\ \bibnamefont
  {{Atwood}}} \emph {et~al.},\ }\href {\doibase 10.1088/0004-637X/697/2/1071}
  {\bibfield  {journal} {\bibinfo  {journal} {\apj}\ }\textbf {\bibinfo
  {volume} {697}},\ \bibinfo {pages} {1071} (\bibinfo {year} {2009})},\ \Eprint
  {http://arxiv.org/abs/0902.1089} {arXiv:0902.1089 [astro-ph.IM]} \BibitemShut
  {NoStop}%
\bibitem [{\citenamefont {{McConnachie}}(2012)}]{McC12}%
  \BibitemOpen
  \bibfield  {author} {\bibinfo {author} {\bibfnamefont {A.~W.}\ \bibnamefont
  {{McConnachie}}},\ }\href {\doibase 10.1088/0004-6256/144/1/4} {\bibfield
  {journal} {\bibinfo  {journal} {\aj}\ }\textbf {\bibinfo {volume} {144}},\
  \bibinfo {eid} {4} (\bibinfo {year} {2012})},\ \Eprint
  {http://arxiv.org/abs/1204.1562} {arXiv:1204.1562 [astro-ph.CO]} \BibitemShut
  {NoStop}%
\bibitem [{\citenamefont {{Zavala}}\ \emph {et~al.}(2013)\citenamefont
  {{Zavala}}, \citenamefont {{Vogelsberger}},\ and\ \citenamefont
  {{Walker}}}]{Zav13}%
  \BibitemOpen
  \bibfield  {author} {\bibinfo {author} {\bibfnamefont {J.}~\bibnamefont
  {{Zavala}}}, \bibinfo {author} {\bibfnamefont {M.}~\bibnamefont
  {{Vogelsberger}}}, \ and\ \bibinfo {author} {\bibfnamefont {M.~G.}\
  \bibnamefont {{Walker}}},\ }\href {\doibase 10.1093/mnrasl/sls053} {\bibfield
   {journal} {\bibinfo  {journal} {\mnras}\ }\textbf {\bibinfo {volume}
  {431}},\ \bibinfo {pages} {L20} (\bibinfo {year} {2013})},\ \Eprint
  {http://arxiv.org/abs/1211.6426} {arXiv:1211.6426 [astro-ph.CO]} \BibitemShut
  {NoStop}%
\bibitem [{\citenamefont {{Drlica-Wagner}}\ \emph
  {et~al.}(2015{\natexlab{a}})\citenamefont {{Drlica-Wagner}} \emph
  {et~al.}}]{Drl15_DES2}%
  \BibitemOpen
  \bibfield  {author} {\bibinfo {author} {\bibfnamefont {A.}~\bibnamefont
  {{Drlica-Wagner}}} \emph {et~al.},\ }\href {\doibase
  10.1088/2041-8205/809/1/L4} {\bibfield  {journal} {\bibinfo  {journal}
  {\apjl}\ }\textbf {\bibinfo {volume} {809}},\ \bibinfo {eid} {L4} (\bibinfo
  {year} {2015}{\natexlab{a}})},\ \Eprint {http://arxiv.org/abs/1503.02632}
  {arXiv:1503.02632 [astro-ph.HE]} \BibitemShut {NoStop}%
\bibitem [{\citenamefont {{Regis}}\ \emph {et~al.}(2015)\citenamefont
  {{Regis}}, \citenamefont {{Richter}}, \citenamefont {{Colafrancesco}},
  \citenamefont {{Profumo}}, \citenamefont {{de Blok}},\ and\ \citenamefont
  {{Massardi}}}]{Reg15}%
  \BibitemOpen
  \bibfield  {author} {\bibinfo {author} {\bibfnamefont {M.}~\bibnamefont
  {{Regis}}}, \bibinfo {author} {\bibfnamefont {L.}~\bibnamefont {{Richter}}},
  \bibinfo {author} {\bibfnamefont {S.}~\bibnamefont {{Colafrancesco}}},
  \bibinfo {author} {\bibfnamefont {S.}~\bibnamefont {{Profumo}}}, \bibinfo
  {author} {\bibfnamefont {W.~J.~G.}\ \bibnamefont {{de Blok}}}, \ and\
  \bibinfo {author} {\bibfnamefont {M.}~\bibnamefont {{Massardi}}},\ }\href
  {\doibase 10.1093/mnras/stv127} {\bibfield  {journal} {\bibinfo  {journal}
  {\mnras}\ }\textbf {\bibinfo {volume} {448}},\ \bibinfo {pages} {3747}
  (\bibinfo {year} {2015})},\ \Eprint {http://arxiv.org/abs/1407.5482}
  {arXiv:1407.5482 [astro-ph.GA]} \BibitemShut {NoStop}%
\bibitem [{\citenamefont {{Regis}}\ \emph {et~al.}(2017)\citenamefont
  {{Regis}}, \citenamefont {{Richter}},\ and\ \citenamefont
  {{Colafrancesco}}}]{Reg17}%
  \BibitemOpen
  \bibfield  {author} {\bibinfo {author} {\bibfnamefont {M.}~\bibnamefont
  {{Regis}}}, \bibinfo {author} {\bibfnamefont {L.}~\bibnamefont {{Richter}}},
  \ and\ \bibinfo {author} {\bibfnamefont {S.}~\bibnamefont
  {{Colafrancesco}}},\ }\href {\doibase 10.1088/1475-7516/2017/07/025}
  {\bibfield  {journal} {\bibinfo  {journal} {\jcap}\ }\textbf {\bibinfo
  {volume} {2017}},\ \bibinfo {eid} {025} (\bibinfo {year} {2017})},\ \Eprint
  {http://arxiv.org/abs/1703.09921} {arXiv:1703.09921 [astro-ph.HE]}
  \BibitemShut {NoStop}%
\bibitem [{\citenamefont {{Caputo}}\ \emph {et~al.}(2018)\citenamefont
  {{Caputo}}, \citenamefont {{Garay}},\ and\ \citenamefont {{Witte}}}]{Cap18}%
  \BibitemOpen
  \bibfield  {author} {\bibinfo {author} {\bibfnamefont {A.}~\bibnamefont
  {{Caputo}}}, \bibinfo {author} {\bibfnamefont {C.~P.}\ \bibnamefont
  {{Garay}}}, \ and\ \bibinfo {author} {\bibfnamefont {S.~J.}\ \bibnamefont
  {{Witte}}},\ }\href {\doibase 10.1103/PhysRevD.98.083024} {\bibfield
  {journal} {\bibinfo  {journal} {\prd}\ }\textbf {\bibinfo {volume} {98}},\
  \bibinfo {eid} {083024} (\bibinfo {year} {2018})},\ \Eprint
  {http://arxiv.org/abs/1805.08780} {arXiv:1805.08780 [astro-ph.CO]}
  \BibitemShut {NoStop}%
\bibitem [{\citenamefont {{Regis}}\ \emph {et~al.}(2020)\citenamefont
  {{Regis}}, \citenamefont {{Taoso}}, \citenamefont {{Vaz}}, \citenamefont
  {{Brinchmann}}, \citenamefont {{Zoutendijk}}, \citenamefont {{Bouch{\'e}}},\
  and\ \citenamefont {{Steinmetz}}}]{Reg20}%
  \BibitemOpen
  \bibfield  {author} {\bibinfo {author} {\bibfnamefont {M.}~\bibnamefont
  {{Regis}}}, \bibinfo {author} {\bibfnamefont {M.}~\bibnamefont {{Taoso}}},
  \bibinfo {author} {\bibfnamefont {D.}~\bibnamefont {{Vaz}}}, \bibinfo
  {author} {\bibfnamefont {J.}~\bibnamefont {{Brinchmann}}}, \bibinfo {author}
  {\bibfnamefont {S.~L.}\ \bibnamefont {{Zoutendijk}}}, \bibinfo {author}
  {\bibfnamefont {N.~F.}\ \bibnamefont {{Bouch{\'e}}}}, \ and\ \bibinfo
  {author} {\bibfnamefont {M.}~\bibnamefont {{Steinmetz}}},\ }\href@noop {}
  {\bibfield  {journal} {\bibinfo  {journal} {arXiv e-prints}\ ,\ \bibinfo
  {eid} {arXiv:2009.01310}} (\bibinfo {year} {2020})},\ \Eprint
  {http://arxiv.org/abs/2009.01310} {arXiv:2009.01310 [astro-ph.CO]}
  \BibitemShut {NoStop}%
\bibitem [{\citenamefont {{Gunn}}\ and\ \citenamefont
  {{Gott}}(1972)}]{GunGot72}%
  \BibitemOpen
  \bibfield  {author} {\bibinfo {author} {\bibfnamefont {J.~E.}\ \bibnamefont
  {{Gunn}}}\ and\ \bibinfo {author} {\bibfnamefont {I.}~\bibnamefont {{Gott}},
  \bibfnamefont {J.~Richard}},\ }\href {\doibase 10.1086/151605} {\bibfield
  {journal} {\bibinfo  {journal} {\apj}\ }\textbf {\bibinfo {volume} {176}},\
  \bibinfo {pages} {1} (\bibinfo {year} {1972})}\BibitemShut {NoStop}%
\bibitem [{\citenamefont {Grcevich}\ and\ \citenamefont
  {Putman}(2009)}]{Grc09}%
  \BibitemOpen
  \bibfield  {author} {\bibinfo {author} {\bibfnamefont {J.}~\bibnamefont
  {Grcevich}}\ and\ \bibinfo {author} {\bibfnamefont {M.~E.}\ \bibnamefont
  {Putman}},\ }\href {\doibase 10.1088/0004-637X/721/1/922} {\bibfield
  {journal} {\bibinfo  {journal} {Astrophys. J.}\ }\textbf {\bibinfo {volume}
  {696}},\ \bibinfo {pages} {385} (\bibinfo {year} {2009})},\ \bibinfo {note}
  {[Erratum: Astrophys.J. 721, 922 (2010)]},\ \Eprint
  {http://arxiv.org/abs/0901.4975} {arXiv:0901.4975 [astro-ph.GA]} \BibitemShut
  {NoStop}%
\bibitem [{\citenamefont {{Walter}}\ \emph {et~al.}(2008)\citenamefont
  {{Walter}}, \citenamefont {{Brinks}}, \citenamefont {{de Blok}},
  \citenamefont {{Bigiel}}, \citenamefont {{Kennicutt}}, \citenamefont
  {{Thornley}},\ and\ \citenamefont {{Leroy}}}]{Wal08_Things}%
  \BibitemOpen
  \bibfield  {author} {\bibinfo {author} {\bibfnamefont {F.}~\bibnamefont
  {{Walter}}}, \bibinfo {author} {\bibfnamefont {E.}~\bibnamefont {{Brinks}}},
  \bibinfo {author} {\bibfnamefont {W.~J.~G.}\ \bibnamefont {{de Blok}}},
  \bibinfo {author} {\bibfnamefont {F.}~\bibnamefont {{Bigiel}}}, \bibinfo
  {author} {\bibfnamefont {J.}~\bibnamefont {{Kennicutt}}, \bibfnamefont
  {Robert~C.}}, \bibinfo {author} {\bibfnamefont {M.~D.}\ \bibnamefont
  {{Thornley}}}, \ and\ \bibinfo {author} {\bibfnamefont {A.}~\bibnamefont
  {{Leroy}}},\ }\href {\doibase 10.1088/0004-6256/136/6/2563} {\bibfield
  {journal} {\bibinfo  {journal} {\aj}\ }\textbf {\bibinfo {volume} {136}},\
  \bibinfo {pages} {2563} (\bibinfo {year} {2008})},\ \Eprint
  {http://arxiv.org/abs/0810.2125} {arXiv:0810.2125 [astro-ph]} \BibitemShut
  {NoStop}%
\bibitem [{\citenamefont {{Begum}}\ \emph {et~al.}(2008)\citenamefont
  {{Begum}}, \citenamefont {{Chengalur}}, \citenamefont {{Karachentsev}},
  \citenamefont {{Sharina}},\ and\ \citenamefont {{Kaisin}}}]{BegChe08_Figgs}%
  \BibitemOpen
  \bibfield  {author} {\bibinfo {author} {\bibfnamefont {A.}~\bibnamefont
  {{Begum}}}, \bibinfo {author} {\bibfnamefont {J.~N.}\ \bibnamefont
  {{Chengalur}}}, \bibinfo {author} {\bibfnamefont {I.~D.}\ \bibnamefont
  {{Karachentsev}}}, \bibinfo {author} {\bibfnamefont {M.~E.}\ \bibnamefont
  {{Sharina}}}, \ and\ \bibinfo {author} {\bibfnamefont {S.~S.}\ \bibnamefont
  {{Kaisin}}},\ }\href {\doibase 10.1111/j.1365-2966.2008.13150.x} {\bibfield
  {journal} {\bibinfo  {journal} {\mnras}\ }\textbf {\bibinfo {volume} {386}},\
  \bibinfo {pages} {1667} (\bibinfo {year} {2008})},\ \Eprint
  {http://arxiv.org/abs/0802.3982} {arXiv:0802.3982 [astro-ph]} \BibitemShut
  {NoStop}%
\bibitem [{\citenamefont {{Cannon}}\ \emph {et~al.}(2011)\citenamefont
  {{Cannon}}, \citenamefont {{Giovanelli}}, \citenamefont {{Haynes}},
  \citenamefont {{Janowiecki}}, \citenamefont {{Parker}}, \citenamefont
  {{Salzer}}, \citenamefont {{Adams}}, \citenamefont {{Engstrom}},
  \citenamefont {{Huang}}, \citenamefont {{McQuinn}}, \citenamefont {{Ott}},
  \citenamefont {{Saintonge}}, \citenamefont {{Skillman}}, \citenamefont
  {{Allan}}, \citenamefont {{Erny}}, \citenamefont {{Fliss}},\ and\
  \citenamefont {{Smith}}}]{Can11_Sheild}%
  \BibitemOpen
  \bibfield  {author} {\bibinfo {author} {\bibfnamefont {J.~M.}\ \bibnamefont
  {{Cannon}}}, \bibinfo {author} {\bibfnamefont {R.}~\bibnamefont
  {{Giovanelli}}}, \bibinfo {author} {\bibfnamefont {M.~P.}\ \bibnamefont
  {{Haynes}}}, \bibinfo {author} {\bibfnamefont {S.}~\bibnamefont
  {{Janowiecki}}}, \bibinfo {author} {\bibfnamefont {A.}~\bibnamefont
  {{Parker}}}, \bibinfo {author} {\bibfnamefont {J.~J.}\ \bibnamefont
  {{Salzer}}}, \bibinfo {author} {\bibfnamefont {E.~A.~K.}\ \bibnamefont
  {{Adams}}}, \bibinfo {author} {\bibfnamefont {E.}~\bibnamefont {{Engstrom}}},
  \bibinfo {author} {\bibfnamefont {S.}~\bibnamefont {{Huang}}}, \bibinfo
  {author} {\bibfnamefont {K.~B.~W.}\ \bibnamefont {{McQuinn}}}, \bibinfo
  {author} {\bibfnamefont {J.}~\bibnamefont {{Ott}}}, \bibinfo {author}
  {\bibfnamefont {A.}~\bibnamefont {{Saintonge}}}, \bibinfo {author}
  {\bibfnamefont {E.~D.}\ \bibnamefont {{Skillman}}}, \bibinfo {author}
  {\bibfnamefont {J.}~\bibnamefont {{Allan}}}, \bibinfo {author} {\bibfnamefont
  {G.}~\bibnamefont {{Erny}}}, \bibinfo {author} {\bibfnamefont
  {P.}~\bibnamefont {{Fliss}}}, \ and\ \bibinfo {author} {\bibfnamefont
  {A.}~\bibnamefont {{Smith}}},\ }\href {\doibase 10.1088/2041-8205/739/1/L22}
  {\bibfield  {journal} {\bibinfo  {journal} {\apjl}\ }\textbf {\bibinfo
  {volume} {739}},\ \bibinfo {eid} {L22} (\bibinfo {year} {2011})},\ \Eprint
  {http://arxiv.org/abs/1105.4505} {arXiv:1105.4505 [astro-ph.CO]} \BibitemShut
  {NoStop}%
\bibitem [{\citenamefont {{Ott}}\ \emph {et~al.}(2012)\citenamefont {{Ott}},
  \citenamefont {{Stilp}}, \citenamefont {{Warren}}, \citenamefont
  {{Skillman}}, \citenamefont {{Dalcanton}}, \citenamefont {{Walter}},
  \citenamefont {{de Blok}}, \citenamefont {{Koribalski}},\ and\ \citenamefont
  {{West}}}]{Ott12_Angst}%
  \BibitemOpen
  \bibfield  {author} {\bibinfo {author} {\bibfnamefont {J.}~\bibnamefont
  {{Ott}}}, \bibinfo {author} {\bibfnamefont {A.~M.}\ \bibnamefont {{Stilp}}},
  \bibinfo {author} {\bibfnamefont {S.~R.}\ \bibnamefont {{Warren}}}, \bibinfo
  {author} {\bibfnamefont {E.~D.}\ \bibnamefont {{Skillman}}}, \bibinfo
  {author} {\bibfnamefont {J.~J.}\ \bibnamefont {{Dalcanton}}}, \bibinfo
  {author} {\bibfnamefont {F.}~\bibnamefont {{Walter}}}, \bibinfo {author}
  {\bibfnamefont {W.~J.~G.}\ \bibnamefont {{de Blok}}}, \bibinfo {author}
  {\bibfnamefont {B.}~\bibnamefont {{Koribalski}}}, \ and\ \bibinfo {author}
  {\bibfnamefont {A.~A.}\ \bibnamefont {{West}}},\ }\href {\doibase
  10.1088/0004-6256/144/4/123} {\bibfield  {journal} {\bibinfo  {journal}
  {\aj}\ }\textbf {\bibinfo {volume} {144}},\ \bibinfo {eid} {123} (\bibinfo
  {year} {2012})},\ \Eprint {http://arxiv.org/abs/1208.3737} {arXiv:1208.3737
  [astro-ph.CO]} \BibitemShut {NoStop}%
\bibitem [{\citenamefont {{Hunter}}\ \emph {et~al.}(2012)\citenamefont
  {{Hunter}}, \citenamefont {{Ficut-Vicas}}, \citenamefont {{Ashley}},
  \citenamefont {{Brinks}}, \citenamefont {{Cigan}}, \citenamefont
  {{Elmegreen}}, \citenamefont {{Heesen}}, \citenamefont {{Herrmann}},
  \citenamefont {{Johnson}}, \citenamefont {{Oh}}, \citenamefont {{Rupen}},
  \citenamefont {{Schruba}}, \citenamefont {{Simpson}}, \citenamefont
  {{Walter}}, \citenamefont {{Westpfahl}}, \citenamefont {{Young}},\ and\
  \citenamefont {{Zhang}}}]{Hun12_LittleThings}%
  \BibitemOpen
  \bibfield  {author} {\bibinfo {author} {\bibfnamefont {D.~A.}\ \bibnamefont
  {{Hunter}}}, \bibinfo {author} {\bibfnamefont {D.}~\bibnamefont
  {{Ficut-Vicas}}}, \bibinfo {author} {\bibfnamefont {T.}~\bibnamefont
  {{Ashley}}}, \bibinfo {author} {\bibfnamefont {E.}~\bibnamefont {{Brinks}}},
  \bibinfo {author} {\bibfnamefont {P.}~\bibnamefont {{Cigan}}}, \bibinfo
  {author} {\bibfnamefont {B.~G.}\ \bibnamefont {{Elmegreen}}}, \bibinfo
  {author} {\bibfnamefont {V.}~\bibnamefont {{Heesen}}}, \bibinfo {author}
  {\bibfnamefont {K.~A.}\ \bibnamefont {{Herrmann}}}, \bibinfo {author}
  {\bibfnamefont {M.}~\bibnamefont {{Johnson}}}, \bibinfo {author}
  {\bibfnamefont {S.-H.}\ \bibnamefont {{Oh}}}, \bibinfo {author}
  {\bibfnamefont {M.~P.}\ \bibnamefont {{Rupen}}}, \bibinfo {author}
  {\bibfnamefont {A.}~\bibnamefont {{Schruba}}}, \bibinfo {author}
  {\bibfnamefont {C.~E.}\ \bibnamefont {{Simpson}}}, \bibinfo {author}
  {\bibfnamefont {F.}~\bibnamefont {{Walter}}}, \bibinfo {author}
  {\bibfnamefont {D.~J.}\ \bibnamefont {{Westpfahl}}}, \bibinfo {author}
  {\bibfnamefont {L.~M.}\ \bibnamefont {{Young}}}, \ and\ \bibinfo {author}
  {\bibfnamefont {H.-X.}\ \bibnamefont {{Zhang}}},\ }\href {\doibase
  10.1088/0004-6256/144/5/134} {\bibfield  {journal} {\bibinfo  {journal}
  {\aj}\ }\textbf {\bibinfo {volume} {144}},\ \bibinfo {eid} {134} (\bibinfo
  {year} {2012})},\ \Eprint {http://arxiv.org/abs/1208.5834} {arXiv:1208.5834
  [astro-ph.GA]} \BibitemShut {NoStop}%
\bibitem [{\citenamefont {{Adams}}\ \emph {et~al.}(2013)\citenamefont
  {{Adams}}, \citenamefont {{Giovanelli}},\ and\ \citenamefont
  {{Haynes}}}]{Ada13}%
  \BibitemOpen
  \bibfield  {author} {\bibinfo {author} {\bibfnamefont {E.~A.~K.}\
  \bibnamefont {{Adams}}}, \bibinfo {author} {\bibfnamefont {R.}~\bibnamefont
  {{Giovanelli}}}, \ and\ \bibinfo {author} {\bibfnamefont {M.~P.}\
  \bibnamefont {{Haynes}}},\ }\href {\doibase 10.1088/0004-637X/768/1/77}
  {\bibfield  {journal} {\bibinfo  {journal} {\apj}\ }\textbf {\bibinfo
  {volume} {768}},\ \bibinfo {eid} {77} (\bibinfo {year} {2013})},\ \Eprint
  {http://arxiv.org/abs/1303.6967} {arXiv:1303.6967 [astro-ph.CO]} \BibitemShut
  {NoStop}%
\bibitem [{\citenamefont {{Saul}}\ \emph {et~al.}(2012)\citenamefont {{Saul}}
  \emph {et~al.}}]{Sau12}%
  \BibitemOpen
  \bibfield  {author} {\bibinfo {author} {\bibfnamefont {D.~R.}\ \bibnamefont
  {{Saul}}} \emph {et~al.},\ }\href {\doibase 10.1088/0004-637X/758/1/44}
  {\bibfield  {journal} {\bibinfo  {journal} {\apj}\ }\textbf {\bibinfo
  {volume} {758}},\ \bibinfo {eid} {44} (\bibinfo {year} {2012})},\ \Eprint
  {http://arxiv.org/abs/1208.4103} {arXiv:1208.4103 [astro-ph.GA]} \BibitemShut
  {NoStop}%
\bibitem [{\citenamefont {{Simon}}(2019)}]{Sim19}%
  \BibitemOpen
  \bibfield  {author} {\bibinfo {author} {\bibfnamefont {J.~D.}\ \bibnamefont
  {{Simon}}},\ }\href {\doibase 10.1146/annurev-astro-091918-104453} {\bibfield
   {journal} {\bibinfo  {journal} {\araa}\ }\textbf {\bibinfo {volume} {57}},\
  \bibinfo {pages} {375} (\bibinfo {year} {2019})},\ \Eprint
  {http://arxiv.org/abs/1901.05465} {arXiv:1901.05465 [astro-ph.GA]}
  \BibitemShut {NoStop}%
\bibitem [{\citenamefont {{Putman}}\ \emph {et~al.}(2021)\citenamefont
  {{Putman}}, \citenamefont {{Zheng}}, \citenamefont {{Price-Whelan}},
  \citenamefont {{Grcevich}}, \citenamefont {{Johnson}}, \citenamefont
  {{Tollerud}},\ and\ \citenamefont {{Peek}}}]{Put21}%
  \BibitemOpen
  \bibfield  {author} {\bibinfo {author} {\bibfnamefont {M.~E.}\ \bibnamefont
  {{Putman}}}, \bibinfo {author} {\bibfnamefont {Y.}~\bibnamefont {{Zheng}}},
  \bibinfo {author} {\bibfnamefont {A.~M.}\ \bibnamefont {{Price-Whelan}}},
  \bibinfo {author} {\bibfnamefont {J.}~\bibnamefont {{Grcevich}}}, \bibinfo
  {author} {\bibfnamefont {A.~C.}\ \bibnamefont {{Johnson}}}, \bibinfo {author}
  {\bibfnamefont {E.}~\bibnamefont {{Tollerud}}}, \ and\ \bibinfo {author}
  {\bibfnamefont {J.~E.~G.}\ \bibnamefont {{Peek}}},\ }\href@noop {} {\bibfield
   {journal} {\bibinfo  {journal} {arXiv e-prints}\ ,\ \bibinfo {eid}
  {arXiv:2101.07809}} (\bibinfo {year} {2021})},\ \Eprint
  {http://arxiv.org/abs/2101.07809} {arXiv:2101.07809 [astro-ph.GA]}
  \BibitemShut {NoStop}%
\bibitem [{\citenamefont {{Dubovsky}}\ \emph {et~al.}(2004)\citenamefont
  {{Dubovsky}}, \citenamefont {{Gorbunov}},\ and\ \citenamefont
  {{Rubtsov}}}]{DubGorRub04}%
  \BibitemOpen
  \bibfield  {author} {\bibinfo {author} {\bibfnamefont {S.~L.}\ \bibnamefont
  {{Dubovsky}}}, \bibinfo {author} {\bibfnamefont {D.~S.}\ \bibnamefont
  {{Gorbunov}}}, \ and\ \bibinfo {author} {\bibfnamefont {G.~I.}\ \bibnamefont
  {{Rubtsov}}},\ }\href {\doibase 10.1134/1.1675909} {\bibfield  {journal}
  {\bibinfo  {journal} {Soviet Journal of Experimental and Theoretical Physics
  Letters}\ }\textbf {\bibinfo {volume} {79}},\ \bibinfo {pages} {1} (\bibinfo
  {year} {2004})},\ \Eprint {http://arxiv.org/abs/hep-ph/0311189}
  {arXiv:hep-ph/0311189 [astro-ph]} \BibitemShut {NoStop}%
\bibitem [{\citenamefont {{McDermott}}\ \emph {et~al.}(2011)\citenamefont
  {{McDermott}}, \citenamefont {{Yu}},\ and\ \citenamefont
  {{Zurek}}}]{McDermott11_millicharge}%
  \BibitemOpen
  \bibfield  {author} {\bibinfo {author} {\bibfnamefont {S.~D.}\ \bibnamefont
  {{McDermott}}}, \bibinfo {author} {\bibfnamefont {H.-B.}\ \bibnamefont
  {{Yu}}}, \ and\ \bibinfo {author} {\bibfnamefont {K.~M.}\ \bibnamefont
  {{Zurek}}},\ }\href {\doibase 10.1103/PhysRevD.83.063509} {\bibfield
  {journal} {\bibinfo  {journal} {\prd}\ }\textbf {\bibinfo {volume} {83}},\
  \bibinfo {eid} {063509} (\bibinfo {year} {2011})},\ \Eprint
  {http://arxiv.org/abs/1011.2907} {arXiv:1011.2907 [hep-ph]} \BibitemShut
  {NoStop}%
\bibitem [{\citenamefont {{Dvorkin}}\ \emph {et~al.}(2014)\citenamefont
  {{Dvorkin}}, \citenamefont {{Blum}},\ and\ \citenamefont
  {{Kamionkowski}}}]{dvorkin14}%
  \BibitemOpen
  \bibfield  {author} {\bibinfo {author} {\bibfnamefont {C.}~\bibnamefont
  {{Dvorkin}}}, \bibinfo {author} {\bibfnamefont {K.}~\bibnamefont {{Blum}}}, \
  and\ \bibinfo {author} {\bibfnamefont {M.}~\bibnamefont {{Kamionkowski}}},\
  }\href {\doibase 10.1103/PhysRevD.89.023519} {\bibfield  {journal} {\bibinfo
  {journal} {\prd}\ }\textbf {\bibinfo {volume} {89}},\ \bibinfo {eid} {023519}
  (\bibinfo {year} {2014})},\ \Eprint {http://arxiv.org/abs/1311.2937}
  {arXiv:1311.2937} \BibitemShut {NoStop}%
\bibitem [{\citenamefont {{Boddy}}\ \emph {et~al.}(2018)\citenamefont
  {{Boddy}}, \citenamefont {{Gluscevic}}, \citenamefont {{Poulin}},
  \citenamefont {{Kovetz}}, \citenamefont {{Kamionkowski}},\ and\ \citenamefont
  {{Barkana}}}]{boddy18}%
  \BibitemOpen
  \bibfield  {author} {\bibinfo {author} {\bibfnamefont {K.~K.}\ \bibnamefont
  {{Boddy}}}, \bibinfo {author} {\bibfnamefont {V.}~\bibnamefont
  {{Gluscevic}}}, \bibinfo {author} {\bibfnamefont {V.}~\bibnamefont
  {{Poulin}}}, \bibinfo {author} {\bibfnamefont {E.~D.}\ \bibnamefont
  {{Kovetz}}}, \bibinfo {author} {\bibfnamefont {M.}~\bibnamefont
  {{Kamionkowski}}}, \ and\ \bibinfo {author} {\bibfnamefont {R.}~\bibnamefont
  {{Barkana}}},\ }\href {\doibase 10.1103/PhysRevD.98.123506} {\bibfield
  {journal} {\bibinfo  {journal} {\prd}\ }\textbf {\bibinfo {volume} {98}},\
  \bibinfo {eid} {123506} (\bibinfo {year} {2018})},\ \Eprint
  {http://arxiv.org/abs/1808.00001} {arXiv:1808.00001 [astro-ph.CO]}
  \BibitemShut {NoStop}%
\bibitem [{\citenamefont {{Kovetz}}\ \emph
  {et~al.}(2018{\natexlab{a}})\citenamefont {{Kovetz}}, \citenamefont
  {{Poulin}}, \citenamefont {{Gluscevic}}, \citenamefont {{Boddy}},
  \citenamefont {{Barkana}},\ and\ \citenamefont
  {{Kamionkowski}}}]{Kov18_MClimits}%
  \BibitemOpen
  \bibfield  {author} {\bibinfo {author} {\bibfnamefont {E.~D.}\ \bibnamefont
  {{Kovetz}}}, \bibinfo {author} {\bibfnamefont {V.}~\bibnamefont {{Poulin}}},
  \bibinfo {author} {\bibfnamefont {V.}~\bibnamefont {{Gluscevic}}}, \bibinfo
  {author} {\bibfnamefont {K.~K.}\ \bibnamefont {{Boddy}}}, \bibinfo {author}
  {\bibfnamefont {R.}~\bibnamefont {{Barkana}}}, \ and\ \bibinfo {author}
  {\bibfnamefont {M.}~\bibnamefont {{Kamionkowski}}},\ }\href {\doibase
  10.1103/PhysRevD.98.103529} {\bibfield  {journal} {\bibinfo  {journal}
  {\prd}\ }\textbf {\bibinfo {volume} {98}},\ \bibinfo {eid} {103529} (\bibinfo
  {year} {2018}{\natexlab{a}})},\ \Eprint {http://arxiv.org/abs/1807.11482}
  {arXiv:1807.11482 [astro-ph.CO]} \BibitemShut {NoStop}%
\bibitem [{\citenamefont {{Xu}}\ \emph {et~al.}(2018)\citenamefont {{Xu}},
  \citenamefont {{Dvorkin}},\ and\ \citenamefont {{Chael}}}]{Xu18_CMB}%
  \BibitemOpen
  \bibfield  {author} {\bibinfo {author} {\bibfnamefont {W.~L.}\ \bibnamefont
  {{Xu}}}, \bibinfo {author} {\bibfnamefont {C.}~\bibnamefont {{Dvorkin}}}, \
  and\ \bibinfo {author} {\bibfnamefont {A.}~\bibnamefont {{Chael}}},\ }\href
  {\doibase 10.1103/PhysRevD.97.103530} {\bibfield  {journal} {\bibinfo
  {journal} {\prd}\ }\textbf {\bibinfo {volume} {97}},\ \bibinfo {eid} {103530}
  (\bibinfo {year} {2018})},\ \Eprint {http://arxiv.org/abs/1802.06788}
  {arXiv:1802.06788 [astro-ph.CO]} \BibitemShut {NoStop}%
\bibitem [{\citenamefont {{Slatyer}}\ and\ \citenamefont
  {{Wu}}(2018)}]{Slatyer18_CMB}%
  \BibitemOpen
  \bibfield  {author} {\bibinfo {author} {\bibfnamefont {T.~R.}\ \bibnamefont
  {{Slatyer}}}\ and\ \bibinfo {author} {\bibfnamefont {C.-L.}\ \bibnamefont
  {{Wu}}},\ }\href {\doibase 10.1103/PhysRevD.98.023013} {\bibfield  {journal}
  {\bibinfo  {journal} {\prd}\ }\textbf {\bibinfo {volume} {98}},\ \bibinfo
  {eid} {023013} (\bibinfo {year} {2018})},\ \Eprint
  {http://arxiv.org/abs/1803.09734} {arXiv:1803.09734 [astro-ph.CO]}
  \BibitemShut {NoStop}%
\bibitem [{\citenamefont {{Davidson}}\ \emph {et~al.}(2000)\citenamefont
  {{Davidson}}, \citenamefont {{Hannestad}},\ and\ \citenamefont
  {{Raffelt}}}]{DavHanRaf00_BBN}%
  \BibitemOpen
  \bibfield  {author} {\bibinfo {author} {\bibfnamefont {S.}~\bibnamefont
  {{Davidson}}}, \bibinfo {author} {\bibfnamefont {S.}~\bibnamefont
  {{Hannestad}}}, \ and\ \bibinfo {author} {\bibfnamefont {G.}~\bibnamefont
  {{Raffelt}}},\ }\href {\doibase 10.1088/1126-6708/2000/05/003} {\bibfield
  {journal} {\bibinfo  {journal} {Journal of High Energy Physics}\ }\textbf
  {\bibinfo {volume} {2000}},\ \bibinfo {eid} {003} (\bibinfo {year} {2000})},\
  \Eprint {http://arxiv.org/abs/hep-ph/0001179} {arXiv:hep-ph/0001179 [hep-ph]}
  \BibitemShut {NoStop}%
\bibitem [{\citenamefont {{Kamada}}\ \emph {et~al.}(2017)\citenamefont
  {{Kamada}}, \citenamefont {{Kohri}}, \citenamefont {{Takahashi}},\ and\
  \citenamefont {{Yoshida}}}]{KamKohTak17}%
  \BibitemOpen
  \bibfield  {author} {\bibinfo {author} {\bibfnamefont {A.}~\bibnamefont
  {{Kamada}}}, \bibinfo {author} {\bibfnamefont {K.}~\bibnamefont {{Kohri}}},
  \bibinfo {author} {\bibfnamefont {T.}~\bibnamefont {{Takahashi}}}, \ and\
  \bibinfo {author} {\bibfnamefont {N.}~\bibnamefont {{Yoshida}}},\ }\href
  {\doibase 10.1103/PhysRevD.95.023502} {\bibfield  {journal} {\bibinfo
  {journal} {\prd}\ }\textbf {\bibinfo {volume} {95}},\ \bibinfo {eid} {023502}
  (\bibinfo {year} {2017})},\ \Eprint {http://arxiv.org/abs/1604.07926}
  {arXiv:1604.07926 [astro-ph.CO]} \BibitemShut {NoStop}%
\bibitem [{\citenamefont {{Tseliakhovich}}\ and\ \citenamefont
  {{Hirata}}(2010)}]{TseHir10}%
  \BibitemOpen
  \bibfield  {author} {\bibinfo {author} {\bibfnamefont {D.}~\bibnamefont
  {{Tseliakhovich}}}\ and\ \bibinfo {author} {\bibfnamefont {C.}~\bibnamefont
  {{Hirata}}},\ }\href {\doibase 10.1103/PhysRevD.82.083520} {\bibfield
  {journal} {\bibinfo  {journal} {\prd}\ }\textbf {\bibinfo {volume} {82}},\
  \bibinfo {eid} {083520} (\bibinfo {year} {2010})},\ \Eprint
  {http://arxiv.org/abs/1005.2416} {arXiv:1005.2416 [astro-ph.CO]} \BibitemShut
  {NoStop}%
\bibitem [{\citenamefont {{Agrawal}}\ \emph {et~al.}(2018)\citenamefont
  {{Agrawal}}, \citenamefont {{Kitajima}}, \citenamefont {{Reece}},
  \citenamefont {{Sekiguchi}},\ and\ \citenamefont
  {{Takahashi}}}]{AgaKitRee18}%
  \BibitemOpen
  \bibfield  {author} {\bibinfo {author} {\bibfnamefont {P.}~\bibnamefont
  {{Agrawal}}}, \bibinfo {author} {\bibfnamefont {N.}~\bibnamefont
  {{Kitajima}}}, \bibinfo {author} {\bibfnamefont {M.}~\bibnamefont {{Reece}}},
  \bibinfo {author} {\bibfnamefont {T.}~\bibnamefont {{Sekiguchi}}}, \ and\
  \bibinfo {author} {\bibfnamefont {F.}~\bibnamefont {{Takahashi}}},\
  }\href@noop {} {\bibfield  {journal} {\bibinfo  {journal} {arXiv e-prints}\
  ,\ \bibinfo {eid} {arXiv:1810.07188}} (\bibinfo {year} {2018})},\ \Eprint
  {http://arxiv.org/abs/1810.07188} {arXiv:1810.07188 [hep-ph]} \BibitemShut
  {NoStop}%
\bibitem [{\citenamefont {{Dror}}\ \emph {et~al.}(2019)\citenamefont {{Dror}},
  \citenamefont {{Harigaya}},\ and\ \citenamefont {{Narayan}}}]{DroHarNar19}%
  \BibitemOpen
  \bibfield  {author} {\bibinfo {author} {\bibfnamefont {J.~A.}\ \bibnamefont
  {{Dror}}}, \bibinfo {author} {\bibfnamefont {K.}~\bibnamefont {{Harigaya}}},
  \ and\ \bibinfo {author} {\bibfnamefont {V.}~\bibnamefont {{Narayan}}},\
  }\href {\doibase 10.1103/PhysRevD.99.035036} {\bibfield  {journal} {\bibinfo
  {journal} {\prd}\ }\textbf {\bibinfo {volume} {99}},\ \bibinfo {eid} {035036}
  (\bibinfo {year} {2019})},\ \Eprint {http://arxiv.org/abs/1810.07195}
  {arXiv:1810.07195 [hep-ph]} \BibitemShut {NoStop}%
\bibitem [{\citenamefont {{Bastero-Gil}}\ \emph {et~al.}(2019)\citenamefont
  {{Bastero-Gil}}, \citenamefont {{Santiago}}, \citenamefont {{Ubaldi}},\ and\
  \citenamefont {{Vega-Morales}}}]{BasSanUba19}%
  \BibitemOpen
  \bibfield  {author} {\bibinfo {author} {\bibfnamefont {M.}~\bibnamefont
  {{Bastero-Gil}}}, \bibinfo {author} {\bibfnamefont {J.}~\bibnamefont
  {{Santiago}}}, \bibinfo {author} {\bibfnamefont {L.}~\bibnamefont
  {{Ubaldi}}}, \ and\ \bibinfo {author} {\bibfnamefont {R.}~\bibnamefont
  {{Vega-Morales}}},\ }\href {\doibase 10.1088/1475-7516/2019/04/015}
  {\bibfield  {journal} {\bibinfo  {journal} {\jcap}\ }\textbf {\bibinfo
  {volume} {2019}},\ \bibinfo {eid} {015} (\bibinfo {year} {2019})},\ \Eprint
  {http://arxiv.org/abs/1810.07208} {arXiv:1810.07208 [hep-ph]} \BibitemShut
  {NoStop}%
\bibitem [{\citenamefont {{Co}}\ \emph {et~al.}(2019)\citenamefont {{Co}},
  \citenamefont {{Pierce}}, \citenamefont {{Zhang}},\ and\ \citenamefont
  {{Zhao}}}]{CoPieZha19}%
  \BibitemOpen
  \bibfield  {author} {\bibinfo {author} {\bibfnamefont {R.~T.}\ \bibnamefont
  {{Co}}}, \bibinfo {author} {\bibfnamefont {A.}~\bibnamefont {{Pierce}}},
  \bibinfo {author} {\bibfnamefont {Z.}~\bibnamefont {{Zhang}}}, \ and\
  \bibinfo {author} {\bibfnamefont {Y.}~\bibnamefont {{Zhao}}},\ }\href
  {\doibase 10.1103/PhysRevD.99.075002} {\bibfield  {journal} {\bibinfo
  {journal} {\prd}\ }\textbf {\bibinfo {volume} {99}},\ \bibinfo {eid} {075002}
  (\bibinfo {year} {2019})},\ \Eprint {http://arxiv.org/abs/1810.07196}
  {arXiv:1810.07196 [hep-ph]} \BibitemShut {NoStop}%
\bibitem [{\citenamefont {{Dubovsky}}\ and\ \citenamefont
  {{Hern{\'a}ndez-Chifflet}}(2015)}]{DubHer15}%
  \BibitemOpen
  \bibfield  {author} {\bibinfo {author} {\bibfnamefont {S.}~\bibnamefont
  {{Dubovsky}}}\ and\ \bibinfo {author} {\bibfnamefont {G.}~\bibnamefont
  {{Hern{\'a}ndez-Chifflet}}},\ }\href {\doibase 10.1088/1475-7516/2015/12/054}
  {\bibfield  {journal} {\bibinfo  {journal} {Journal of Cosmology and
  Astro-Particle Physics}\ }\textbf {\bibinfo {volume} {2015}},\ \bibinfo {eid}
  {054} (\bibinfo {year} {2015})},\ \Eprint {http://arxiv.org/abs/1509.00039}
  {arXiv:1509.00039 [hep-ph]} \BibitemShut {NoStop}%
\bibitem [{\citenamefont {{Arias}}\ \emph {et~al.}(2012)\citenamefont
  {{Arias}}, \citenamefont {{Cadamuro}}, \citenamefont {{Goodsell}},
  \citenamefont {{Jaeckel}}, \citenamefont {{Redondo}},\ and\ \citenamefont
  {{Ringwald}}}]{DarkPhotonCMB_Ari12}%
  \BibitemOpen
  \bibfield  {author} {\bibinfo {author} {\bibfnamefont {P.}~\bibnamefont
  {{Arias}}}, \bibinfo {author} {\bibfnamefont {D.}~\bibnamefont {{Cadamuro}}},
  \bibinfo {author} {\bibfnamefont {M.}~\bibnamefont {{Goodsell}}}, \bibinfo
  {author} {\bibfnamefont {J.}~\bibnamefont {{Jaeckel}}}, \bibinfo {author}
  {\bibfnamefont {J.}~\bibnamefont {{Redondo}}}, \ and\ \bibinfo {author}
  {\bibfnamefont {A.}~\bibnamefont {{Ringwald}}},\ }\href {\doibase
  10.1088/1475-7516/2012/06/013} {\bibfield  {journal} {\bibinfo  {journal}
  {Journal of Cosmology and Astro-Particle Physics}\ }\textbf {\bibinfo
  {volume} {2012}},\ \bibinfo {eid} {013} (\bibinfo {year} {2012})},\ \Eprint
  {http://arxiv.org/abs/1201.5902} {arXiv:1201.5902 [hep-ph]} \BibitemShut
  {NoStop}%
\bibitem [{\citenamefont {{An}}\ \emph {et~al.}(2013)\citenamefont {{An}},
  \citenamefont {{Pospelov}},\ and\ \citenamefont
  {{Pradler}}}]{AnPosPra13_HPDM}%
  \BibitemOpen
  \bibfield  {author} {\bibinfo {author} {\bibfnamefont {H.}~\bibnamefont
  {{An}}}, \bibinfo {author} {\bibfnamefont {M.}~\bibnamefont {{Pospelov}}}, \
  and\ \bibinfo {author} {\bibfnamefont {J.}~\bibnamefont {{Pradler}}},\ }\href
  {\doibase 10.1016/j.physletb.2013.07.008} {\bibfield  {journal} {\bibinfo
  {journal} {Physics Letters B}\ }\textbf {\bibinfo {volume} {725}},\ \bibinfo
  {pages} {190} (\bibinfo {year} {2013})},\ \Eprint
  {http://arxiv.org/abs/1302.3884} {arXiv:1302.3884 [hep-ph]} \BibitemShut
  {NoStop}%
\bibitem [{\citenamefont {{Jaeckel}}\ and\ \citenamefont
  {{Ringwald}}(2010)}]{DarkPhotonCMB_Jae10}%
  \BibitemOpen
  \bibfield  {author} {\bibinfo {author} {\bibfnamefont {J.}~\bibnamefont
  {{Jaeckel}}}\ and\ \bibinfo {author} {\bibfnamefont {A.}~\bibnamefont
  {{Ringwald}}},\ }\href {\doibase 10.1146/annurev.nucl.012809.104433}
  {\bibfield  {journal} {\bibinfo  {journal} {Annual Review of Nuclear and
  Particle Science}\ }\textbf {\bibinfo {volume} {60}},\ \bibinfo {pages} {405}
  (\bibinfo {year} {2010})},\ \Eprint {http://arxiv.org/abs/1002.0329}
  {arXiv:1002.0329 [hep-ph]} \BibitemShut {NoStop}%
\bibitem [{\citenamefont {{Graham}}\ \emph {et~al.}(2016)\citenamefont
  {{Graham}}, \citenamefont {{Mardon}},\ and\ \citenamefont
  {{Rajendran}}}]{GraMad16}%
  \BibitemOpen
  \bibfield  {author} {\bibinfo {author} {\bibfnamefont {P.~W.}\ \bibnamefont
  {{Graham}}}, \bibinfo {author} {\bibfnamefont {J.}~\bibnamefont {{Mardon}}},
  \ and\ \bibinfo {author} {\bibfnamefont {S.}~\bibnamefont {{Rajendran}}},\
  }\href {\doibase 10.1103/PhysRevD.93.103520} {\bibfield  {journal} {\bibinfo
  {journal} {\prd}\ }\textbf {\bibinfo {volume} {93}},\ \bibinfo {eid} {103520}
  (\bibinfo {year} {2016})},\ \Eprint {http://arxiv.org/abs/1504.02102}
  {arXiv:1504.02102 [hep-ph]} \BibitemShut {NoStop}%
\bibitem [{\citenamefont {{An}}\ \emph {et~al.}(2015)\citenamefont {{An}},
  \citenamefont {{Pospelov}}, \citenamefont {{Pradler}},\ and\ \citenamefont
  {{Ritz}}}]{AnPosPra15_HPDM}%
  \BibitemOpen
  \bibfield  {author} {\bibinfo {author} {\bibfnamefont {H.}~\bibnamefont
  {{An}}}, \bibinfo {author} {\bibfnamefont {M.}~\bibnamefont {{Pospelov}}},
  \bibinfo {author} {\bibfnamefont {J.}~\bibnamefont {{Pradler}}}, \ and\
  \bibinfo {author} {\bibfnamefont {A.}~\bibnamefont {{Ritz}}},\ }\href
  {\doibase 10.1016/j.physletb.2015.06.018} {\bibfield  {journal} {\bibinfo
  {journal} {Physics Letters B}\ }\textbf {\bibinfo {volume} {747}},\ \bibinfo
  {pages} {331} (\bibinfo {year} {2015})},\ \Eprint
  {http://arxiv.org/abs/1412.8378} {arXiv:1412.8378 [hep-ph]} \BibitemShut
  {NoStop}%
\bibitem [{\citenamefont {{Nelson}}\ and\ \citenamefont
  {{Scholtz}}(2011)}]{NelSch11}%
  \BibitemOpen
  \bibfield  {author} {\bibinfo {author} {\bibfnamefont {A.~E.}\ \bibnamefont
  {{Nelson}}}\ and\ \bibinfo {author} {\bibfnamefont {J.}~\bibnamefont
  {{Scholtz}}},\ }\href {\doibase 10.1103/PhysRevD.84.103501} {\bibfield
  {journal} {\bibinfo  {journal} {\prd}\ }\textbf {\bibinfo {volume} {84}},\
  \bibinfo {eid} {103501} (\bibinfo {year} {2011})},\ \Eprint
  {http://arxiv.org/abs/1105.2812} {arXiv:1105.2812 [hep-ph]} \BibitemShut
  {NoStop}%
\bibitem [{\citenamefont {{Chaudhuri}}\ \emph {et~al.}(2015)\citenamefont
  {{Chaudhuri}}, \citenamefont {{Graham}}, \citenamefont {{Irwin}},
  \citenamefont {{Mardon}}, \citenamefont {{Rajendran}},\ and\ \citenamefont
  {{Zhao}}}]{ChaGraIrw15}%
  \BibitemOpen
  \bibfield  {author} {\bibinfo {author} {\bibfnamefont {S.}~\bibnamefont
  {{Chaudhuri}}}, \bibinfo {author} {\bibfnamefont {P.~W.}\ \bibnamefont
  {{Graham}}}, \bibinfo {author} {\bibfnamefont {K.}~\bibnamefont {{Irwin}}},
  \bibinfo {author} {\bibfnamefont {J.}~\bibnamefont {{Mardon}}}, \bibinfo
  {author} {\bibfnamefont {S.}~\bibnamefont {{Rajendran}}}, \ and\ \bibinfo
  {author} {\bibfnamefont {Y.}~\bibnamefont {{Zhao}}},\ }\href {\doibase
  10.1103/PhysRevD.92.075012} {\bibfield  {journal} {\bibinfo  {journal}
  {\prd}\ }\textbf {\bibinfo {volume} {92}},\ \bibinfo {eid} {075012} (\bibinfo
  {year} {2015})},\ \Eprint {http://arxiv.org/abs/1411.7382} {arXiv:1411.7382
  [hep-ph]} \BibitemShut {NoStop}%
\bibitem [{\citenamefont {{Pignol}}\ \emph {et~al.}(2015)\citenamefont
  {{Pignol}}, \citenamefont {{Clement}}, \citenamefont {{Guigue}},
  \citenamefont {{Rebreyend}},\ and\ \citenamefont
  {{Voirin}}}]{PigCleGui15_HPDM}%
  \BibitemOpen
  \bibfield  {author} {\bibinfo {author} {\bibfnamefont {G.}~\bibnamefont
  {{Pignol}}}, \bibinfo {author} {\bibfnamefont {B.}~\bibnamefont {{Clement}}},
  \bibinfo {author} {\bibfnamefont {M.}~\bibnamefont {{Guigue}}}, \bibinfo
  {author} {\bibfnamefont {D.}~\bibnamefont {{Rebreyend}}}, \ and\ \bibinfo
  {author} {\bibfnamefont {B.}~\bibnamefont {{Voirin}}},\ }\href@noop {}
  {\bibfield  {journal} {\bibinfo  {journal} {arXiv e-prints}\ ,\ \bibinfo
  {eid} {arXiv:1507.06875}} (\bibinfo {year} {2015})},\ \Eprint
  {http://arxiv.org/abs/1507.06875} {arXiv:1507.06875 [astro-ph.CO]}
  \BibitemShut {NoStop}%
\bibitem [{\citenamefont {{Kovetz}}\ \emph
  {et~al.}(2018{\natexlab{b}})\citenamefont {{Kovetz}}, \citenamefont
  {{Cholis}},\ and\ \citenamefont {{Kaplan}}}]{Kov18_DPDM}%
  \BibitemOpen
  \bibfield  {author} {\bibinfo {author} {\bibfnamefont {E.~D.}\ \bibnamefont
  {{Kovetz}}}, \bibinfo {author} {\bibfnamefont {I.}~\bibnamefont {{Cholis}}},
  \ and\ \bibinfo {author} {\bibfnamefont {D.~E.}\ \bibnamefont {{Kaplan}}},\
  }\href@noop {} {\bibfield  {journal} {\bibinfo  {journal} {arXiv e-prints}\
  ,\ \bibinfo {eid} {arXiv:1809.01139}} (\bibinfo {year}
  {2018}{\natexlab{b}})},\ \Eprint {http://arxiv.org/abs/1809.01139}
  {arXiv:1809.01139 [astro-ph.CO]} \BibitemShut {NoStop}%
\bibitem [{\citenamefont {{Arvanitaki}}\ and\ \citenamefont
  {{Dubovsky}}(2011)}]{ArvDub11}%
  \BibitemOpen
  \bibfield  {author} {\bibinfo {author} {\bibfnamefont {A.}~\bibnamefont
  {{Arvanitaki}}}\ and\ \bibinfo {author} {\bibfnamefont {S.}~\bibnamefont
  {{Dubovsky}}},\ }\href {\doibase 10.1103/PhysRevD.83.044026} {\bibfield
  {journal} {\bibinfo  {journal} {\prd}\ }\textbf {\bibinfo {volume} {83}},\
  \bibinfo {eid} {044026} (\bibinfo {year} {2011})},\ \Eprint
  {http://arxiv.org/abs/1004.3558} {arXiv:1004.3558 [hep-th]} \BibitemShut
  {NoStop}%
\bibitem [{\citenamefont {{Bhoonah}}\ \emph
  {et~al.}(2018{\natexlab{a}})\citenamefont {{Bhoonah}}, \citenamefont
  {{Bramante}}, \citenamefont {{Elahi}},\ and\ \citenamefont
  {{Schon}}}]{BhoBramante19}%
  \BibitemOpen
  \bibfield  {author} {\bibinfo {author} {\bibfnamefont {A.}~\bibnamefont
  {{Bhoonah}}}, \bibinfo {author} {\bibfnamefont {J.}~\bibnamefont
  {{Bramante}}}, \bibinfo {author} {\bibfnamefont {F.}~\bibnamefont {{Elahi}}},
  \ and\ \bibinfo {author} {\bibfnamefont {S.}~\bibnamefont {{Schon}}},\
  }\href@noop {} {\bibfield  {journal} {\bibinfo  {journal} {arXiv e-prints}\ }
  (\bibinfo {year} {2018}{\natexlab{a}})},\ \Eprint
  {http://arxiv.org/abs/1812.10919} {arXiv:1812.10919 [hep-ph]} \BibitemShut
  {NoStop}%
\bibitem [{\citenamefont {{Long}}\ and\ \citenamefont
  {{Wang}}(2019)}]{LonWan19}%
  \BibitemOpen
  \bibfield  {author} {\bibinfo {author} {\bibfnamefont {A.~J.}\ \bibnamefont
  {{Long}}}\ and\ \bibinfo {author} {\bibfnamefont {L.-T.}\ \bibnamefont
  {{Wang}}},\ }\href {\doibase 10.1103/PhysRevD.99.063529} {\bibfield
  {journal} {\bibinfo  {journal} {\prd}\ }\textbf {\bibinfo {volume} {99}},\
  \bibinfo {eid} {063529} (\bibinfo {year} {2019})},\ \Eprint
  {http://arxiv.org/abs/1901.03312} {arXiv:1901.03312 [hep-ph]} \BibitemShut
  {NoStop}%
\bibitem [{\citenamefont {{McDermott}}\ and\ \citenamefont
  {{Witte}}(2020)}]{McDWit20}%
  \BibitemOpen
  \bibfield  {author} {\bibinfo {author} {\bibfnamefont {S.~D.}\ \bibnamefont
  {{McDermott}}}\ and\ \bibinfo {author} {\bibfnamefont {S.~J.}\ \bibnamefont
  {{Witte}}},\ }\href {\doibase 10.1103/PhysRevD.101.063030} {\bibfield
  {journal} {\bibinfo  {journal} {\prd}\ }\textbf {\bibinfo {volume} {101}},\
  \bibinfo {eid} {063030} (\bibinfo {year} {2020})},\ \Eprint
  {http://arxiv.org/abs/1911.05086} {arXiv:1911.05086 [hep-ph]} \BibitemShut
  {NoStop}%
\bibitem [{\citenamefont {{Caputo}}\ \emph
  {et~al.}(2020{\natexlab{a}})\citenamefont {{Caputo}}, \citenamefont {{Liu}},
  \citenamefont {{Mishra-Sharma}},\ and\ \citenamefont
  {{Ruderman}}}]{CapLiuMis20}%
  \BibitemOpen
  \bibfield  {author} {\bibinfo {author} {\bibfnamefont {A.}~\bibnamefont
  {{Caputo}}}, \bibinfo {author} {\bibfnamefont {H.}~\bibnamefont {{Liu}}},
  \bibinfo {author} {\bibfnamefont {S.}~\bibnamefont {{Mishra-Sharma}}}, \ and\
  \bibinfo {author} {\bibfnamefont {J.~T.}\ \bibnamefont {{Ruderman}}},\
  }\href@noop {} {\bibfield  {journal} {\bibinfo  {journal} {arXiv e-prints}\
  ,\ \bibinfo {eid} {arXiv:2002.05165}} (\bibinfo {year}
  {2020}{\natexlab{a}})},\ \Eprint {http://arxiv.org/abs/2002.05165}
  {arXiv:2002.05165 [astro-ph.CO]} \BibitemShut {NoStop}%
\bibitem [{\citenamefont {{Caputo}}\ \emph
  {et~al.}(2020{\natexlab{b}})\citenamefont {{Caputo}}, \citenamefont {{Liu}},
  \citenamefont {{Mishra-Sharma}},\ and\ \citenamefont
  {{Ruderman}}}]{CapLiuMis20b}%
  \BibitemOpen
  \bibfield  {author} {\bibinfo {author} {\bibfnamefont {A.}~\bibnamefont
  {{Caputo}}}, \bibinfo {author} {\bibfnamefont {H.}~\bibnamefont {{Liu}}},
  \bibinfo {author} {\bibfnamefont {S.}~\bibnamefont {{Mishra-Sharma}}}, \ and\
  \bibinfo {author} {\bibfnamefont {J.~T.}\ \bibnamefont {{Ruderman}}},\
  }\href@noop {} {\bibfield  {journal} {\bibinfo  {journal} {arXiv e-prints}\
  ,\ \bibinfo {eid} {arXiv:2004.06733}} (\bibinfo {year}
  {2020}{\natexlab{b}})},\ \Eprint {http://arxiv.org/abs/2004.06733}
  {arXiv:2004.06733 [astro-ph.CO]} \BibitemShut {NoStop}%
\bibitem [{\citenamefont {{Graham}}\ \emph {et~al.}(2018)\citenamefont
  {{Graham}}, \citenamefont {{Kaplan}}, \citenamefont {{Mardon}}, \citenamefont
  {{Rajendran}}, \citenamefont {{Terrano}}, \citenamefont {{Trahms}},\ and\
  \citenamefont {{Wilkason}}}]{GraKapMar18}%
  \BibitemOpen
  \bibfield  {author} {\bibinfo {author} {\bibfnamefont {P.~W.}\ \bibnamefont
  {{Graham}}}, \bibinfo {author} {\bibfnamefont {D.~E.}\ \bibnamefont
  {{Kaplan}}}, \bibinfo {author} {\bibfnamefont {J.}~\bibnamefont {{Mardon}}},
  \bibinfo {author} {\bibfnamefont {S.}~\bibnamefont {{Rajendran}}}, \bibinfo
  {author} {\bibfnamefont {W.~A.}\ \bibnamefont {{Terrano}}}, \bibinfo {author}
  {\bibfnamefont {L.}~\bibnamefont {{Trahms}}}, \ and\ \bibinfo {author}
  {\bibfnamefont {T.}~\bibnamefont {{Wilkason}}},\ }\href {\doibase
  10.1103/PhysRevD.97.055006} {\bibfield  {journal} {\bibinfo  {journal}
  {\prd}\ }\textbf {\bibinfo {volume} {97}},\ \bibinfo {eid} {055006} (\bibinfo
  {year} {2018})},\ \Eprint {http://arxiv.org/abs/1709.07852} {arXiv:1709.07852
  [hep-ph]} \BibitemShut {NoStop}%
\bibitem [{\citenamefont {{Witte}}\ \emph {et~al.}(2020)\citenamefont
  {{Witte}}, \citenamefont {{Rosauro-Alcaraz}}, \citenamefont {{McDermott}},\
  and\ \citenamefont {{Poulin}}}]{WitRos20}%
  \BibitemOpen
  \bibfield  {author} {\bibinfo {author} {\bibfnamefont {S.~J.}\ \bibnamefont
  {{Witte}}}, \bibinfo {author} {\bibfnamefont {S.}~\bibnamefont
  {{Rosauro-Alcaraz}}}, \bibinfo {author} {\bibfnamefont {S.~D.}\ \bibnamefont
  {{McDermott}}}, \ and\ \bibinfo {author} {\bibfnamefont {V.}~\bibnamefont
  {{Poulin}}},\ }\href@noop {} {\bibfield  {journal} {\bibinfo  {journal}
  {arXiv e-prints}\ ,\ \bibinfo {eid} {arXiv:2003.13698}} (\bibinfo {year}
  {2020})},\ \Eprint {http://arxiv.org/abs/2003.13698} {arXiv:2003.13698
  [astro-ph.CO]} \BibitemShut {NoStop}%
\bibitem [{\citenamefont {Silva-Feaver}\ \emph {et~al.}(2017)\citenamefont
  {Silva-Feaver} \emph {et~al.}}]{Silva-Feaver:2016qhh}%
  \BibitemOpen
  \bibfield  {author} {\bibinfo {author} {\bibfnamefont {M.}~\bibnamefont
  {Silva-Feaver}} \emph {et~al.},\ }\bibfield  {booktitle} {\emph {\bibinfo
  {booktitle} {{Proceedings, Applied Superconductivity Conference (ASC 2016):
  Denver, Colorado, September 4-9, 2016}}},\ }\href {\doibase
  10.1109/TASC.2016.2631425} {\bibfield  {journal} {\bibinfo  {journal} {IEEE
  Trans. Appl. Supercond.}\ }\textbf {\bibinfo {volume} {27}},\ \bibinfo
  {pages} {1400204} (\bibinfo {year} {2017})},\ \Eprint
  {http://arxiv.org/abs/1610.09344} {arXiv:1610.09344 [astro-ph.IM]}
  \BibitemShut {NoStop}%
\bibitem [{\citenamefont {Harnik}(2019)}]{HarnikSRF}%
  \BibitemOpen
  \bibfield  {author} {\bibinfo {author} {\bibfnamefont {R.}~\bibnamefont
  {Harnik}},\ }\href
  {https://indico.fnal.gov/event/19433/session/2/contribution/1/material/slides/0.pdf}
  {\enquote {\bibinfo {title} {Dark {SRF} - theory},}\ } (\bibinfo {year}
  {2019}),\ \bibinfo {note} {2019 January PAC Meeting}\BibitemShut {NoStop}%
\bibitem [{\citenamefont {Grassellino}(2019)}]{GrassellinoSRF}%
  \BibitemOpen
  \bibfield  {author} {\bibinfo {author} {\bibfnamefont {A.}~\bibnamefont
  {Grassellino}},\ }\href
  {https://indico.fnal.gov/event/19433/session/2/contribution/2/material/slides/0.pdf}
  {\enquote {\bibinfo {title} {Dark {SRF} - experiment},}\ } (\bibinfo {year}
  {2019}),\ \bibinfo {note} {2019 January PAC Meeting}\BibitemShut {NoStop}%
\bibitem [{\citenamefont {{Pierce}}\ \emph {et~al.}(2018)\citenamefont
  {{Pierce}}, \citenamefont {{Riles}},\ and\ \citenamefont
  {{Zhao}}}]{PieRilZha18}%
  \BibitemOpen
  \bibfield  {author} {\bibinfo {author} {\bibfnamefont {A.}~\bibnamefont
  {{Pierce}}}, \bibinfo {author} {\bibfnamefont {K.}~\bibnamefont {{Riles}}}, \
  and\ \bibinfo {author} {\bibfnamefont {Y.}~\bibnamefont {{Zhao}}},\ }\href
  {\doibase 10.1103/PhysRevLett.121.061102} {\bibfield  {journal} {\bibinfo
  {journal} {\prl}\ }\textbf {\bibinfo {volume} {121}},\ \bibinfo {eid}
  {061102} (\bibinfo {year} {2018})},\ \Eprint
  {http://arxiv.org/abs/1801.10161} {arXiv:1801.10161 [hep-ph]} \BibitemShut
  {NoStop}%
\bibitem [{\citenamefont {{Baryakhtar}}\ \emph {et~al.}(2018)\citenamefont
  {{Baryakhtar}}, \citenamefont {{Huang}},\ and\ \citenamefont
  {{Lasenby}}}]{BarHuaLas18}%
  \BibitemOpen
  \bibfield  {author} {\bibinfo {author} {\bibfnamefont {M.}~\bibnamefont
  {{Baryakhtar}}}, \bibinfo {author} {\bibfnamefont {J.}~\bibnamefont
  {{Huang}}}, \ and\ \bibinfo {author} {\bibfnamefont {R.}~\bibnamefont
  {{Lasenby}}},\ }\href {\doibase 10.1103/PhysRevD.98.035006} {\bibfield
  {journal} {\bibinfo  {journal} {\prd}\ }\textbf {\bibinfo {volume} {98}},\
  \bibinfo {eid} {035006} (\bibinfo {year} {2018})},\ \Eprint
  {http://arxiv.org/abs/1803.11455} {arXiv:1803.11455 [hep-ph]} \BibitemShut
  {NoStop}%
\bibitem [{\citenamefont {Battaglieri}\ and\ \citenamefont
  {et~al.}(2017)}]{osti_1409838}%
  \BibitemOpen
  \bibfield  {author} {\bibinfo {author} {\bibfnamefont {M.}~\bibnamefont
  {Battaglieri}}\ and\ \bibinfo {author} {\bibnamefont {et~al.}},\ }\href
  {https://www.osti.gov/biblio/1409838} {\  (\bibinfo {year}
  {2017})}\BibitemShut {NoStop}%
\bibitem [{\citenamefont {{Buckley}}\ and\ \citenamefont
  {{Peter}}(2018)}]{BucPet18}%
  \BibitemOpen
  \bibfield  {author} {\bibinfo {author} {\bibfnamefont {M.~R.}\ \bibnamefont
  {{Buckley}}}\ and\ \bibinfo {author} {\bibfnamefont {A.~H.~G.}\ \bibnamefont
  {{Peter}}},\ }\href {\doibase 10.1016/j.physrep.2018.07.003} {\bibfield
  {journal} {\bibinfo  {journal} {\physrep}\ }\textbf {\bibinfo {volume}
  {761}},\ \bibinfo {pages} {1} (\bibinfo {year} {2018})},\ \Eprint
  {http://arxiv.org/abs/1712.06615} {arXiv:1712.06615 [astro-ph.CO]}
  \BibitemShut {NoStop}%
\bibitem [{\citenamefont {{Koribalski}}\ \emph {et~al.}(2020)\citenamefont
  {{Koribalski}} \emph {et~al.}}]{Kor20_Wallaby}%
  \BibitemOpen
  \bibfield  {author} {\bibinfo {author} {\bibfnamefont {B.~S.}\ \bibnamefont
  {{Koribalski}}} \emph {et~al.},\ }\href {\doibase 10.1007/s10509-020-03831-4}
  {\bibfield  {journal} {\bibinfo  {journal} {\apss}\ }\textbf {\bibinfo
  {volume} {365}},\ \bibinfo {eid} {118} (\bibinfo {year} {2020})},\ \Eprint
  {http://arxiv.org/abs/2002.07311} {arXiv:2002.07311 [astro-ph.GA]}
  \BibitemShut {NoStop}%
\bibitem [{\citenamefont {Bourke}\ \emph {et~al.}(2015)\citenamefont {Bourke}
  \emph {et~al.}}]{SKA}%
  \BibitemOpen
  \bibinfo {editor} {\bibfnamefont {T.~L.}\ \bibnamefont {Bourke}} \emph
  {et~al.},\ eds.,\ \href@noop {} {\emph {\bibinfo {title} {{Proceedings,
  Advancing Astrophysics with the Square Kilometre Array (AASKA14)}: {Giardini
  Naxos, Italy, June 9-13, 2014}}}},\ Vol.\ \bibinfo {volume} {AASKA14}\
  (\bibinfo  {publisher} {SISSA},\ \bibinfo {year} {2015})\BibitemShut
  {NoStop}%
\bibitem [{\citenamefont {{Koposov}}\ \emph {et~al.}(2015)\citenamefont
  {{Koposov}}, \citenamefont {{Belokurov}}, \citenamefont {{Torrealba}},\ and\
  \citenamefont {{Evans}}}]{Kop15_DES}%
  \BibitemOpen
  \bibfield  {author} {\bibinfo {author} {\bibfnamefont {S.~E.}\ \bibnamefont
  {{Koposov}}}, \bibinfo {author} {\bibfnamefont {V.}~\bibnamefont
  {{Belokurov}}}, \bibinfo {author} {\bibfnamefont {G.}~\bibnamefont
  {{Torrealba}}}, \ and\ \bibinfo {author} {\bibfnamefont {N.~W.}\ \bibnamefont
  {{Evans}}},\ }\href {\doibase 10.1088/0004-637X/805/2/130} {\bibfield
  {journal} {\bibinfo  {journal} {\apj}\ }\textbf {\bibinfo {volume} {805}},\
  \bibinfo {eid} {130} (\bibinfo {year} {2015})},\ \Eprint
  {http://arxiv.org/abs/1503.02079} {arXiv:1503.02079 [astro-ph.GA]}
  \BibitemShut {NoStop}%
\bibitem [{\citenamefont {{Bechtol}}\ \emph {et~al.}(2015)\citenamefont
  {{Bechtol}} \emph {et~al.}}]{Bec15_DES}%
  \BibitemOpen
  \bibfield  {author} {\bibinfo {author} {\bibfnamefont {K.}~\bibnamefont
  {{Bechtol}}} \emph {et~al.},\ }\href {\doibase 10.1088/0004-637X/807/1/50}
  {\bibfield  {journal} {\bibinfo  {journal} {\apj}\ }\textbf {\bibinfo
  {volume} {807}},\ \bibinfo {eid} {50} (\bibinfo {year} {2015})},\ \Eprint
  {http://arxiv.org/abs/1503.02584} {arXiv:1503.02584 [astro-ph.GA]}
  \BibitemShut {NoStop}%
\bibitem [{\citenamefont {{Drlica-Wagner}}\ \emph
  {et~al.}(2015{\natexlab{b}})\citenamefont {{Drlica-Wagner}} \emph
  {et~al.}}]{Drl15_DES}%
  \BibitemOpen
  \bibfield  {author} {\bibinfo {author} {\bibfnamefont {A.}~\bibnamefont
  {{Drlica-Wagner}}} \emph {et~al.},\ }\href {\doibase
  10.1088/0004-637X/813/2/109} {\bibfield  {journal} {\bibinfo  {journal}
  {\apj}\ }\textbf {\bibinfo {volume} {813}},\ \bibinfo {eid} {109} (\bibinfo
  {year} {2015}{\natexlab{b}})},\ \Eprint {http://arxiv.org/abs/1508.03622}
  {arXiv:1508.03622 [astro-ph.GA]} \BibitemShut {NoStop}%
\bibitem [{\citenamefont {Drlica-Wagner}\ \emph {et~al.}(2020)\citenamefont
  {Drlica-Wagner} \emph {et~al.}}]{Drl20}%
  \BibitemOpen
  \bibfield  {author} {\bibinfo {author} {\bibfnamefont {A.}~\bibnamefont
  {Drlica-Wagner}} \emph {et~al.} (\bibinfo {collaboration} {DES}),\ }\href
  {\doibase 10.3847/1538-4357/ab7eb9} {\bibfield  {journal} {\bibinfo
  {journal} {Astrophys. J.}\ }\textbf {\bibinfo {volume} {893}},\ \bibinfo
  {pages} {1} (\bibinfo {year} {2020})},\ \Eprint
  {http://arxiv.org/abs/1912.03302} {arXiv:1912.03302 [astro-ph.GA]}
  \BibitemShut {NoStop}%
\bibitem [{\citenamefont {{Geha}}\ \emph {et~al.}(2017)\citenamefont {{Geha}},
  \citenamefont {{Wechsler}}, \citenamefont {{Mao}}, \citenamefont
  {{Tollerud}}, \citenamefont {{Weiner}}, \citenamefont {{Bernstein}},
  \citenamefont {{Hoyle}}, \citenamefont {{Marchi}}, \citenamefont
  {{Marshall}}, \citenamefont {{Mu{\~n}oz}},\ and\ \citenamefont
  {{Lu}}}]{Geh17}%
  \BibitemOpen
  \bibfield  {author} {\bibinfo {author} {\bibfnamefont {M.}~\bibnamefont
  {{Geha}}}, \bibinfo {author} {\bibfnamefont {R.~H.}\ \bibnamefont
  {{Wechsler}}}, \bibinfo {author} {\bibfnamefont {Y.-Y.}\ \bibnamefont
  {{Mao}}}, \bibinfo {author} {\bibfnamefont {E.~J.}\ \bibnamefont
  {{Tollerud}}}, \bibinfo {author} {\bibfnamefont {B.}~\bibnamefont
  {{Weiner}}}, \bibinfo {author} {\bibfnamefont {R.}~\bibnamefont
  {{Bernstein}}}, \bibinfo {author} {\bibfnamefont {B.}~\bibnamefont
  {{Hoyle}}}, \bibinfo {author} {\bibfnamefont {S.}~\bibnamefont {{Marchi}}},
  \bibinfo {author} {\bibfnamefont {P.~J.}\ \bibnamefont {{Marshall}}},
  \bibinfo {author} {\bibfnamefont {R.}~\bibnamefont {{Mu{\~n}oz}}}, \ and\
  \bibinfo {author} {\bibfnamefont {Y.}~\bibnamefont {{Lu}}},\ }\href {\doibase
  10.3847/1538-4357/aa8626} {\bibfield  {journal} {\bibinfo  {journal} {\apj}\
  }\textbf {\bibinfo {volume} {847}},\ \bibinfo {eid} {4} (\bibinfo {year}
  {2017})},\ \Eprint {http://arxiv.org/abs/1705.06743} {arXiv:1705.06743
  [astro-ph.GA]} \BibitemShut {NoStop}%
\bibitem [{\citenamefont {{Mao}}\ \emph {et~al.}(2021)\citenamefont {{Mao}},
  \citenamefont {{Geha}}, \citenamefont {{Wechsler}}, \citenamefont {{Weiner}},
  \citenamefont {{Tollerud}}, \citenamefont {{Nadler}},\ and\ \citenamefont
  {{Kallivayalil}}}]{Mao21}%
  \BibitemOpen
  \bibfield  {author} {\bibinfo {author} {\bibfnamefont {Y.-Y.}\ \bibnamefont
  {{Mao}}}, \bibinfo {author} {\bibfnamefont {M.}~\bibnamefont {{Geha}}},
  \bibinfo {author} {\bibfnamefont {R.~H.}\ \bibnamefont {{Wechsler}}},
  \bibinfo {author} {\bibfnamefont {B.}~\bibnamefont {{Weiner}}}, \bibinfo
  {author} {\bibfnamefont {E.~J.}\ \bibnamefont {{Tollerud}}}, \bibinfo
  {author} {\bibfnamefont {E.~O.}\ \bibnamefont {{Nadler}}}, \ and\ \bibinfo
  {author} {\bibfnamefont {N.}~\bibnamefont {{Kallivayalil}}},\ }\href
  {\doibase 10.3847/1538-4357/abce58} {\bibfield  {journal} {\bibinfo
  {journal} {\apj}\ }\textbf {\bibinfo {volume} {907}},\ \bibinfo {eid} {85}
  (\bibinfo {year} {2021})},\ \Eprint {http://arxiv.org/abs/2008.12783}
  {arXiv:2008.12783 [astro-ph.GA]} \BibitemShut {NoStop}%
\bibitem [{\citenamefont {Aghamousa}\ \emph {et~al.}(2016)\citenamefont
  {Aghamousa} \emph {et~al.}}]{DESI}%
  \BibitemOpen
  \bibfield  {author} {\bibinfo {author} {\bibfnamefont {A.}~\bibnamefont
  {Aghamousa}} \emph {et~al.} (\bibinfo {collaboration} {DESI}),\ }\href@noop
  {} {\  (\bibinfo {year} {2016})},\ \Eprint {http://arxiv.org/abs/1611.00036}
  {arXiv:1611.00036 [astro-ph.IM]} \BibitemShut {NoStop}%
\bibitem [{\citenamefont {{Aihara}}\ \emph {et~al.}(2018)\citenamefont
  {{Aihara}} \emph {et~al.}}]{Aih18}%
  \BibitemOpen
  \bibfield  {author} {\bibinfo {author} {\bibfnamefont {H.}~\bibnamefont
  {{Aihara}}} \emph {et~al.},\ }\href {\doibase 10.1093/pasj/psx066} {\bibfield
   {journal} {\bibinfo  {journal} {\pasj}\ }\textbf {\bibinfo {volume} {70}},\
  \bibinfo {eid} {S4} (\bibinfo {year} {2018})},\ \Eprint
  {http://arxiv.org/abs/1704.05858} {arXiv:1704.05858 [astro-ph.IM]}
  \BibitemShut {NoStop}%
\bibitem [{\citenamefont {{McConnachie}}\ \emph {et~al.}(2016)\citenamefont
  {{McConnachie}} \emph {et~al.}}]{McC16}%
  \BibitemOpen
  \bibfield  {author} {\bibinfo {author} {\bibfnamefont {A.~W.}\ \bibnamefont
  {{McConnachie}}} \emph {et~al.},\ }\href@noop {} {\bibfield  {journal}
  {\bibinfo  {journal} {arXiv e-prints}\ ,\ \bibinfo {eid} {arXiv:1606.00060}}
  (\bibinfo {year} {2016})},\ \Eprint {http://arxiv.org/abs/1606.00060}
  {arXiv:1606.00060 [astro-ph.IM]} \BibitemShut {NoStop}%
\bibitem [{\citenamefont {{LSST Dark Energy Science
  Collaboration}}(2012)}]{LSST}%
  \BibitemOpen
  \bibfield  {author} {\bibinfo {author} {\bibnamefont {{LSST Dark Energy
  Science Collaboration}}},\ }\href@noop {} {\bibfield  {journal} {\bibinfo
  {journal} {arXiv e-prints}\ ,\ \bibinfo {eid} {arXiv:1211.0310}} (\bibinfo
  {year} {2012})},\ \Eprint {http://arxiv.org/abs/1211.0310} {arXiv:1211.0310
  [astro-ph.CO]} \BibitemShut {NoStop}%
\bibitem [{\citenamefont {{Drlica-Wagner}}\ \emph {et~al.}(2019)\citenamefont
  {{Drlica-Wagner}} \emph {et~al.}}]{Drl19_LSST}%
  \BibitemOpen
  \bibfield  {author} {\bibinfo {author} {\bibfnamefont {A.}~\bibnamefont
  {{Drlica-Wagner}}} \emph {et~al.},\ }\href@noop {} {\bibfield  {journal}
  {\bibinfo  {journal} {arXiv e-prints}\ ,\ \bibinfo {eid} {arXiv:1902.01055}}
  (\bibinfo {year} {2019})},\ \Eprint {http://arxiv.org/abs/1902.01055}
  {arXiv:1902.01055 [astro-ph.CO]} \BibitemShut {NoStop}%
\bibitem [{\citenamefont {Dor\'e}\ \emph {et~al.}(2018)\citenamefont {Dor\'e}
  \emph {et~al.}}]{WFIRST}%
  \BibitemOpen
  \bibfield  {author} {\bibinfo {author} {\bibfnamefont {O.}~\bibnamefont
  {Dor\'e}} \emph {et~al.} (\bibinfo {collaboration} {WFIRST}),\ }\href@noop {}
  {\  (\bibinfo {year} {2018})},\ \Eprint {http://arxiv.org/abs/1804.03628}
  {arXiv:1804.03628 [astro-ph.CO]} \BibitemShut {NoStop}%
\bibitem [{\citenamefont {{Chivukula}}\ \emph {et~al.}(1990)\citenamefont
  {{Chivukula}}, \citenamefont {{Cohen}}, \citenamefont {{Dimopoulos}},\ and\
  \citenamefont {{Walker}}}]{Chi90}%
  \BibitemOpen
  \bibfield  {author} {\bibinfo {author} {\bibfnamefont {R.~S.}\ \bibnamefont
  {{Chivukula}}}, \bibinfo {author} {\bibfnamefont {A.~G.}\ \bibnamefont
  {{Cohen}}}, \bibinfo {author} {\bibfnamefont {S.}~\bibnamefont
  {{Dimopoulos}}}, \ and\ \bibinfo {author} {\bibfnamefont {T.~P.}\
  \bibnamefont {{Walker}}},\ }\href {\doibase 10.1103/PhysRevLett.65.957}
  {\bibfield  {journal} {\bibinfo  {journal} {\prl}\ }\textbf {\bibinfo
  {volume} {65}},\ \bibinfo {pages} {957} (\bibinfo {year} {1990})}\BibitemShut
  {NoStop}%
\bibitem [{\citenamefont {{Bhoonah}}\ \emph
  {et~al.}(2018{\natexlab{b}})\citenamefont {{Bhoonah}}, \citenamefont
  {{Bramante}}, \citenamefont {{Elahi}},\ and\ \citenamefont
  {{Schon}}}]{BhoBramante18}%
  \BibitemOpen
  \bibfield  {author} {\bibinfo {author} {\bibfnamefont {A.}~\bibnamefont
  {{Bhoonah}}}, \bibinfo {author} {\bibfnamefont {J.}~\bibnamefont
  {{Bramante}}}, \bibinfo {author} {\bibfnamefont {F.}~\bibnamefont {{Elahi}}},
  \ and\ \bibinfo {author} {\bibfnamefont {S.}~\bibnamefont {{Schon}}},\ }\href
  {\doibase 10.1103/PhysRevLett.121.131101} {\bibfield  {journal} {\bibinfo
  {journal} {\prl}\ }\textbf {\bibinfo {volume} {121}},\ \bibinfo {eid}
  {131101} (\bibinfo {year} {2018}{\natexlab{b}})},\ \Eprint
  {http://arxiv.org/abs/1806.06857} {arXiv:1806.06857 [hep-ph]} \BibitemShut
  {NoStop}%
\bibitem [{\citenamefont {{Bhoonah}}\ \emph {et~al.}(2020)\citenamefont
  {{Bhoonah}}, \citenamefont {{Bramante}}, \citenamefont {{Schon}},\ and\
  \citenamefont {{Song}}}]{BhoBra20}%
  \BibitemOpen
  \bibfield  {author} {\bibinfo {author} {\bibfnamefont {A.}~\bibnamefont
  {{Bhoonah}}}, \bibinfo {author} {\bibfnamefont {J.}~\bibnamefont
  {{Bramante}}}, \bibinfo {author} {\bibfnamefont {S.}~\bibnamefont {{Schon}}},
  \ and\ \bibinfo {author} {\bibfnamefont {N.}~\bibnamefont {{Song}}},\
  }\href@noop {} {\bibfield  {journal} {\bibinfo  {journal} {arXiv e-prints}\
  ,\ \bibinfo {eid} {arXiv:2010.07240}} (\bibinfo {year} {2020})},\ \Eprint
  {http://arxiv.org/abs/2010.07240} {arXiv:2010.07240 [hep-ph]} \BibitemShut
  {NoStop}%
\bibitem [{\citenamefont {{McClure-Griffiths}}\ \emph
  {et~al.}(2013)\citenamefont {{McClure-Griffiths}}, \citenamefont {{Green}},
  \citenamefont {{Hill}}, \citenamefont {{Lockman}}, \citenamefont {{Dickey}},
  \citenamefont {{Gaensler}},\ and\ \citenamefont {{Green}}}]{McClureLock13}%
  \BibitemOpen
  \bibfield  {author} {\bibinfo {author} {\bibfnamefont {N.~M.}\ \bibnamefont
  {{McClure-Griffiths}}}, \bibinfo {author} {\bibfnamefont {J.~A.}\
  \bibnamefont {{Green}}}, \bibinfo {author} {\bibfnamefont {A.~S.}\
  \bibnamefont {{Hill}}}, \bibinfo {author} {\bibfnamefont {F.~J.}\
  \bibnamefont {{Lockman}}}, \bibinfo {author} {\bibfnamefont {J.~M.}\
  \bibnamefont {{Dickey}}}, \bibinfo {author} {\bibfnamefont {B.~M.}\
  \bibnamefont {{Gaensler}}}, \ and\ \bibinfo {author} {\bibfnamefont {A.~J.}\
  \bibnamefont {{Green}}},\ }\href {\doibase 10.1088/2041-8205/770/1/L4}
  {\bibfield  {journal} {\bibinfo  {journal} {\apjl}\ }\textbf {\bibinfo
  {volume} {770}},\ \bibinfo {eid} {L4} (\bibinfo {year} {2013})},\ \Eprint
  {http://arxiv.org/abs/1304.7538} {arXiv:1304.7538} \BibitemShut {NoStop}%
\bibitem [{\citenamefont {{Di Teodoro}}\ \emph {et~al.}(2018)\citenamefont {{Di
  Teodoro}}, \citenamefont {{McClure-Griffiths}}, \citenamefont {{Lockman}},
  \citenamefont {{Denbo}}, \citenamefont {{Endsley}}, \citenamefont {{Ford}},\
  and\ \citenamefont {{Harrington}}}]{DiTeodoroLock18}%
  \BibitemOpen
  \bibfield  {author} {\bibinfo {author} {\bibfnamefont {E.~M.}\ \bibnamefont
  {{Di Teodoro}}}, \bibinfo {author} {\bibfnamefont {N.~M.}\ \bibnamefont
  {{McClure-Griffiths}}}, \bibinfo {author} {\bibfnamefont {F.~J.}\
  \bibnamefont {{Lockman}}}, \bibinfo {author} {\bibfnamefont {S.~R.}\
  \bibnamefont {{Denbo}}}, \bibinfo {author} {\bibfnamefont {R.}~\bibnamefont
  {{Endsley}}}, \bibinfo {author} {\bibfnamefont {H.~A.}\ \bibnamefont
  {{Ford}}}, \ and\ \bibinfo {author} {\bibfnamefont {K.}~\bibnamefont
  {{Harrington}}},\ }\href {\doibase 10.3847/1538-4357/aaad6a} {\bibfield
  {journal} {\bibinfo  {journal} {\apj}\ }\textbf {\bibinfo {volume} {855}},\
  \bibinfo {eid} {33} (\bibinfo {year} {2018})},\ \Eprint
  {http://arxiv.org/abs/1802.02152} {arXiv:1802.02152} \BibitemShut {NoStop}%
\bibitem [{\citenamefont {{Cooper}}\ \emph {et~al.}(2008)\citenamefont
  {{Cooper}}, \citenamefont {{Bicknell}}, \citenamefont {{Sutherland}},\ and\
  \citenamefont {{Bland-Hawthorn}}}]{cooper08_WindSim}%
  \BibitemOpen
  \bibfield  {author} {\bibinfo {author} {\bibfnamefont {J.~L.}\ \bibnamefont
  {{Cooper}}}, \bibinfo {author} {\bibfnamefont {G.~V.}\ \bibnamefont
  {{Bicknell}}}, \bibinfo {author} {\bibfnamefont {R.~S.}\ \bibnamefont
  {{Sutherland}}}, \ and\ \bibinfo {author} {\bibfnamefont {J.}~\bibnamefont
  {{Bland-Hawthorn}}},\ }\href {\doibase 10.1086/524918} {\bibfield  {journal}
  {\bibinfo  {journal} {\apj}\ }\textbf {\bibinfo {volume} {674}},\ \bibinfo
  {pages} {157} (\bibinfo {year} {2008})}\BibitemShut {NoStop}%
\bibitem [{\citenamefont {{Scannapieco}}\ and\ \citenamefont
  {{Br{\"u}ggen}}(2015)}]{ScaBru15_WindSim}%
  \BibitemOpen
  \bibfield  {author} {\bibinfo {author} {\bibfnamefont {E.}~\bibnamefont
  {{Scannapieco}}}\ and\ \bibinfo {author} {\bibfnamefont {M.}~\bibnamefont
  {{Br{\"u}ggen}}},\ }\href {\doibase 10.1088/0004-637X/805/2/158} {\bibfield
  {journal} {\bibinfo  {journal} {\apj}\ }\textbf {\bibinfo {volume} {805}},\
  \bibinfo {eid} {158} (\bibinfo {year} {2015})},\ \Eprint
  {http://arxiv.org/abs/1503.06800} {arXiv:1503.06800 [astro-ph.GA]}
  \BibitemShut {NoStop}%
\bibitem [{\citenamefont {{Schneider}}\ and\ \citenamefont
  {{Robertson}}(2017)}]{SchRob17_WindSim}%
  \BibitemOpen
  \bibfield  {author} {\bibinfo {author} {\bibfnamefont {E.~E.}\ \bibnamefont
  {{Schneider}}}\ and\ \bibinfo {author} {\bibfnamefont {B.~E.}\ \bibnamefont
  {{Robertson}}},\ }\href {\doibase 10.3847/1538-4357/834/2/144} {\bibfield
  {journal} {\bibinfo  {journal} {\apj}\ }\textbf {\bibinfo {volume} {834}},\
  \bibinfo {eid} {144} (\bibinfo {year} {2017})},\ \Eprint
  {http://arxiv.org/abs/1607.01788} {arXiv:1607.01788 [astro-ph.GA]}
  \BibitemShut {NoStop}%
\bibitem [{\citenamefont {{Armillotta}}\ \emph {et~al.}(2017)\citenamefont
  {{Armillotta}}, \citenamefont {{Fraternali}}, \citenamefont {{Werk}},
  \citenamefont {{Prochaska}},\ and\ \citenamefont
  {{Marinacci}}}]{ArmFra17_WindSim}%
  \BibitemOpen
  \bibfield  {author} {\bibinfo {author} {\bibfnamefont {L.}~\bibnamefont
  {{Armillotta}}}, \bibinfo {author} {\bibfnamefont {F.}~\bibnamefont
  {{Fraternali}}}, \bibinfo {author} {\bibfnamefont {J.~K.}\ \bibnamefont
  {{Werk}}}, \bibinfo {author} {\bibfnamefont {J.~X.}\ \bibnamefont
  {{Prochaska}}}, \ and\ \bibinfo {author} {\bibfnamefont {F.}~\bibnamefont
  {{Marinacci}}},\ }\href {\doibase 10.1093/mnras/stx1239} {\bibfield
  {journal} {\bibinfo  {journal} {\mnras}\ }\textbf {\bibinfo {volume} {470}},\
  \bibinfo {pages} {114} (\bibinfo {year} {2017})},\ \Eprint
  {http://arxiv.org/abs/1608.05416} {arXiv:1608.05416 [astro-ph.GA]}
  \BibitemShut {NoStop}%
\bibitem [{\citenamefont {{Melioli}}\ \emph {et~al.}(2013)\citenamefont
  {{Melioli}}, \citenamefont {{de Gouveia Dal Pino}},\ and\ \citenamefont
  {{Geraissate}}}]{MelGou13_WindSim}%
  \BibitemOpen
  \bibfield  {author} {\bibinfo {author} {\bibfnamefont {C.}~\bibnamefont
  {{Melioli}}}, \bibinfo {author} {\bibfnamefont {E.~M.}\ \bibnamefont {{de
  Gouveia Dal Pino}}}, \ and\ \bibinfo {author} {\bibfnamefont {F.~G.}\
  \bibnamefont {{Geraissate}}},\ }\href {\doibase 10.1093/mnras/stt126}
  {\bibfield  {journal} {\bibinfo  {journal} {\mnras}\ }\textbf {\bibinfo
  {volume} {430}},\ \bibinfo {pages} {3235} (\bibinfo {year} {2013})},\ \Eprint
  {http://arxiv.org/abs/1301.5005} {arXiv:1301.5005 [astro-ph.CO]} \BibitemShut
  {NoStop}%
\bibitem [{\citenamefont {{McCourt}}\ \emph {et~al.}(2018)\citenamefont
  {{McCourt}}, \citenamefont {{Oh}}, \citenamefont {{O'Leary}},\ and\
  \citenamefont {{Madigan}}}]{McCourt18_WindSim}%
  \BibitemOpen
  \bibfield  {author} {\bibinfo {author} {\bibfnamefont {M.}~\bibnamefont
  {{McCourt}}}, \bibinfo {author} {\bibfnamefont {S.~P.}\ \bibnamefont {{Oh}}},
  \bibinfo {author} {\bibfnamefont {R.}~\bibnamefont {{O'Leary}}}, \ and\
  \bibinfo {author} {\bibfnamefont {A.-M.}\ \bibnamefont {{Madigan}}},\ }\href
  {\doibase 10.1093/mnras/stx2687} {\bibfield  {journal} {\bibinfo  {journal}
  {\mnras}\ }\textbf {\bibinfo {volume} {473}},\ \bibinfo {pages} {5407}
  (\bibinfo {year} {2018})},\ \Eprint {http://arxiv.org/abs/1610.01164}
  {arXiv:1610.01164} \BibitemShut {NoStop}%
\bibitem [{\citenamefont {{Pidopryhora}}\ \emph {et~al.}(2015)\citenamefont
  {{Pidopryhora}}, \citenamefont {{Lockman}}, \citenamefont {{Dickey}},\ and\
  \citenamefont {{Rupen}}}]{PidLock15}%
  \BibitemOpen
  \bibfield  {author} {\bibinfo {author} {\bibfnamefont {Y.}~\bibnamefont
  {{Pidopryhora}}}, \bibinfo {author} {\bibfnamefont {F.~J.}\ \bibnamefont
  {{Lockman}}}, \bibinfo {author} {\bibfnamefont {J.~M.}\ \bibnamefont
  {{Dickey}}}, \ and\ \bibinfo {author} {\bibfnamefont {M.~P.}\ \bibnamefont
  {{Rupen}}},\ }\href {\doibase 10.1088/0067-0049/219/2/16} {\bibfield
  {journal} {\bibinfo  {journal} {\apjs}\ }\textbf {\bibinfo {volume} {219}},\
  \bibinfo {eid} {16} (\bibinfo {year} {2015})},\ \Eprint
  {http://arxiv.org/abs/1506.03873} {arXiv:1506.03873} \BibitemShut {NoStop}%
\bibitem [{\citenamefont {{Smith}}\ \emph {et~al.}(2017)\citenamefont
  {{Smith}}, \citenamefont {{Bryan}}, \citenamefont {{Glover}}, \citenamefont
  {{Goldbaum}}, \citenamefont {{Turk}}, \citenamefont {{Regan}}, \citenamefont
  {{Wise}}, \citenamefont {{Schive}}, \citenamefont {{Abel}}, \citenamefont
  {{Emerick}}, \citenamefont {{O'Shea}}, \citenamefont {{Anninos}},
  \citenamefont {{Hummels}},\ and\ \citenamefont {{Khochfar}}}]{grackle}%
  \BibitemOpen
  \bibfield  {author} {\bibinfo {author} {\bibfnamefont {B.~D.}\ \bibnamefont
  {{Smith}}}, \bibinfo {author} {\bibfnamefont {G.~L.}\ \bibnamefont
  {{Bryan}}}, \bibinfo {author} {\bibfnamefont {S.~C.~O.}\ \bibnamefont
  {{Glover}}}, \bibinfo {author} {\bibfnamefont {N.~J.}\ \bibnamefont
  {{Goldbaum}}}, \bibinfo {author} {\bibfnamefont {M.~J.}\ \bibnamefont
  {{Turk}}}, \bibinfo {author} {\bibfnamefont {J.}~\bibnamefont {{Regan}}},
  \bibinfo {author} {\bibfnamefont {J.~H.}\ \bibnamefont {{Wise}}}, \bibinfo
  {author} {\bibfnamefont {H.-Y.}\ \bibnamefont {{Schive}}}, \bibinfo {author}
  {\bibfnamefont {T.}~\bibnamefont {{Abel}}}, \bibinfo {author} {\bibfnamefont
  {A.}~\bibnamefont {{Emerick}}}, \bibinfo {author} {\bibfnamefont {B.~W.}\
  \bibnamefont {{O'Shea}}}, \bibinfo {author} {\bibfnamefont {P.}~\bibnamefont
  {{Anninos}}}, \bibinfo {author} {\bibfnamefont {C.~B.}\ \bibnamefont
  {{Hummels}}}, \ and\ \bibinfo {author} {\bibfnamefont {S.}~\bibnamefont
  {{Khochfar}}},\ }\href {\doibase 10.1093/mnras/stw3291} {\bibfield  {journal}
  {\bibinfo  {journal} {\mnras}\ }\textbf {\bibinfo {volume} {466}},\ \bibinfo
  {pages} {2217} (\bibinfo {year} {2017})},\ \Eprint
  {http://arxiv.org/abs/1610.09591} {arXiv:1610.09591} \BibitemShut {NoStop}%
\bibitem [{\citenamefont {{Ryan-Weber}}\ \emph {et~al.}(2008)\citenamefont
  {{Ryan-Weber}}, \citenamefont {{Begum}}, \citenamefont {{Oosterloo}},
  \citenamefont {{Pal}}, \citenamefont {{Irwin}}, \citenamefont {{Belokurov}},
  \citenamefont {{Evans}},\ and\ \citenamefont {{Zucker}}}]{leoObsOld08}%
  \BibitemOpen
  \bibfield  {author} {\bibinfo {author} {\bibfnamefont {E.~V.}\ \bibnamefont
  {{Ryan-Weber}}}, \bibinfo {author} {\bibfnamefont {A.}~\bibnamefont
  {{Begum}}}, \bibinfo {author} {\bibfnamefont {T.}~\bibnamefont
  {{Oosterloo}}}, \bibinfo {author} {\bibfnamefont {S.}~\bibnamefont {{Pal}}},
  \bibinfo {author} {\bibfnamefont {M.~J.}\ \bibnamefont {{Irwin}}}, \bibinfo
  {author} {\bibfnamefont {V.}~\bibnamefont {{Belokurov}}}, \bibinfo {author}
  {\bibfnamefont {N.~W.}\ \bibnamefont {{Evans}}}, \ and\ \bibinfo {author}
  {\bibfnamefont {D.~B.}\ \bibnamefont {{Zucker}}},\ }\href {\doibase
  10.1111/j.1365-2966.2007.12734.x} {\bibfield  {journal} {\bibinfo  {journal}
  {\mnras}\ }\textbf {\bibinfo {volume} {384}},\ \bibinfo {pages} {535}
  (\bibinfo {year} {2008})},\ \Eprint {http://arxiv.org/abs/0711.2979}
  {arXiv:0711.2979} \BibitemShut {NoStop}%
\bibitem [{\citenamefont {{Kirby}}\ \emph {et~al.}(2013)\citenamefont
  {{Kirby}}, \citenamefont {{Cohen}}, \citenamefont {{Guhathakurta}},
  \citenamefont {{Cheng}}, \citenamefont {{Bullock}},\ and\ \citenamefont
  {{Gallazzi}}}]{Kir13}%
  \BibitemOpen
  \bibfield  {author} {\bibinfo {author} {\bibfnamefont {E.~N.}\ \bibnamefont
  {{Kirby}}}, \bibinfo {author} {\bibfnamefont {J.~G.}\ \bibnamefont
  {{Cohen}}}, \bibinfo {author} {\bibfnamefont {P.}~\bibnamefont
  {{Guhathakurta}}}, \bibinfo {author} {\bibfnamefont {L.}~\bibnamefont
  {{Cheng}}}, \bibinfo {author} {\bibfnamefont {J.~S.}\ \bibnamefont
  {{Bullock}}}, \ and\ \bibinfo {author} {\bibfnamefont {A.}~\bibnamefont
  {{Gallazzi}}},\ }\href {\doibase 10.1088/0004-637X/779/2/102} {\bibfield
  {journal} {\bibinfo  {journal} {\apj}\ }\textbf {\bibinfo {volume} {779}},\
  \bibinfo {eid} {102} (\bibinfo {year} {2013})},\ \Eprint
  {http://arxiv.org/abs/1310.0814} {arXiv:1310.0814 [astro-ph.GA]} \BibitemShut
  {NoStop}%
\bibitem [{\citenamefont {{Weisz}}\ \emph {et~al.}(2012)\citenamefont
  {{Weisz}}, \citenamefont {{Zucker}}, \citenamefont {{Dolphin}}, \citenamefont
  {{Martin}}, \citenamefont {{de Jong}}, \citenamefont {{Holtzman}},
  \citenamefont {{Dalcanton}}, \citenamefont {{Gilbert}}, \citenamefont
  {{Williams}}, \citenamefont {{Bell}}, \citenamefont {{Belokurov}},\ and\
  \citenamefont {{Evans}}}]{weisz12}%
  \BibitemOpen
  \bibfield  {author} {\bibinfo {author} {\bibfnamefont {D.~R.}\ \bibnamefont
  {{Weisz}}}, \bibinfo {author} {\bibfnamefont {D.~B.}\ \bibnamefont
  {{Zucker}}}, \bibinfo {author} {\bibfnamefont {A.~E.}\ \bibnamefont
  {{Dolphin}}}, \bibinfo {author} {\bibfnamefont {N.~F.}\ \bibnamefont
  {{Martin}}}, \bibinfo {author} {\bibfnamefont {J.~T.~A.}\ \bibnamefont {{de
  Jong}}}, \bibinfo {author} {\bibfnamefont {J.~A.}\ \bibnamefont
  {{Holtzman}}}, \bibinfo {author} {\bibfnamefont {J.~J.}\ \bibnamefont
  {{Dalcanton}}}, \bibinfo {author} {\bibfnamefont {K.~M.}\ \bibnamefont
  {{Gilbert}}}, \bibinfo {author} {\bibfnamefont {B.~F.}\ \bibnamefont
  {{Williams}}}, \bibinfo {author} {\bibfnamefont {E.~F.}\ \bibnamefont
  {{Bell}}}, \bibinfo {author} {\bibfnamefont {V.}~\bibnamefont {{Belokurov}}},
  \ and\ \bibinfo {author} {\bibfnamefont {N.~W.}\ \bibnamefont {{Evans}}},\
  }\href {\doibase 10.1088/0004-637X/748/2/88} {\bibfield  {journal} {\bibinfo
  {journal} {\apj}\ }\textbf {\bibinfo {volume} {748}},\ \bibinfo {eid} {88}
  (\bibinfo {year} {2012})},\ \Eprint {http://arxiv.org/abs/1201.4859}
  {arXiv:1201.4859} \BibitemShut {NoStop}%
\bibitem [{\citenamefont {Faerman}\ \emph {et~al.}(2013)\citenamefont
  {Faerman}, \citenamefont {Sternberg},\ and\ \citenamefont
  {McKee}}]{faerman13}%
  \BibitemOpen
  \bibfield  {author} {\bibinfo {author} {\bibfnamefont {Y.}~\bibnamefont
  {Faerman}}, \bibinfo {author} {\bibfnamefont {A.}~\bibnamefont {Sternberg}},
  \ and\ \bibinfo {author} {\bibfnamefont {C.~F.}\ \bibnamefont {McKee}},\
  }\href {\doibase 10.1088/0004-637X/777/2/119} {\bibfield  {journal} {\bibinfo
   {journal} {Astrophys. J.}\ }\textbf {\bibinfo {volume} {777}},\ \bibinfo
  {pages} {119} (\bibinfo {year} {2013})},\ \Eprint
  {http://arxiv.org/abs/1309.0815} {arXiv:1309.0815 [astro-ph.CO]} \BibitemShut
  {NoStop}%
\bibitem [{\citenamefont {{Simon}}\ and\ \citenamefont
  {{Geha}}(2007)}]{SimonGeha07}%
  \BibitemOpen
  \bibfield  {author} {\bibinfo {author} {\bibfnamefont {J.~D.}\ \bibnamefont
  {{Simon}}}\ and\ \bibinfo {author} {\bibfnamefont {M.}~\bibnamefont
  {{Geha}}},\ }\href {\doibase 10.1086/521816} {\bibfield  {journal} {\bibinfo
  {journal} {\apj}\ }\textbf {\bibinfo {volume} {670}},\ \bibinfo {pages} {313}
  (\bibinfo {year} {2007})},\ \Eprint {http://arxiv.org/abs/0706.0516}
  {arXiv:0706.0516} \BibitemShut {NoStop}%
\bibitem [{\citenamefont {{Adams}}\ and\ \citenamefont
  {{Oosterloo}}(2018)}]{adams17}%
  \BibitemOpen
  \bibfield  {author} {\bibinfo {author} {\bibfnamefont {E.~A.~K.}\
  \bibnamefont {{Adams}}}\ and\ \bibinfo {author} {\bibfnamefont {T.~A.}\
  \bibnamefont {{Oosterloo}}},\ }\href {\doibase 10.1051/0004-6361/201732017}
  {\bibfield  {journal} {\bibinfo  {journal} {\aap}\ }\textbf {\bibinfo
  {volume} {612}},\ \bibinfo {eid} {A26} (\bibinfo {year} {2018})},\ \Eprint
  {http://arxiv.org/abs/1712.06636} {arXiv:1712.06636} \BibitemShut {NoStop}%
\bibitem [{\citenamefont {{Chuzhoy}}\ and\ \citenamefont
  {{Kolb}}(2009)}]{chuzhoy09}%
  \BibitemOpen
  \bibfield  {author} {\bibinfo {author} {\bibfnamefont {L.}~\bibnamefont
  {{Chuzhoy}}}\ and\ \bibinfo {author} {\bibfnamefont {E.~W.}\ \bibnamefont
  {{Kolb}}},\ }\href {\doibase 10.1088/1475-7516/2009/07/014} {\bibfield
  {journal} {\bibinfo  {journal} {\jcap}\ }\textbf {\bibinfo {volume} {7}},\
  \bibinfo {eid} {014} (\bibinfo {year} {2009})},\ \Eprint
  {http://arxiv.org/abs/0809.0436} {arXiv:0809.0436} \BibitemShut {NoStop}%
\bibitem [{\citenamefont {Munoz}\ and\ \citenamefont {Loeb}(2018)}]{MunLoeb18}%
  \BibitemOpen
  \bibfield  {author} {\bibinfo {author} {\bibfnamefont {J.~B.}\ \bibnamefont
  {Munoz}}\ and\ \bibinfo {author} {\bibfnamefont {A.}~\bibnamefont {Loeb}},\
  }\href {\doibase 10.1038/s41586-018-0151-x} {\bibfield  {journal} {\bibinfo
  {journal} {Nature}\ }\textbf {\bibinfo {volume} {557}},\ \bibinfo {pages}
  {684} (\bibinfo {year} {2018})},\ \Eprint {http://arxiv.org/abs/1802.10094}
  {arXiv:1802.10094 [astro-ph.CO]} \BibitemShut {NoStop}%
\bibitem [{\citenamefont {{Kadota}}\ \emph {et~al.}(2016)\citenamefont
  {{Kadota}}, \citenamefont {{Sekiguchi}},\ and\ \citenamefont
  {{Tashiro}}}]{KadSek16_MilliConstraint}%
  \BibitemOpen
  \bibfield  {author} {\bibinfo {author} {\bibfnamefont {K.}~\bibnamefont
  {{Kadota}}}, \bibinfo {author} {\bibfnamefont {T.}~\bibnamefont
  {{Sekiguchi}}}, \ and\ \bibinfo {author} {\bibfnamefont {H.}~\bibnamefont
  {{Tashiro}}},\ }\href@noop {} {\bibfield  {journal} {\bibinfo  {journal}
  {arXiv e-prints}\ ,\ \bibinfo {eid} {arXiv:1602.04009}} (\bibinfo {year}
  {2016})},\ \Eprint {http://arxiv.org/abs/1602.04009} {arXiv:1602.04009
  [astro-ph.CO]} \BibitemShut {NoStop}%
\bibitem [{\citenamefont {{Dunsky}}\ \emph {et~al.}(2018)\citenamefont
  {{Dunsky}}, \citenamefont {{Hall}},\ and\ \citenamefont
  {{Harigaya}}}]{DunHal18_MilliConstraint}%
  \BibitemOpen
  \bibfield  {author} {\bibinfo {author} {\bibfnamefont {D.}~\bibnamefont
  {{Dunsky}}}, \bibinfo {author} {\bibfnamefont {L.~J.}\ \bibnamefont
  {{Hall}}}, \ and\ \bibinfo {author} {\bibfnamefont {K.}~\bibnamefont
  {{Harigaya}}},\ }\href@noop {} {\bibfield  {journal} {\bibinfo  {journal}
  {arXiv e-prints}\ ,\ \bibinfo {eid} {arXiv:1812.11116}} (\bibinfo {year}
  {2018})},\ \Eprint {http://arxiv.org/abs/1812.11116} {arXiv:1812.11116
  [astro-ph.HE]} \BibitemShut {NoStop}%
\bibitem [{\citenamefont {{Lockman}}(2002)}]{Lock02}%
  \BibitemOpen
  \bibfield  {author} {\bibinfo {author} {\bibfnamefont {F.~J.}\ \bibnamefont
  {{Lockman}}},\ }\href {\doibase 10.1086/345495} {\bibfield  {journal}
  {\bibinfo  {journal} {\apjl}\ }\textbf {\bibinfo {volume} {580}},\ \bibinfo
  {pages} {L47} (\bibinfo {year} {2002})},\ \Eprint
  {http://arxiv.org/abs/astro-ph/0210424} {astro-ph/0210424} \BibitemShut
  {NoStop}%
\bibitem [{\citenamefont {{Ford}}\ \emph {et~al.}(2008)\citenamefont {{Ford}},
  \citenamefont {{McClure-Griffiths}}, \citenamefont {{Lockman}}, \citenamefont
  {{Bailin}}, \citenamefont {{Calabretta}}, \citenamefont {{Kalberla}},
  \citenamefont {{Murphy}},\ and\ \citenamefont {{Pisano}}}]{Ford08}%
  \BibitemOpen
  \bibfield  {author} {\bibinfo {author} {\bibfnamefont {H.~A.}\ \bibnamefont
  {{Ford}}}, \bibinfo {author} {\bibfnamefont {N.~M.}\ \bibnamefont
  {{McClure-Griffiths}}}, \bibinfo {author} {\bibfnamefont {F.~J.}\
  \bibnamefont {{Lockman}}}, \bibinfo {author} {\bibfnamefont {J.}~\bibnamefont
  {{Bailin}}}, \bibinfo {author} {\bibfnamefont {M.~R.}\ \bibnamefont
  {{Calabretta}}}, \bibinfo {author} {\bibfnamefont {P.~M.~W.}\ \bibnamefont
  {{Kalberla}}}, \bibinfo {author} {\bibfnamefont {T.}~\bibnamefont
  {{Murphy}}}, \ and\ \bibinfo {author} {\bibfnamefont {D.~J.}\ \bibnamefont
  {{Pisano}}},\ }\href {\doibase 10.1086/592188} {\bibfield  {journal}
  {\bibinfo  {journal} {\apj}\ }\textbf {\bibinfo {volume} {688}},\ \bibinfo
  {pages} {290} (\bibinfo {year} {2008})},\ \Eprint
  {http://arxiv.org/abs/0807.3550} {arXiv:0807.3550} \BibitemShut {NoStop}%
\bibitem [{\citenamefont {{Ford}}\ \emph {et~al.}(2010)\citenamefont {{Ford}},
  \citenamefont {{Lockman}},\ and\ \citenamefont
  {{McClure-Griffiths}}}]{Ford10}%
  \BibitemOpen
  \bibfield  {author} {\bibinfo {author} {\bibfnamefont {H.~A.}\ \bibnamefont
  {{Ford}}}, \bibinfo {author} {\bibfnamefont {F.~J.}\ \bibnamefont
  {{Lockman}}}, \ and\ \bibinfo {author} {\bibfnamefont {N.~M.}\ \bibnamefont
  {{McClure-Griffiths}}},\ }\href {\doibase 10.1088/0004-637X/722/1/367}
  {\bibfield  {journal} {\bibinfo  {journal} {\apj}\ }\textbf {\bibinfo
  {volume} {722}},\ \bibinfo {pages} {367} (\bibinfo {year} {2010})},\ \Eprint
  {http://arxiv.org/abs/1008.2760} {arXiv:1008.2760} \BibitemShut {NoStop}%
\bibitem [{\citenamefont {{Lehner}}\ \emph {et~al.}(2004)\citenamefont
  {{Lehner}}, \citenamefont {{Wakker}},\ and\ \citenamefont
  {{Savage}}}]{Leh04}%
  \BibitemOpen
  \bibfield  {author} {\bibinfo {author} {\bibfnamefont {N.}~\bibnamefont
  {{Lehner}}}, \bibinfo {author} {\bibfnamefont {B.~P.}\ \bibnamefont
  {{Wakker}}}, \ and\ \bibinfo {author} {\bibfnamefont {B.~D.}\ \bibnamefont
  {{Savage}}},\ }\href {\doibase 10.1086/424730} {\bibfield  {journal}
  {\bibinfo  {journal} {\apj}\ }\textbf {\bibinfo {volume} {615}},\ \bibinfo
  {pages} {767} (\bibinfo {year} {2004})},\ \Eprint
  {http://arxiv.org/abs/astro-ph/0407363} {arXiv:astro-ph/0407363 [astro-ph]}
  \BibitemShut {NoStop}%
\bibitem [{\citenamefont {{Wakker}}(2001)}]{CloudMetal_Wak01}%
  \BibitemOpen
  \bibfield  {author} {\bibinfo {author} {\bibfnamefont {B.~P.}\ \bibnamefont
  {{Wakker}}},\ }\href {\doibase 10.1086/321783} {\bibfield  {journal}
  {\bibinfo  {journal} {\apjs}\ }\textbf {\bibinfo {volume} {136}},\ \bibinfo
  {pages} {463} (\bibinfo {year} {2001})},\ \Eprint
  {http://arxiv.org/abs/astro-ph/0102147} {astro-ph/0102147} \BibitemShut
  {NoStop}%
\bibitem [{\citenamefont {{Richter}}\ \emph {et~al.}(2001)\citenamefont
  {{Richter}}, \citenamefont {{Sembach}}, \citenamefont {{Wakker}},
  \citenamefont {{Savage}}, \citenamefont {{Tripp}}, \citenamefont {{Murphy}},
  \citenamefont {{Kalberla}},\ and\ \citenamefont
  {{Jenkins}}}]{CloudMetal_Rich01}%
  \BibitemOpen
  \bibfield  {author} {\bibinfo {author} {\bibfnamefont {P.}~\bibnamefont
  {{Richter}}}, \bibinfo {author} {\bibfnamefont {K.~R.}\ \bibnamefont
  {{Sembach}}}, \bibinfo {author} {\bibfnamefont {B.~P.}\ \bibnamefont
  {{Wakker}}}, \bibinfo {author} {\bibfnamefont {B.~D.}\ \bibnamefont
  {{Savage}}}, \bibinfo {author} {\bibfnamefont {T.~M.}\ \bibnamefont
  {{Tripp}}}, \bibinfo {author} {\bibfnamefont {E.~M.}\ \bibnamefont
  {{Murphy}}}, \bibinfo {author} {\bibfnamefont {P.~M.~W.}\ \bibnamefont
  {{Kalberla}}}, \ and\ \bibinfo {author} {\bibfnamefont {E.~B.}\ \bibnamefont
  {{Jenkins}}},\ }\href {\doibase 10.1086/322401} {\bibfield  {journal}
  {\bibinfo  {journal} {\apj}\ }\textbf {\bibinfo {volume} {559}},\ \bibinfo
  {pages} {318} (\bibinfo {year} {2001})},\ \Eprint
  {http://arxiv.org/abs/astro-ph/0105466} {astro-ph/0105466} \BibitemShut
  {NoStop}%
\bibitem [{\citenamefont {{Richter}}\ \emph {et~al.}(2003)\citenamefont
  {{Richter}}, \citenamefont {{Wakker}}, \citenamefont {{Savage}},\ and\
  \citenamefont {{Sembach}}}]{CloudMetal_Rich01b}%
  \BibitemOpen
  \bibfield  {author} {\bibinfo {author} {\bibfnamefont {P.}~\bibnamefont
  {{Richter}}}, \bibinfo {author} {\bibfnamefont {B.~P.}\ \bibnamefont
  {{Wakker}}}, \bibinfo {author} {\bibfnamefont {B.~D.}\ \bibnamefont
  {{Savage}}}, \ and\ \bibinfo {author} {\bibfnamefont {K.~R.}\ \bibnamefont
  {{Sembach}}},\ }\href {\doibase 10.1086/346204} {\bibfield  {journal}
  {\bibinfo  {journal} {\apj}\ }\textbf {\bibinfo {volume} {586}},\ \bibinfo
  {pages} {230} (\bibinfo {year} {2003})},\ \Eprint
  {http://arxiv.org/abs/astro-ph/0211356} {astro-ph/0211356} \BibitemShut
  {NoStop}%
\bibitem [{\citenamefont {{Sembach}}\ \emph {et~al.}(2004)\citenamefont
  {{Sembach}}, \citenamefont {{Wakker}}, \citenamefont {{Tripp}}, \citenamefont
  {{Richter}}, \citenamefont {{Kruk}}, \citenamefont {{Blair}}, \citenamefont
  {{Moos}}, \citenamefont {{Savage}}, \citenamefont {{Shull}}, \citenamefont
  {{York}}, \citenamefont {{Sonneborn}}, \citenamefont {{H{\'e}brard}},
  \citenamefont {{Ferlet}}, \citenamefont {{Vidal-Madjar}}, \citenamefont
  {{Friedman}},\ and\ \citenamefont {{Jenkins}}}]{CloudMetal_Semb04}%
  \BibitemOpen
  \bibfield  {author} {\bibinfo {author} {\bibfnamefont {K.~R.}\ \bibnamefont
  {{Sembach}}}, \bibinfo {author} {\bibfnamefont {B.~P.}\ \bibnamefont
  {{Wakker}}}, \bibinfo {author} {\bibfnamefont {T.~M.}\ \bibnamefont
  {{Tripp}}}, \bibinfo {author} {\bibfnamefont {P.}~\bibnamefont {{Richter}}},
  \bibinfo {author} {\bibfnamefont {J.~W.}\ \bibnamefont {{Kruk}}}, \bibinfo
  {author} {\bibfnamefont {W.~P.}\ \bibnamefont {{Blair}}}, \bibinfo {author}
  {\bibfnamefont {H.~W.}\ \bibnamefont {{Moos}}}, \bibinfo {author}
  {\bibfnamefont {B.~D.}\ \bibnamefont {{Savage}}}, \bibinfo {author}
  {\bibfnamefont {J.~M.}\ \bibnamefont {{Shull}}}, \bibinfo {author}
  {\bibfnamefont {D.~G.}\ \bibnamefont {{York}}}, \bibinfo {author}
  {\bibfnamefont {G.}~\bibnamefont {{Sonneborn}}}, \bibinfo {author}
  {\bibfnamefont {G.}~\bibnamefont {{H{\'e}brard}}}, \bibinfo {author}
  {\bibfnamefont {R.}~\bibnamefont {{Ferlet}}}, \bibinfo {author}
  {\bibfnamefont {A.}~\bibnamefont {{Vidal-Madjar}}}, \bibinfo {author}
  {\bibfnamefont {S.~D.}\ \bibnamefont {{Friedman}}}, \ and\ \bibinfo {author}
  {\bibfnamefont {E.~B.}\ \bibnamefont {{Jenkins}}},\ }\href {\doibase
  10.1086/381025} {\bibfield  {journal} {\bibinfo  {journal} {\apjs}\ }\textbf
  {\bibinfo {volume} {150}},\ \bibinfo {pages} {387} (\bibinfo {year}
  {2004})},\ \Eprint {http://arxiv.org/abs/astro-ph/0311177} {astro-ph/0311177}
  \BibitemShut {NoStop}%
\bibitem [{\citenamefont {{Hernandez}}\ \emph {et~al.}(2013)\citenamefont
  {{Hernandez}}, \citenamefont {{Wakker}}, \citenamefont {{Benjamin}},
  \citenamefont {{French}}, \citenamefont {{Kerp}}, \citenamefont {{Lockman}},
  \citenamefont {{O'Toole}},\ and\ \citenamefont
  {{Winkel}}}]{CloudMetal_HerLock13}%
  \BibitemOpen
  \bibfield  {author} {\bibinfo {author} {\bibfnamefont {A.~K.}\ \bibnamefont
  {{Hernandez}}}, \bibinfo {author} {\bibfnamefont {B.~P.}\ \bibnamefont
  {{Wakker}}}, \bibinfo {author} {\bibfnamefont {R.~A.}\ \bibnamefont
  {{Benjamin}}}, \bibinfo {author} {\bibfnamefont {D.}~\bibnamefont
  {{French}}}, \bibinfo {author} {\bibfnamefont {J.}~\bibnamefont {{Kerp}}},
  \bibinfo {author} {\bibfnamefont {F.~J.}\ \bibnamefont {{Lockman}}}, \bibinfo
  {author} {\bibfnamefont {S.}~\bibnamefont {{O'Toole}}}, \ and\ \bibinfo
  {author} {\bibfnamefont {B.}~\bibnamefont {{Winkel}}},\ }\href {\doibase
  10.1088/0004-637X/777/1/19} {\bibfield  {journal} {\bibinfo  {journal}
  {\apj}\ }\textbf {\bibinfo {volume} {777}},\ \bibinfo {eid} {19} (\bibinfo
  {year} {2013})},\ \Eprint {http://arxiv.org/abs/1308.6313} {arXiv:1308.6313
  [astro-ph.GA]} \BibitemShut {NoStop}%
\bibitem [{\citenamefont {{Maio}}\ \emph {et~al.}(2007)\citenamefont {{Maio}},
  \citenamefont {{Dolag}}, \citenamefont {{Ciardi}},\ and\ \citenamefont
  {{Tornatore}}}]{Maio07}%
  \BibitemOpen
  \bibfield  {author} {\bibinfo {author} {\bibfnamefont {U.}~\bibnamefont
  {{Maio}}}, \bibinfo {author} {\bibfnamefont {K.}~\bibnamefont {{Dolag}}},
  \bibinfo {author} {\bibfnamefont {B.}~\bibnamefont {{Ciardi}}}, \ and\
  \bibinfo {author} {\bibfnamefont {L.}~\bibnamefont {{Tornatore}}},\ }\href
  {\doibase 10.1111/j.1365-2966.2007.12016.x} {\bibfield  {journal} {\bibinfo
  {journal} {\mnras}\ }\textbf {\bibinfo {volume} {379}},\ \bibinfo {pages}
  {963} (\bibinfo {year} {2007})},\ \Eprint {http://arxiv.org/abs/0704.2182}
  {arXiv:0704.2182 [astro-ph]} \BibitemShut {NoStop}%
\bibitem [{\citenamefont {{De Rijcke}}\ \emph {et~al.}(2013)\citenamefont {{De
  Rijcke}}, \citenamefont {{Schroyen}}, \citenamefont {{Vandenbroucke}},
  \citenamefont {{Jachowicz}}, \citenamefont {{Decroos}}, \citenamefont
  {{Cloet-Osselaer}},\ and\ \citenamefont {{Koleva}}}]{BhooCool13}%
  \BibitemOpen
  \bibfield  {author} {\bibinfo {author} {\bibfnamefont {S.}~\bibnamefont {{De
  Rijcke}}}, \bibinfo {author} {\bibfnamefont {J.}~\bibnamefont {{Schroyen}}},
  \bibinfo {author} {\bibfnamefont {B.}~\bibnamefont {{Vandenbroucke}}},
  \bibinfo {author} {\bibfnamefont {N.}~\bibnamefont {{Jachowicz}}}, \bibinfo
  {author} {\bibfnamefont {J.}~\bibnamefont {{Decroos}}}, \bibinfo {author}
  {\bibfnamefont {A.}~\bibnamefont {{Cloet-Osselaer}}}, \ and\ \bibinfo
  {author} {\bibfnamefont {M.}~\bibnamefont {{Koleva}}},\ }\href {\doibase
  10.1093/mnras/stt942} {\bibfield  {journal} {\bibinfo  {journal} {\mnras}\
  }\textbf {\bibinfo {volume} {433}},\ \bibinfo {pages} {3005} (\bibinfo {year}
  {2013})},\ \Eprint {http://arxiv.org/abs/1306.4860} {arXiv:1306.4860}
  \BibitemShut {NoStop}%
\bibitem [{\citenamefont {Neufeld}\ \emph {et~al.}(2018)\citenamefont
  {Neufeld}, \citenamefont {Farrar},\ and\ \citenamefont {McKee}}]{nfm18}%
  \BibitemOpen
  \bibfield  {author} {\bibinfo {author} {\bibfnamefont {D.~A.}\ \bibnamefont
  {Neufeld}}, \bibinfo {author} {\bibfnamefont {G.~R.}\ \bibnamefont {Farrar}},
  \ and\ \bibinfo {author} {\bibfnamefont {C.~F.}\ \bibnamefont {McKee}},\
  }\href {\doibase 10.3847/1538-4357/aad6a4} {\bibfield  {journal} {\bibinfo
  {journal} {Astrophys. J.}\ }\textbf {\bibinfo {volume} {866}},\ \bibinfo
  {pages} {111} (\bibinfo {year} {2018})},\ \Eprint
  {http://arxiv.org/abs/1805.08794} {arXiv:1805.08794 [astro-ph.CO]}
  \BibitemShut {NoStop}%
\bibitem [{\citenamefont {{Xu}}\ and\ \citenamefont
  {{Farrar}}(2020)}]{XFinprep20}%
  \BibitemOpen
  \bibfield  {author} {\bibinfo {author} {\bibfnamefont {X.}~\bibnamefont
  {{Xu}}}\ and\ \bibinfo {author} {\bibfnamefont {G.~R.}\ \bibnamefont
  {{Farrar}}},\ }\href@noop {} {\bibfield  {journal} {\bibinfo  {journal}
  {arXiv e-prints}\ ,\ \bibinfo {eid} {arXiv:2101.00142}} (\bibinfo {year}
  {2020})},\ \Eprint {http://arxiv.org/abs/2101.00142} {arXiv:2101.00142
  [hep-ph]} \BibitemShut {NoStop}%
\bibitem [{\citenamefont {{Mahdawi}}\ and\ \citenamefont
  {{Farrar}}(2018)}]{mahdawi18}%
  \BibitemOpen
  \bibfield  {author} {\bibinfo {author} {\bibfnamefont {M.~S.}\ \bibnamefont
  {{Mahdawi}}}\ and\ \bibinfo {author} {\bibfnamefont {G.~R.}\ \bibnamefont
  {{Farrar}}},\ }\href {\doibase 10.1088/1475-7516/2018/10/007} {\bibfield
  {journal} {\bibinfo  {journal} {\jcap}\ }\textbf {\bibinfo {volume} {10}},\
  \bibinfo {eid} {007} (\bibinfo {year} {2018})},\ \Eprint
  {http://arxiv.org/abs/1804.03073} {arXiv:1804.03073 [hep-ph]} \BibitemShut
  {NoStop}%
\bibitem [{\citenamefont {{Crisler}}\ \emph {et~al.}(2018)\citenamefont
  {{Crisler}}, \citenamefont {{Essig}}, \citenamefont {{Estrada}},
  \citenamefont {{Fernandez}}, \citenamefont {{Tiffenberg}}, \citenamefont
  {{Haro}}, \citenamefont {{Volansky}}, \citenamefont {{Yu}},\ and\
  \citenamefont {{Sensei Collaboration}}}]{SENSEI18}%
  \BibitemOpen
  \bibfield  {author} {\bibinfo {author} {\bibfnamefont {M.}~\bibnamefont
  {{Crisler}}}, \bibinfo {author} {\bibfnamefont {R.}~\bibnamefont {{Essig}}},
  \bibinfo {author} {\bibfnamefont {J.}~\bibnamefont {{Estrada}}}, \bibinfo
  {author} {\bibfnamefont {G.}~\bibnamefont {{Fernandez}}}, \bibinfo {author}
  {\bibfnamefont {J.}~\bibnamefont {{Tiffenberg}}}, \bibinfo {author}
  {\bibfnamefont {M.~S.}\ \bibnamefont {{Haro}}}, \bibinfo {author}
  {\bibfnamefont {T.}~\bibnamefont {{Volansky}}}, \bibinfo {author}
  {\bibfnamefont {T.-T.}\ \bibnamefont {{Yu}}}, \ and\ \bibinfo {author}
  {\bibnamefont {{Sensei Collaboration}}},\ }\href {\doibase
  10.1103/PhysRevLett.121.061803} {\bibfield  {journal} {\bibinfo  {journal}
  {\prl}\ }\textbf {\bibinfo {volume} {121}},\ \bibinfo {eid} {061803}
  (\bibinfo {year} {2018})},\ \Eprint {http://arxiv.org/abs/1804.00088}
  {arXiv:1804.00088 [hep-ex]} \BibitemShut {NoStop}%
\bibitem [{\citenamefont {Prinz}\ \emph {et~al.}(1998)\citenamefont {Prinz},
  \citenamefont {Baggs}, \citenamefont {Ballam}, \citenamefont {Ecklund},
  \citenamefont {Fertig}, \citenamefont {Jaros}, \citenamefont {Kase},
  \citenamefont {Kulikov}, \citenamefont {Langeveld}, \citenamefont {Leonard},
  \citenamefont {Marvin}, \citenamefont {Nakashima}, \citenamefont {Nelson},
  \citenamefont {Odian}, \citenamefont {Pertsova}, \citenamefont {Putallaz},\
  and\ \citenamefont {Weinstein}}]{SLAC98}%
  \BibitemOpen
  \bibfield  {author} {\bibinfo {author} {\bibfnamefont {A.~A.}\ \bibnamefont
  {Prinz}}, \bibinfo {author} {\bibfnamefont {R.}~\bibnamefont {Baggs}},
  \bibinfo {author} {\bibfnamefont {J.}~\bibnamefont {Ballam}}, \bibinfo
  {author} {\bibfnamefont {S.}~\bibnamefont {Ecklund}}, \bibinfo {author}
  {\bibfnamefont {C.}~\bibnamefont {Fertig}}, \bibinfo {author} {\bibfnamefont
  {J.~A.}\ \bibnamefont {Jaros}}, \bibinfo {author} {\bibfnamefont
  {K.}~\bibnamefont {Kase}}, \bibinfo {author} {\bibfnamefont {A.}~\bibnamefont
  {Kulikov}}, \bibinfo {author} {\bibfnamefont {W.~G.~J.}\ \bibnamefont
  {Langeveld}}, \bibinfo {author} {\bibfnamefont {R.}~\bibnamefont {Leonard}},
  \bibinfo {author} {\bibfnamefont {T.}~\bibnamefont {Marvin}}, \bibinfo
  {author} {\bibfnamefont {T.}~\bibnamefont {Nakashima}}, \bibinfo {author}
  {\bibfnamefont {W.~R.}\ \bibnamefont {Nelson}}, \bibinfo {author}
  {\bibfnamefont {A.}~\bibnamefont {Odian}}, \bibinfo {author} {\bibfnamefont
  {M.}~\bibnamefont {Pertsova}}, \bibinfo {author} {\bibfnamefont
  {G.}~\bibnamefont {Putallaz}}, \ and\ \bibinfo {author} {\bibfnamefont
  {A.}~\bibnamefont {Weinstein}},\ }\href {\doibase
  10.1103/PhysRevLett.81.1175} {\bibfield  {journal} {\bibinfo  {journal}
  {Phys. Rev. Lett.}\ }\textbf {\bibinfo {volume} {81}},\ \bibinfo {pages}
  {1175} (\bibinfo {year} {1998})}\BibitemShut {NoStop}%
\bibitem [{\citenamefont {{Essig}}\ \emph {et~al.}(2012)\citenamefont
  {{Essig}}, \citenamefont {{Manalaysay}}, \citenamefont {{Mardon}},
  \citenamefont {{Sorensen}},\ and\ \citenamefont {{Volansky}}}]{Xenon12}%
  \BibitemOpen
  \bibfield  {author} {\bibinfo {author} {\bibfnamefont {R.}~\bibnamefont
  {{Essig}}}, \bibinfo {author} {\bibfnamefont {A.}~\bibnamefont
  {{Manalaysay}}}, \bibinfo {author} {\bibfnamefont {J.}~\bibnamefont
  {{Mardon}}}, \bibinfo {author} {\bibfnamefont {P.}~\bibnamefont
  {{Sorensen}}}, \ and\ \bibinfo {author} {\bibfnamefont {T.}~\bibnamefont
  {{Volansky}}},\ }\href {\doibase 10.1103/PhysRevLett.109.021301} {\bibfield
  {journal} {\bibinfo  {journal} {\prl}\ }\textbf {\bibinfo {volume} {109}},\
  \bibinfo {eid} {021301} (\bibinfo {year} {2012})},\ \Eprint
  {http://arxiv.org/abs/1206.2644} {arXiv:1206.2644 [astro-ph.CO]} \BibitemShut
  {NoStop}%
\bibitem [{\citenamefont {{Essig}}\ \emph {et~al.}(2017)\citenamefont
  {{Essig}}, \citenamefont {{Volansky}},\ and\ \citenamefont {{Yu}}}]{Xenon17}%
  \BibitemOpen
  \bibfield  {author} {\bibinfo {author} {\bibfnamefont {R.}~\bibnamefont
  {{Essig}}}, \bibinfo {author} {\bibfnamefont {T.}~\bibnamefont {{Volansky}}},
  \ and\ \bibinfo {author} {\bibfnamefont {T.-T.}\ \bibnamefont {{Yu}}},\
  }\href {\doibase 10.1103/PhysRevD.96.043017} {\bibfield  {journal} {\bibinfo
  {journal} {\prd}\ }\textbf {\bibinfo {volume} {96}},\ \bibinfo {eid} {043017}
  (\bibinfo {year} {2017})},\ \Eprint {http://arxiv.org/abs/1703.00910}
  {arXiv:1703.00910 [hep-ph]} \BibitemShut {NoStop}%
\bibitem [{\citenamefont {McCammon}\ \emph {et~al.}(2002)\citenamefont
  {McCammon} \emph {et~al.}}]{XQC02}%
  \BibitemOpen
  \bibfield  {author} {\bibinfo {author} {\bibfnamefont {D.}~\bibnamefont
  {McCammon}} \emph {et~al.},\ }\href {\doibase 10.1086/341727} {\bibfield
  {journal} {\bibinfo  {journal} {Astrophys. J.}\ }\textbf {\bibinfo {volume}
  {576}},\ \bibinfo {pages} {188} (\bibinfo {year} {2002})},\ \Eprint
  {http://arxiv.org/abs/astro-ph/0205012} {arXiv:astro-ph/0205012 [astro-ph]}
  \BibitemShut {NoStop}%
\bibitem [{\citenamefont {Angloher}\ \emph {et~al.}(2017)\citenamefont
  {Angloher} \emph {et~al.}}]{CRESST17}%
  \BibitemOpen
  \bibfield  {author} {\bibinfo {author} {\bibfnamefont {G.}~\bibnamefont
  {Angloher}} \emph {et~al.} (\bibinfo {collaboration} {CRESST}),\ }\href
  {\doibase 10.1140/epjc/s10052-017-5223-9} {\bibfield  {journal} {\bibinfo
  {journal} {Eur. Phys. J.}\ }\textbf {\bibinfo {volume} {C77}},\ \bibinfo
  {pages} {637} (\bibinfo {year} {2017})},\ \Eprint
  {http://arxiv.org/abs/1707.06749} {arXiv:1707.06749 [astro-ph.CO]}
  \BibitemShut {NoStop}%
\bibitem [{\citenamefont {Barkana}\ \emph {et~al.}(2018)\citenamefont
  {Barkana}, \citenamefont {Outmezguine}, \citenamefont {Redigolo},\ and\
  \citenamefont {Volansky}}]{barkana+18}%
  \BibitemOpen
  \bibfield  {author} {\bibinfo {author} {\bibfnamefont {R.}~\bibnamefont
  {Barkana}}, \bibinfo {author} {\bibfnamefont {N.~J.}\ \bibnamefont
  {Outmezguine}}, \bibinfo {author} {\bibfnamefont {D.}~\bibnamefont
  {Redigolo}}, \ and\ \bibinfo {author} {\bibfnamefont {T.}~\bibnamefont
  {Volansky}},\ }\href {\doibase 10.1103/PhysRevD.98.103005} {\bibfield
  {journal} {\bibinfo  {journal} {Phys. Rev.}\ }\textbf {\bibinfo {volume}
  {D98}},\ \bibinfo {pages} {103005} (\bibinfo {year} {2018})},\ \Eprint
  {http://arxiv.org/abs/1803.03091} {arXiv:1803.03091 [hep-ph]} \BibitemShut
  {NoStop}%
\bibitem [{\citenamefont {{Berlin}}\ \emph {et~al.}(2018)\citenamefont
  {{Berlin}}, \citenamefont {{Hooper}}, \citenamefont {{Krnjaic}},\ and\
  \citenamefont {{McDermott}}}]{Berlin18_MClimit}%
  \BibitemOpen
  \bibfield  {author} {\bibinfo {author} {\bibfnamefont {A.}~\bibnamefont
  {{Berlin}}}, \bibinfo {author} {\bibfnamefont {D.}~\bibnamefont {{Hooper}}},
  \bibinfo {author} {\bibfnamefont {G.}~\bibnamefont {{Krnjaic}}}, \ and\
  \bibinfo {author} {\bibfnamefont {S.~D.}\ \bibnamefont {{McDermott}}},\
  }\href {\doibase 10.1103/PhysRevLett.121.011102} {\bibfield  {journal}
  {\bibinfo  {journal} {\prl}\ }\textbf {\bibinfo {volume} {121}},\ \bibinfo
  {eid} {011102} (\bibinfo {year} {2018})},\ \Eprint
  {http://arxiv.org/abs/1803.02804} {arXiv:1803.02804 [hep-ph]} \BibitemShut
  {NoStop}%
\bibitem [{\citenamefont {{Liu}}\ \emph {et~al.}(2019)\citenamefont {{Liu}},
  \citenamefont {{Outmezguine}}, \citenamefont {{Redigolo}},\ and\
  \citenamefont {{Volansky}}}]{LiuOutRedVol19}%
  \BibitemOpen
  \bibfield  {author} {\bibinfo {author} {\bibfnamefont {H.}~\bibnamefont
  {{Liu}}}, \bibinfo {author} {\bibfnamefont {N.~J.}\ \bibnamefont
  {{Outmezguine}}}, \bibinfo {author} {\bibfnamefont {D.}~\bibnamefont
  {{Redigolo}}}, \ and\ \bibinfo {author} {\bibfnamefont {T.}~\bibnamefont
  {{Volansky}}},\ }\href@noop {} {\bibfield  {journal} {\bibinfo  {journal}
  {arXiv e-prints}\ ,\ \bibinfo {eid} {arXiv:1908.06986}} (\bibinfo {year}
  {2019})},\ \Eprint {http://arxiv.org/abs/1908.06986} {arXiv:1908.06986
  [hep-ph]} \BibitemShut {NoStop}%
\bibitem [{\citenamefont {{Xu}}\ and\ \citenamefont
  {{Farrar}}(2021)}]{XFinprep20b}%
  \BibitemOpen
  \bibfield  {author} {\bibinfo {author} {\bibfnamefont {X.}~\bibnamefont
  {{Xu}}}\ and\ \bibinfo {author} {\bibfnamefont {G.~R.}\ \bibnamefont
  {{Farrar}}},\ }\href@noop {} {\bibfield  {journal} {\bibinfo  {journal} {in
  preparation}\ } (\bibinfo {year} {2021})}\BibitemShut {NoStop}%
\bibitem [{\citenamefont {{Necib}}\ \emph {et~al.}(2018)\citenamefont
  {{Necib}}, \citenamefont {{Lisanti}},\ and\ \citenamefont
  {{Belokurov}}}]{necib+18}%
  \BibitemOpen
  \bibfield  {author} {\bibinfo {author} {\bibfnamefont {L.}~\bibnamefont
  {{Necib}}}, \bibinfo {author} {\bibfnamefont {M.}~\bibnamefont {{Lisanti}}},
  \ and\ \bibinfo {author} {\bibfnamefont {V.}~\bibnamefont {{Belokurov}}},\
  }\href@noop {} {\bibfield  {journal} {\bibinfo  {journal} {ArXiv e-prints}\ }
  (\bibinfo {year} {2018})},\ \Eprint {http://arxiv.org/abs/1807.02519}
  {arXiv:1807.02519} \BibitemShut {NoStop}%
\bibitem [{\citenamefont {{Antoja}}\ \emph {et~al.}(2018)\citenamefont
  {{Antoja}}, \citenamefont {{Helmi}}, \citenamefont {{Romero-G{\'o}mez}},
  \citenamefont {{Katz}}, \citenamefont {{Babusiaux}}, \citenamefont
  {{Drimmel}}, \citenamefont {{Evans}}, \citenamefont {{Figueras}},
  \citenamefont {{Poggio}},\ and\ \citenamefont {{Reyl{\'e}}}}]{AntHel18}%
  \BibitemOpen
  \bibfield  {author} {\bibinfo {author} {\bibfnamefont {T.}~\bibnamefont
  {{Antoja}}}, \bibinfo {author} {\bibfnamefont {A.}~\bibnamefont {{Helmi}}},
  \bibinfo {author} {\bibfnamefont {M.}~\bibnamefont {{Romero-G{\'o}mez}}},
  \bibinfo {author} {\bibfnamefont {D.}~\bibnamefont {{Katz}}}, \bibinfo
  {author} {\bibfnamefont {C.}~\bibnamefont {{Babusiaux}}}, \bibinfo {author}
  {\bibfnamefont {R.}~\bibnamefont {{Drimmel}}}, \bibinfo {author}
  {\bibfnamefont {D.~W.}\ \bibnamefont {{Evans}}}, \bibinfo {author}
  {\bibfnamefont {F.}~\bibnamefont {{Figueras}}}, \bibinfo {author}
  {\bibfnamefont {E.}~\bibnamefont {{Poggio}}}, \ and\ \bibinfo {author}
  {\bibfnamefont {C.}~\bibnamefont {{Reyl{\'e}}}},\ }\href {\doibase
  10.1038/s41586-018-0510-7} {\bibfield  {journal} {\bibinfo  {journal} {\nat}\
  }\textbf {\bibinfo {volume} {561}},\ \bibinfo {pages} {360} (\bibinfo {year}
  {2018})},\ \Eprint {http://arxiv.org/abs/1804.10196} {arXiv:1804.10196
  [astro-ph.GA]} \BibitemShut {NoStop}%
\bibitem [{\citenamefont {{Mandal}}\ \emph {et~al.}(2019)\citenamefont
  {{Mandal}}, \citenamefont {{Majumdar}}, \citenamefont {{Rentala}},\ and\
  \citenamefont {{Thakur}}}]{ManMajRen19}%
  \BibitemOpen
  \bibfield  {author} {\bibinfo {author} {\bibfnamefont {S.}~\bibnamefont
  {{Mandal}}}, \bibinfo {author} {\bibfnamefont {S.}~\bibnamefont
  {{Majumdar}}}, \bibinfo {author} {\bibfnamefont {V.}~\bibnamefont
  {{Rentala}}}, \ and\ \bibinfo {author} {\bibfnamefont {R.~B.}\ \bibnamefont
  {{Thakur}}},\ }\href {\doibase 10.1103/PhysRevD.100.023002} {\bibfield
  {journal} {\bibinfo  {journal} {\prd}\ }\textbf {\bibinfo {volume} {100}},\
  \bibinfo {eid} {023002} (\bibinfo {year} {2019})},\ \Eprint
  {http://arxiv.org/abs/1806.06872} {arXiv:1806.06872 [hep-ph]} \BibitemShut
  {NoStop}%
\bibitem [{\citenamefont {{Lu}}\ \emph {et~al.}(2021)\citenamefont {{Lu}},
  \citenamefont {{Takhistov}}, \citenamefont {{Gelmini}}, \citenamefont
  {{Hayashi}}, \citenamefont {{Inoue}},\ and\ \citenamefont
  {{Kusenko}}}]{LuTak21}%
  \BibitemOpen
  \bibfield  {author} {\bibinfo {author} {\bibfnamefont {P.}~\bibnamefont
  {{Lu}}}, \bibinfo {author} {\bibfnamefont {V.}~\bibnamefont {{Takhistov}}},
  \bibinfo {author} {\bibfnamefont {G.~B.}\ \bibnamefont {{Gelmini}}}, \bibinfo
  {author} {\bibfnamefont {K.}~\bibnamefont {{Hayashi}}}, \bibinfo {author}
  {\bibfnamefont {Y.}~\bibnamefont {{Inoue}}}, \ and\ \bibinfo {author}
  {\bibfnamefont {A.}~\bibnamefont {{Kusenko}}},\ }\href {\doibase
  10.3847/2041-8213/abdcb6} {\bibfield  {journal} {\bibinfo  {journal} {\apjl}\
  }\textbf {\bibinfo {volume} {908}},\ \bibinfo {eid} {L23} (\bibinfo {year}
  {2021})},\ \Eprint {http://arxiv.org/abs/2007.02213} {arXiv:2007.02213
  [astro-ph.CO]} \BibitemShut {NoStop}%
\bibitem [{\citenamefont {{Kim}}(2020)}]{Kim20}%
  \BibitemOpen
  \bibfield  {author} {\bibinfo {author} {\bibfnamefont {H.}~\bibnamefont
  {{Kim}}},\ }\href@noop {} {\bibfield  {journal} {\bibinfo  {journal} {arXiv
  e-prints}\ ,\ \bibinfo {eid} {arXiv:2007.07739}} (\bibinfo {year} {2020})},\
  \Eprint {http://arxiv.org/abs/2007.07739} {arXiv:2007.07739 [hep-ph]}
  \BibitemShut {NoStop}%
\bibitem [{\citenamefont {{Laha}}\ \emph {et~al.}(2020)\citenamefont {{Laha}},
  \citenamefont {{Lu}},\ and\ \citenamefont {{Takhistov}}}]{LahLu20}%
  \BibitemOpen
  \bibfield  {author} {\bibinfo {author} {\bibfnamefont {R.}~\bibnamefont
  {{Laha}}}, \bibinfo {author} {\bibfnamefont {P.}~\bibnamefont {{Lu}}}, \ and\
  \bibinfo {author} {\bibfnamefont {V.}~\bibnamefont {{Takhistov}}},\
  }\href@noop {} {\bibfield  {journal} {\bibinfo  {journal} {arXiv e-prints}\
  ,\ \bibinfo {eid} {arXiv:2009.11837}} (\bibinfo {year} {2020})},\ \Eprint
  {http://arxiv.org/abs/2009.11837} {arXiv:2009.11837 [astro-ph.CO]}
  \BibitemShut {NoStop}%
\bibitem [{\citenamefont {{Takhistov}}\ \emph {et~al.}(2021)\citenamefont
  {{Takhistov}}, \citenamefont {{Lu}}, \citenamefont {{Gelmini}}, \citenamefont
  {{Hayashi}}, \citenamefont {{Inoue}},\ and\ \citenamefont
  {{Kusenko}}}]{TakLu21}%
  \BibitemOpen
  \bibfield  {author} {\bibinfo {author} {\bibfnamefont {V.}~\bibnamefont
  {{Takhistov}}}, \bibinfo {author} {\bibfnamefont {P.}~\bibnamefont {{Lu}}},
  \bibinfo {author} {\bibfnamefont {G.~B.}\ \bibnamefont {{Gelmini}}}, \bibinfo
  {author} {\bibfnamefont {K.}~\bibnamefont {{Hayashi}}}, \bibinfo {author}
  {\bibfnamefont {Y.}~\bibnamefont {{Inoue}}}, \ and\ \bibinfo {author}
  {\bibfnamefont {A.}~\bibnamefont {{Kusenko}}},\ }\href@noop {} {\bibfield
  {journal} {\bibinfo  {journal} {arXiv e-prints}\ ,\ \bibinfo {eid}
  {arXiv:2105.06099}} (\bibinfo {year} {2021})},\ \Eprint
  {http://arxiv.org/abs/2105.06099} {arXiv:2105.06099 [astro-ph.GA]}
  \BibitemShut {NoStop}%
\bibitem [{\citenamefont {{Wadekar}}\ \emph {et~al.}(2021)\citenamefont
  {{Wadekar}}, \citenamefont {{Farrar}},\ and\ \citenamefont
  {{Liu}}}]{WadFarinprep20}%
  \BibitemOpen
  \bibfield  {author} {\bibinfo {author} {\bibfnamefont {D.}~\bibnamefont
  {{Wadekar}}}, \bibinfo {author} {\bibfnamefont {G.~R.}\ \bibnamefont
  {{Farrar}}}, \ and\ \bibinfo {author} {\bibfnamefont {H.}~\bibnamefont
  {{Liu}}},\ }\href@noop {} {\bibfield  {journal} {\bibinfo  {journal} {in
  preparation}\ } (\bibinfo {year} {2021})}\BibitemShut {NoStop}%
\bibitem [{\citenamefont {{Begum}}\ \emph {et~al.}(2006)\citenamefont
  {{Begum}}, \citenamefont {{Chengalur}}, \citenamefont {{Karachentsev}},
  \citenamefont {{Kaisin}},\ and\ \citenamefont {{Sharina}}}]{BegChe06}%
  \BibitemOpen
  \bibfield  {author} {\bibinfo {author} {\bibfnamefont {A.}~\bibnamefont
  {{Begum}}}, \bibinfo {author} {\bibfnamefont {J.~N.}\ \bibnamefont
  {{Chengalur}}}, \bibinfo {author} {\bibfnamefont {I.~D.}\ \bibnamefont
  {{Karachentsev}}}, \bibinfo {author} {\bibfnamefont {S.~S.}\ \bibnamefont
  {{Kaisin}}}, \ and\ \bibinfo {author} {\bibfnamefont {M.~E.}\ \bibnamefont
  {{Sharina}}},\ }\href {\doibase 10.1111/j.1365-2966.2005.09817.x} {\bibfield
  {journal} {\bibinfo  {journal} {\mnras}\ }\textbf {\bibinfo {volume} {365}},\
  \bibinfo {pages} {1220} (\bibinfo {year} {2006})},\ \Eprint
  {http://arxiv.org/abs/astro-ph/0511253} {astro-ph/0511253} \BibitemShut
  {NoStop}%
\bibitem [{\citenamefont {{Patra}}(2018)}]{Patra18}%
  \BibitemOpen
  \bibfield  {author} {\bibinfo {author} {\bibfnamefont {N.~N.}\ \bibnamefont
  {{Patra}}},\ }\href {\doibase 10.1093/mnras/sty2167} {\bibfield  {journal}
  {\bibinfo  {journal} {\mnras}\ }\textbf {\bibinfo {volume} {480}},\ \bibinfo
  {pages} {4369} (\bibinfo {year} {2018})},\ \Eprint
  {http://arxiv.org/abs/1802.04478} {arXiv:1802.04478} \BibitemShut {NoStop}%
\bibitem [{\citenamefont {{Haardt}}\ and\ \citenamefont
  {{Madau}}(2012)}]{hm12}%
  \BibitemOpen
  \bibfield  {author} {\bibinfo {author} {\bibfnamefont {F.}~\bibnamefont
  {{Haardt}}}\ and\ \bibinfo {author} {\bibfnamefont {P.}~\bibnamefont
  {{Madau}}},\ }\href {\doibase 10.1088/0004-637X/746/2/125} {\bibfield
  {journal} {\bibinfo  {journal} {\apj}\ }\textbf {\bibinfo {volume} {746}},\
  \bibinfo {eid} {125} (\bibinfo {year} {2012})},\ \Eprint
  {http://arxiv.org/abs/1105.2039} {arXiv:1105.2039} \BibitemShut {NoStop}%
\bibitem [{\citenamefont {{Strigari}}\ \emph {et~al.}(2008)\citenamefont
  {{Strigari}}, \citenamefont {{Bullock}}, \citenamefont {{Kaplinghat}},
  \citenamefont {{Simon}}, \citenamefont {{Geha}}, \citenamefont {{Willman}},\
  and\ \citenamefont {{Walker}}}]{StrBulKap08}%
  \BibitemOpen
  \bibfield  {author} {\bibinfo {author} {\bibfnamefont {L.~E.}\ \bibnamefont
  {{Strigari}}}, \bibinfo {author} {\bibfnamefont {J.~S.}\ \bibnamefont
  {{Bullock}}}, \bibinfo {author} {\bibfnamefont {M.}~\bibnamefont
  {{Kaplinghat}}}, \bibinfo {author} {\bibfnamefont {J.~D.}\ \bibnamefont
  {{Simon}}}, \bibinfo {author} {\bibfnamefont {M.}~\bibnamefont {{Geha}}},
  \bibinfo {author} {\bibfnamefont {B.}~\bibnamefont {{Willman}}}, \ and\
  \bibinfo {author} {\bibfnamefont {M.~G.}\ \bibnamefont {{Walker}}},\ }\href
  {\doibase 10.1038/nature07222} {\bibfield  {journal} {\bibinfo  {journal}
  {\nat}\ }\textbf {\bibinfo {volume} {454}},\ \bibinfo {pages} {1096}
  (\bibinfo {year} {2008})},\ \Eprint {http://arxiv.org/abs/0808.3772}
  {arXiv:0808.3772 [astro-ph]} \BibitemShut {NoStop}%
\bibitem [{\citenamefont {{Nesti}}\ and\ \citenamefont
  {{Salucci}}(2013)}]{Nes13_BurkertFit}%
  \BibitemOpen
  \bibfield  {author} {\bibinfo {author} {\bibfnamefont {F.}~\bibnamefont
  {{Nesti}}}\ and\ \bibinfo {author} {\bibfnamefont {P.}~\bibnamefont
  {{Salucci}}},\ }\href {\doibase 10.1088/1475-7516/2013/07/016} {\bibfield
  {journal} {\bibinfo  {journal} {Journal of Cosmology and Astro-Particle
  Physics}\ }\textbf {\bibinfo {volume} {2013}},\ \bibinfo {eid} {016}
  (\bibinfo {year} {2013})},\ \Eprint {http://arxiv.org/abs/1304.5127}
  {arXiv:1304.5127 [astro-ph.GA]} \BibitemShut {NoStop}%
\bibitem [{\citenamefont {{Bovy}}(2015)}]{Galpy}%
  \BibitemOpen
  \bibfield  {author} {\bibinfo {author} {\bibfnamefont {J.}~\bibnamefont
  {{Bovy}}},\ }\href {\doibase 10.1088/0067-0049/216/2/29} {\bibfield
  {journal} {\bibinfo  {journal} {\apjs}\ }\textbf {\bibinfo {volume} {216}},\
  \bibinfo {eid} {29} (\bibinfo {year} {2015})},\ \Eprint
  {http://arxiv.org/abs/1412.3451} {arXiv:1412.3451} \BibitemShut {NoStop}%
\bibitem [{\citenamefont {{Hoeft}}\ \emph {et~al.}(2004)\citenamefont
  {{Hoeft}}, \citenamefont {{M{\"u}cket}},\ and\ \citenamefont
  {{Gottl{\"o}ber}}}]{Hoeft04_DMdispersion}%
  \BibitemOpen
  \bibfield  {author} {\bibinfo {author} {\bibfnamefont {M.}~\bibnamefont
  {{Hoeft}}}, \bibinfo {author} {\bibfnamefont {J.~P.}\ \bibnamefont
  {{M{\"u}cket}}}, \ and\ \bibinfo {author} {\bibfnamefont {S.}~\bibnamefont
  {{Gottl{\"o}ber}}},\ }\href {\doibase 10.1086/380990} {\bibfield  {journal}
  {\bibinfo  {journal} {\apj}\ }\textbf {\bibinfo {volume} {602}},\ \bibinfo
  {pages} {162} (\bibinfo {year} {2004})},\ \Eprint
  {http://arxiv.org/abs/astro-ph/0311083} {arXiv:astro-ph/0311083 [astro-ph]}
  \BibitemShut {NoStop}%
\bibitem [{\citenamefont {{McClure-Griffiths}}\ \emph
  {et~al.}(2012)\citenamefont {{McClure-Griffiths}}, \citenamefont {{Dickey}},
  \citenamefont {{Gaensler}}, \citenamefont {{Green}}, \citenamefont
  {{Green}},\ and\ \citenamefont {{Haverkorn}}}]{MCGLock12}%
  \BibitemOpen
  \bibfield  {author} {\bibinfo {author} {\bibfnamefont {N.~M.}\ \bibnamefont
  {{McClure-Griffiths}}}, \bibinfo {author} {\bibfnamefont {J.~M.}\
  \bibnamefont {{Dickey}}}, \bibinfo {author} {\bibfnamefont {B.~M.}\
  \bibnamefont {{Gaensler}}}, \bibinfo {author} {\bibfnamefont {A.~J.}\
  \bibnamefont {{Green}}}, \bibinfo {author} {\bibfnamefont {J.~A.}\
  \bibnamefont {{Green}}}, \ and\ \bibinfo {author} {\bibfnamefont
  {M.}~\bibnamefont {{Haverkorn}}},\ }\href {\doibase
  10.1088/0067-0049/199/1/12} {\bibfield  {journal} {\bibinfo  {journal}
  {\apjs}\ }\textbf {\bibinfo {volume} {199}},\ \bibinfo {eid} {12} (\bibinfo
  {year} {2012})},\ \bibinfo {note} {online data at
  \url{http://www.atnf.csiro.au/research/HI/sgps/GalacticCenter/Data.html}},\
  \Eprint {http://arxiv.org/abs/1201.2438} {arXiv:1201.2438} \BibitemShut
  {NoStop}%
\bibitem [{\citenamefont {{Farrar}}\ \emph
  {et~al.}(2020{\natexlab{a}})\citenamefont {{Farrar}}, \citenamefont
  {{Lockman}}, \citenamefont {{McClure-Griffiths}},\ and\ \citenamefont
  {{Wadekar}}}]{comment}%
  \BibitemOpen
  \bibfield  {author} {\bibinfo {author} {\bibfnamefont {G.}~\bibnamefont
  {{Farrar}}}, \bibinfo {author} {\bibfnamefont {F.~J.}\ \bibnamefont
  {{Lockman}}}, \bibinfo {author} {\bibfnamefont {N.~M.}\ \bibnamefont
  {{McClure-Griffiths}}}, \ and\ \bibinfo {author} {\bibfnamefont
  {D.}~\bibnamefont {{Wadekar}}},\ }\href {\doibase
  10.1103/PhysRevLett.124.029001} {\bibfield  {journal} {\bibinfo  {journal}
  {\prl}\ }\textbf {\bibinfo {volume} {124}},\ \bibinfo {eid} {029001}
  (\bibinfo {year} {2020}{\natexlab{a}})},\ \Eprint
  {http://arxiv.org/abs/1903.12191} {arXiv:1903.12191 [hep-ph]} \BibitemShut
  {NoStop}%
\bibitem [{\citenamefont {{Gronke}}\ and\ \citenamefont
  {{Oh}}(2018)}]{Gronke18_WindSim}%
  \BibitemOpen
  \bibfield  {author} {\bibinfo {author} {\bibfnamefont {M.}~\bibnamefont
  {{Gronke}}}\ and\ \bibinfo {author} {\bibfnamefont {S.~P.}\ \bibnamefont
  {{Oh}}},\ }\href {\doibase 10.1093/mnrasl/sly131} {\bibfield  {journal}
  {\bibinfo  {journal} {\mnras}\ }\textbf {\bibinfo {volume} {480}},\ \bibinfo
  {pages} {L111} (\bibinfo {year} {2018})},\ \Eprint
  {http://arxiv.org/abs/1806.02728} {arXiv:1806.02728} \BibitemShut {NoStop}%
\bibitem [{\citenamefont {{Sparre}}\ \emph {et~al.}(2019)\citenamefont
  {{Sparre}}, \citenamefont {{Pfrommer}},\ and\ \citenamefont
  {{Vogelsberger}}}]{Sparre18_WindSim}%
  \BibitemOpen
  \bibfield  {author} {\bibinfo {author} {\bibfnamefont {M.}~\bibnamefont
  {{Sparre}}}, \bibinfo {author} {\bibfnamefont {C.}~\bibnamefont
  {{Pfrommer}}}, \ and\ \bibinfo {author} {\bibfnamefont {M.}~\bibnamefont
  {{Vogelsberger}}},\ }\href {\doibase 10.1093/mnras/sty3063} {\bibfield
  {journal} {\bibinfo  {journal} {\mnras}\ }\textbf {\bibinfo {volume} {482}},\
  \bibinfo {pages} {5401} (\bibinfo {year} {2019})},\ \Eprint
  {http://arxiv.org/abs/1807.07971} {arXiv:1807.07971} \BibitemShut {NoStop}%
\bibitem [{\citenamefont {{Ferland}}\ \emph {et~al.}(2017)\citenamefont
  {{Ferland}}, \citenamefont {{Chatzikos}}, \citenamefont {{Guzm{\'a}n}},
  \citenamefont {{Lykins}}, \citenamefont {{van Hoof}}, \citenamefont
  {{Williams}}, \citenamefont {{Abel}}, \citenamefont {{Badnell}},
  \citenamefont {{Keenan}}, \citenamefont {{Porter}},\ and\ \citenamefont
  {{Stancil}}}]{cloudy}%
  \BibitemOpen
  \bibfield  {author} {\bibinfo {author} {\bibfnamefont {G.~J.}\ \bibnamefont
  {{Ferland}}}, \bibinfo {author} {\bibfnamefont {M.}~\bibnamefont
  {{Chatzikos}}}, \bibinfo {author} {\bibfnamefont {F.}~\bibnamefont
  {{Guzm{\'a}n}}}, \bibinfo {author} {\bibfnamefont {M.~L.}\ \bibnamefont
  {{Lykins}}}, \bibinfo {author} {\bibfnamefont {P.~A.~M.}\ \bibnamefont {{van
  Hoof}}}, \bibinfo {author} {\bibfnamefont {R.~J.~R.}\ \bibnamefont
  {{Williams}}}, \bibinfo {author} {\bibfnamefont {N.~P.}\ \bibnamefont
  {{Abel}}}, \bibinfo {author} {\bibfnamefont {N.~R.}\ \bibnamefont
  {{Badnell}}}, \bibinfo {author} {\bibfnamefont {F.~P.}\ \bibnamefont
  {{Keenan}}}, \bibinfo {author} {\bibfnamefont {R.~L.}\ \bibnamefont
  {{Porter}}}, \ and\ \bibinfo {author} {\bibfnamefont {P.~C.}\ \bibnamefont
  {{Stancil}}},\ }\href@noop {} {\bibfield  {journal} {\bibinfo  {journal}
  {\rmxaa}\ }\textbf {\bibinfo {volume} {53}},\ \bibinfo {pages} {385}
  (\bibinfo {year} {2017})},\ \Eprint {http://arxiv.org/abs/1705.10877}
  {arXiv:1705.10877 [astro-ph.GA]} \BibitemShut {NoStop}%
\bibitem [{\citenamefont {{Rahmati}}\ \emph {et~al.}(2013)\citenamefont
  {{Rahmati}}, \citenamefont {{Pawlik}}, \citenamefont {{Rai{\v c}evi{\'c}}},\
  and\ \citenamefont {{Schaye}}}]{Rahmati13}%
  \BibitemOpen
  \bibfield  {author} {\bibinfo {author} {\bibfnamefont {A.}~\bibnamefont
  {{Rahmati}}}, \bibinfo {author} {\bibfnamefont {A.~H.}\ \bibnamefont
  {{Pawlik}}}, \bibinfo {author} {\bibfnamefont {M.}~\bibnamefont {{Rai{\v
  c}evi{\'c}}}}, \ and\ \bibinfo {author} {\bibfnamefont {J.}~\bibnamefont
  {{Schaye}}},\ }\href {\doibase 10.1093/mnras/stt066} {\bibfield  {journal}
  {\bibinfo  {journal} {\mnras}\ }\textbf {\bibinfo {volume} {430}},\ \bibinfo
  {pages} {2427} (\bibinfo {year} {2013})},\ \Eprint
  {http://arxiv.org/abs/1210.7808} {arXiv:1210.7808 [astro-ph.CO]} \BibitemShut
  {NoStop}%
\bibitem [{\citenamefont {{Wolfire}}\ \emph {et~al.}(1995)\citenamefont
  {{Wolfire}}, \citenamefont {{Hollenbach}}, \citenamefont {{McKee}},
  \citenamefont {{Tielens}},\ and\ \citenamefont {{Bakes}}}]{WolMck95}%
  \BibitemOpen
  \bibfield  {author} {\bibinfo {author} {\bibfnamefont {M.~G.}\ \bibnamefont
  {{Wolfire}}}, \bibinfo {author} {\bibfnamefont {D.}~\bibnamefont
  {{Hollenbach}}}, \bibinfo {author} {\bibfnamefont {C.~F.}\ \bibnamefont
  {{McKee}}}, \bibinfo {author} {\bibfnamefont {A.~G.~G.~M.}\ \bibnamefont
  {{Tielens}}}, \ and\ \bibinfo {author} {\bibfnamefont {E.~L.~O.}\
  \bibnamefont {{Bakes}}},\ }\href {\doibase 10.1086/175510} {\bibfield
  {journal} {\bibinfo  {journal} {\apj}\ }\textbf {\bibinfo {volume} {443}},\
  \bibinfo {pages} {152} (\bibinfo {year} {1995})}\BibitemShut {NoStop}%
\bibitem [{\citenamefont {{Wolfire}}\ \emph {et~al.}(2003)\citenamefont
  {{Wolfire}}, \citenamefont {{McKee}}, \citenamefont {{Hollenbach}},\ and\
  \citenamefont {{Tielens}}}]{WolMckee03}%
  \BibitemOpen
  \bibfield  {author} {\bibinfo {author} {\bibfnamefont {M.~G.}\ \bibnamefont
  {{Wolfire}}}, \bibinfo {author} {\bibfnamefont {C.~F.}\ \bibnamefont
  {{McKee}}}, \bibinfo {author} {\bibfnamefont {D.}~\bibnamefont
  {{Hollenbach}}}, \ and\ \bibinfo {author} {\bibfnamefont {A.~G.~G.~M.}\
  \bibnamefont {{Tielens}}},\ }\href {\doibase 10.1086/368016} {\bibfield
  {journal} {\bibinfo  {journal} {\apj}\ }\textbf {\bibinfo {volume} {587}},\
  \bibinfo {pages} {278} (\bibinfo {year} {2003})},\ \Eprint
  {http://arxiv.org/abs/astro-ph/0207098} {arXiv:astro-ph/0207098 [astro-ph]}
  \BibitemShut {NoStop}%
\bibitem [{\citenamefont {{Madau}}\ and\ \citenamefont
  {{Efstathiou}}(1999)}]{MadEfs99}%
  \BibitemOpen
  \bibfield  {author} {\bibinfo {author} {\bibfnamefont {P.}~\bibnamefont
  {{Madau}}}\ and\ \bibinfo {author} {\bibfnamefont {G.}~\bibnamefont
  {{Efstathiou}}},\ }\href {\doibase 10.1086/312022} {\bibfield  {journal}
  {\bibinfo  {journal} {\apjl}\ }\textbf {\bibinfo {volume} {517}},\ \bibinfo
  {pages} {L9} (\bibinfo {year} {1999})},\ \Eprint
  {http://arxiv.org/abs/astro-ph/9902080} {arXiv:astro-ph/9902080 [astro-ph]}
  \BibitemShut {NoStop}%
\bibitem [{\citenamefont {{Cantalupo}}(2010)}]{Can10}%
  \BibitemOpen
  \bibfield  {author} {\bibinfo {author} {\bibfnamefont {S.}~\bibnamefont
  {{Cantalupo}}},\ }\href {\doibase 10.1111/j.1745-3933.2010.00806.x}
  {\bibfield  {journal} {\bibinfo  {journal} {\mnras}\ }\textbf {\bibinfo
  {volume} {403}},\ \bibinfo {pages} {L16} (\bibinfo {year} {2010})},\ \Eprint
  {http://arxiv.org/abs/0912.4149} {arXiv:0912.4149} \BibitemShut {NoStop}%
\bibitem [{\citenamefont {{Bruzual}}\ and\ \citenamefont
  {{Charlot}}(2003)}]{BruChar03}%
  \BibitemOpen
  \bibfield  {author} {\bibinfo {author} {\bibfnamefont {G.}~\bibnamefont
  {{Bruzual}}}\ and\ \bibinfo {author} {\bibfnamefont {S.}~\bibnamefont
  {{Charlot}}},\ }\href {\doibase 10.1046/j.1365-8711.2003.06897.x} {\bibfield
  {journal} {\bibinfo  {journal} {\mnras}\ }\textbf {\bibinfo {volume} {344}},\
  \bibinfo {pages} {1000} (\bibinfo {year} {2003})},\ \Eprint
  {http://arxiv.org/abs/astro-ph/0309134} {astro-ph/0309134} \BibitemShut
  {NoStop}%
\bibitem [{\citenamefont {Kannan}\ \emph {et~al.}(2014)\citenamefont {Kannan}
  \emph {et~al.}}]{KanMac14}%
  \BibitemOpen
  \bibfield  {author} {\bibinfo {author} {\bibfnamefont {R.}~\bibnamefont
  {Kannan}} \emph {et~al.},\ }\href {\doibase 10.1093/mnras/stt2098} {\bibfield
   {journal} {\bibinfo  {journal} {Mon. Not. Roy. Astron. Soc.}\ }\textbf
  {\bibinfo {volume} {437}},\ \bibinfo {pages} {2882} (\bibinfo {year}
  {2014})},\ \Eprint {http://arxiv.org/abs/1310.6748} {arXiv:1310.6748
  [astro-ph.GA]} \BibitemShut {NoStop}%
\bibitem [{\citenamefont {{Dvorkin}}\ \emph {et~al.}(2019)\citenamefont
  {{Dvorkin}}, \citenamefont {{Lin}},\ and\ \citenamefont
  {{Schutz}}}]{DvoLinSch19}%
  \BibitemOpen
  \bibfield  {author} {\bibinfo {author} {\bibfnamefont {C.}~\bibnamefont
  {{Dvorkin}}}, \bibinfo {author} {\bibfnamefont {T.}~\bibnamefont {{Lin}}}, \
  and\ \bibinfo {author} {\bibfnamefont {K.}~\bibnamefont {{Schutz}}},\
  }\href@noop {} {\bibfield  {journal} {\bibinfo  {journal} {arXiv e-prints}\
  ,\ \bibinfo {eid} {arXiv:1902.08623}} (\bibinfo {year} {2019})},\ \Eprint
  {http://arxiv.org/abs/1902.08623} {arXiv:1902.08623 [hep-ph]} \BibitemShut
  {NoStop}%
\bibitem [{\citenamefont {Raffelt}(1986)}]{Raffelt86}%
  \BibitemOpen
  \bibfield  {author} {\bibinfo {author} {\bibfnamefont {G.~G.}\ \bibnamefont
  {Raffelt}},\ }\href {\doibase 10.1103/PhysRevD.33.897} {\bibfield  {journal}
  {\bibinfo  {journal} {Phys. Rev. D}\ }\textbf {\bibinfo {volume} {33}},\
  \bibinfo {pages} {897} (\bibinfo {year} {1986})}\BibitemShut {NoStop}%
\bibitem [{\citenamefont {{Kouvaris}}\ and\ \citenamefont
  {{Shoemaker}}(2014)}]{KouShoe14}%
  \BibitemOpen
  \bibfield  {author} {\bibinfo {author} {\bibfnamefont {C.}~\bibnamefont
  {{Kouvaris}}}\ and\ \bibinfo {author} {\bibfnamefont {I.~M.}\ \bibnamefont
  {{Shoemaker}}},\ }\href {\doibase 10.1103/PhysRevD.90.095011} {\bibfield
  {journal} {\bibinfo  {journal} {\prd}\ }\textbf {\bibinfo {volume} {90}},\
  \bibinfo {eid} {095011} (\bibinfo {year} {2014})},\ \Eprint
  {http://arxiv.org/abs/1405.1729} {arXiv:1405.1729 [hep-ph]} \BibitemShut
  {NoStop}%
\bibitem [{\citenamefont {{Sigurdson}}\ \emph {et~al.}(2004)\citenamefont
  {{Sigurdson}}, \citenamefont {{Doran}}, \citenamefont {{Kurylov}},
  \citenamefont {{Caldwell}},\ and\ \citenamefont
  {{Kamionkowski}}}]{sigurdson_DMdipole}%
  \BibitemOpen
  \bibfield  {author} {\bibinfo {author} {\bibfnamefont {K.}~\bibnamefont
  {{Sigurdson}}}, \bibinfo {author} {\bibfnamefont {M.}~\bibnamefont
  {{Doran}}}, \bibinfo {author} {\bibfnamefont {A.}~\bibnamefont {{Kurylov}}},
  \bibinfo {author} {\bibfnamefont {R.~R.}\ \bibnamefont {{Caldwell}}}, \ and\
  \bibinfo {author} {\bibfnamefont {M.}~\bibnamefont {{Kamionkowski}}},\ }\href
  {\doibase 10.1103/PhysRevD.70.083501} {\bibfield  {journal} {\bibinfo
  {journal} {\prd}\ }\textbf {\bibinfo {volume} {70}},\ \bibinfo {eid} {083501}
  (\bibinfo {year} {2004})},\ \Eprint {http://arxiv.org/abs/astro-ph/0406355}
  {astro-ph/0406355} \BibitemShut {NoStop}%
\bibitem [{\citenamefont {{Tulin}}\ \emph {et~al.}(2013)\citenamefont
  {{Tulin}}, \citenamefont {{Yu}},\ and\ \citenamefont {{Zurek}}}]{TulYuZur13}%
  \BibitemOpen
  \bibfield  {author} {\bibinfo {author} {\bibfnamefont {S.}~\bibnamefont
  {{Tulin}}}, \bibinfo {author} {\bibfnamefont {H.-B.}\ \bibnamefont {{Yu}}}, \
  and\ \bibinfo {author} {\bibfnamefont {K.~M.}\ \bibnamefont {{Zurek}}},\
  }\href {\doibase 10.1103/PhysRevD.87.115007} {\bibfield  {journal} {\bibinfo
  {journal} {\prd}\ }\textbf {\bibinfo {volume} {87}},\ \bibinfo {eid} {115007}
  (\bibinfo {year} {2013})},\ \Eprint {http://arxiv.org/abs/1302.3898}
  {arXiv:1302.3898 [hep-ph]} \BibitemShut {NoStop}%
\bibitem [{\citenamefont {Khrapak}\ \emph {et~al.}(2003)\citenamefont
  {Khrapak}, \citenamefont {Ivlev}, \citenamefont {Morfill},\ and\
  \citenamefont {Zhdanov}}]{Khr03}%
  \BibitemOpen
  \bibfield  {author} {\bibinfo {author} {\bibfnamefont {S.~A.}\ \bibnamefont
  {Khrapak}}, \bibinfo {author} {\bibfnamefont {A.~V.}\ \bibnamefont {Ivlev}},
  \bibinfo {author} {\bibfnamefont {G.~E.}\ \bibnamefont {Morfill}}, \ and\
  \bibinfo {author} {\bibfnamefont {S.~K.}\ \bibnamefont {Zhdanov}},\ }\href
  {\doibase 10.1103/PhysRevLett.90.225002} {\bibfield  {journal} {\bibinfo
  {journal} {Phys. Rev. Lett.}\ }\textbf {\bibinfo {volume} {90}},\ \bibinfo
  {pages} {225002} (\bibinfo {year} {2003})}\BibitemShut {NoStop}%
\bibitem [{\citenamefont {{Farrar}}\ \emph
  {et~al.}(2020{\natexlab{b}})\citenamefont {{Farrar}}, \citenamefont
  {{Wang}},\ and\ \citenamefont {{Xu}}}]{FarWanXu20}%
  \BibitemOpen
  \bibfield  {author} {\bibinfo {author} {\bibfnamefont {G.~R.}\ \bibnamefont
  {{Farrar}}}, \bibinfo {author} {\bibfnamefont {Z.}~\bibnamefont {{Wang}}}, \
  and\ \bibinfo {author} {\bibfnamefont {X.}~\bibnamefont {{Xu}}},\ }\href@noop
  {} {\bibfield  {journal} {\bibinfo  {journal} {arXiv e-prints}\ ,\ \bibinfo
  {eid} {arXiv:2007.10378}} (\bibinfo {year} {2020}{\natexlab{b}})},\ \Eprint
  {http://arxiv.org/abs/2007.10378} {arXiv:2007.10378 [hep-ph]} \BibitemShut
  {NoStop}%
\bibitem [{\citenamefont {Helm}(1956)}]{Helm56}%
  \BibitemOpen
  \bibfield  {author} {\bibinfo {author} {\bibfnamefont {R.~H.}\ \bibnamefont
  {Helm}},\ }\href {\doibase 10.1103/PhysRev.104.1466} {\bibfield  {journal}
  {\bibinfo  {journal} {Phys. Rev.}\ }\textbf {\bibinfo {volume} {104}},\
  \bibinfo {pages} {1466} (\bibinfo {year} {1956})}\BibitemShut {NoStop}%
\bibitem [{\citenamefont {{Ali-Ha{\"\i}moud}}\ \emph
  {et~al.}(2015)\citenamefont {{Ali-Ha{\"\i}moud}}, \citenamefont {{Chluba}},\
  and\ \citenamefont {{Kamionkowski}}}]{HaiChlKam15}%
  \BibitemOpen
  \bibfield  {author} {\bibinfo {author} {\bibfnamefont {Y.}~\bibnamefont
  {{Ali-Ha{\"\i}moud}}}, \bibinfo {author} {\bibfnamefont {J.}~\bibnamefont
  {{Chluba}}}, \ and\ \bibinfo {author} {\bibfnamefont {M.}~\bibnamefont
  {{Kamionkowski}}},\ }\href {\doibase 10.1103/PhysRevLett.115.071304}
  {\bibfield  {journal} {\bibinfo  {journal} {\prl}\ }\textbf {\bibinfo
  {volume} {115}},\ \bibinfo {eid} {071304} (\bibinfo {year} {2015})},\ \Eprint
  {http://arxiv.org/abs/1506.04745} {arXiv:1506.04745 [astro-ph.CO]}
  \BibitemShut {NoStop}%
\bibitem [{\citenamefont {{Cappiello}}\ \emph {et~al.}(2018)\citenamefont
  {{Cappiello}}, \citenamefont {{Ng}},\ and\ \citenamefont
  {{Beacom}}}]{CapNgBea18}%
  \BibitemOpen
  \bibfield  {author} {\bibinfo {author} {\bibfnamefont {C.~V.}\ \bibnamefont
  {{Cappiello}}}, \bibinfo {author} {\bibfnamefont {K.~C.~Y.}\ \bibnamefont
  {{Ng}}}, \ and\ \bibinfo {author} {\bibfnamefont {J.~F.}\ \bibnamefont
  {{Beacom}}},\ }\href@noop {} {\bibfield  {journal} {\bibinfo  {journal}
  {arXiv e-prints}\ ,\ \bibinfo {eid} {arXiv:1810.07705}} (\bibinfo {year}
  {2018})},\ \Eprint {http://arxiv.org/abs/1810.07705} {arXiv:1810.07705
  [hep-ph]} \BibitemShut {NoStop}%
\bibitem [{\citenamefont {{Lodders}}(2010)}]{solarAbun_Lod10}%
  \BibitemOpen
  \bibfield  {author} {\bibinfo {author} {\bibfnamefont {K.}~\bibnamefont
  {{Lodders}}},\ }\href {\doibase 10.1007/978-3-642-10352-0_8} {\bibfield
  {journal} {\bibinfo  {journal} {Astrophysics and Space Science Proceedings}\
  }\textbf {\bibinfo {volume} {16}},\ \bibinfo {pages} {379} (\bibinfo {year}
  {2010})},\ \Eprint {http://arxiv.org/abs/1010.2746} {arXiv:1010.2746
  [astro-ph.SR]} \BibitemShut {NoStop}%
\bibitem [{\citenamefont {{Mu{\~n}oz}}\ \emph {et~al.}(2015)\citenamefont
  {{Mu{\~n}oz}}, \citenamefont {{Kovetz}},\ and\ \citenamefont
  {{Ali-Ha{\"i}moud}}}]{MunYacine15}%
  \BibitemOpen
  \bibfield  {author} {\bibinfo {author} {\bibfnamefont {J.~B.}\ \bibnamefont
  {{Mu{\~n}oz}}}, \bibinfo {author} {\bibfnamefont {E.~D.}\ \bibnamefont
  {{Kovetz}}}, \ and\ \bibinfo {author} {\bibfnamefont {Y.}~\bibnamefont
  {{Ali-Ha{\"i}moud}}},\ }\href {\doibase 10.1103/PhysRevD.92.083528}
  {\bibfield  {journal} {\bibinfo  {journal} {\prd}\ }\textbf {\bibinfo
  {volume} {92}},\ \bibinfo {eid} {083528} (\bibinfo {year} {2015})},\ \Eprint
  {http://arxiv.org/abs/1509.00029} {arXiv:1509.00029} \BibitemShut {NoStop}%
\end{thebibliography}%

\end{document}